\documentclass[fleqn,usenatbib]{mnras}
\usepackage{newtxtext,newtxmath}
\usepackage{mathtools} 
\usepackage{amsmath}
\usepackage{comment}
\usepackage{mathastext}
\usepackage{placeins}
\usepackage{longtable}
\usepackage{subcaption}
\usepackage{textcomp}
\usepackage{hyperref}
\usepackage[normalem]{ulem}
 \usepackage[flushleft]{threeparttable}
\usepackage{float}
\usepackage[T1]{fontenc}
\DeclareRobustCommand{\VAN}[3]{#2}
\let\VANthebibliography\thebibliography
\def\thebibliography{\DeclareRobustCommand{\VAN}[3]{##3}\VANthebibliography}

\usepackage{graphicx}	% Including figure files
\usepackage{microtype}
\usepackage{mathtools}
\usepackage[dvipsnames]{xcolor}
\newcommand{\usmg}{USMg\textsc{ii} }

\newcommand{\zem}{$z_{em}$ }
\newcommand{\zabs}{$z_{abs}$}
\newcommand{\kms}{$km s^{-1}$}

\newcommand{\HI}{\mbox{H\,{\sc i}}}

\newcommand{\OII}{\mbox{[O\,{\sc ii}]}}
\newcommand{\OIII}{\mbox{[O\,{\sc iii}]}}

\newcommand{\CI}{\mbox{C\,{\sc i}}}

\newcommand{\CIV}{\mbox{C\,{\sc iv}}}

\newcommand{\MgII}{\mbox{Mg\,{\sc ii}}}
\newcommand{\MgI}{\mbox{Mg\,{\sc i}}}

\newcommand{\ang}{\textup{\AA} }

\newcommand{\FeII}{\mbox{Fe\,{\sc ii}}}
\newcommand{\MnII}{\mbox{Mn\,{\sc ii}}}
\newcommand{\CaII}{\mbox{Ca\,{\sc ii}}}
\newcommand{\BAGPIPES}{\textsc{BAGPIPES}}
\newcommand{\Magiicat}{\textsc{MAGIICAT}}

\newcommand{\RD}[1]{{\color{purple}[RD: #1]}}
\title[\usmg galaxies at $z\sim 0.5$]{Host Galaxies of Ultra Strong Mg \textsc{ii} absorbers at z $\sim$ 0.5}

\author[Guha et al.]{
Labanya Kumar Guha $^{1}$\thanks{E-mail: labanya@iucaa.in (LKG)},
Raghunathan Srianand $^{1}$, 
Rajeshwari Dutta$^{2,6}$,
Ravi Joshi $^{3}$, \newauthor{
Pasquier Noterdaeme $^{4,5}$ \&
Patrick Petitjean $^4$}
\\
\\
$^{1}$ IUCAA, Postbag 4, Ganeshkhind, Pune 411007, India\\
$^{2}$ Dipartimento di Fisica G. Occhialini, Università degli Studi di Milano-Bicocca, Piazza della Scienza 3, 20126 Milano, Italy\\
$^{3}$ Indian Institute of Astrophysics, Koramangala II Block, Bangalore 560 034, India.\\
$^4$ Institut d'Astrophysique de Paris, Sorbonne Universit\'es and CNRS, 98bis boulevard Arago, 75014 Paris, France\\
$^5$ Franco-Chilean Laboratory for Astronomy, IRL3386, Camino el Observatorio 1515, Las Condes, Santiago, Chile\\
$^6$ INAF - Osservatorio Astronomico di Brera, via Bianchi 46, 23087 Merate (LC), Italy
}
\date{Accepted XXX. Received YYY; in original form ZZZ}

\pubyear{2021}

\begin{document}
\label{firstpage}
\pagerange{\pageref{firstpage}--\pageref{lastpage}}
\maketitle

%\iffalse
\begin{abstract}
{From a sample of 109
candidate Ultra-Strong Mg~{\sc ii} (USMg{\sc ii}; having rest equivalent width of Mg~{\sc ii}, $W_{2796}>3.0$\AA)
systems at z=0.4-0.6, we confirm 27 and identify host galaxies of 20 systems based on associated nebular line emission from our SALT observations or from SDSS fiber spectra. The measured impact parameter, \OII\ luminosity, star formation rate, B-band luminosity and stellar mass are in the ranges $7.3\le D[kpc]\le79$, $0.2\le L_{[O~\textsc{ii}]}[ 10^{41}~erg s^{-1}]$ $\le 4.5$, $2.59\le SFR[M_\odot yr^{-1} ]\le 33.51$, $0.15L_B^*\le L_B\le1.63L_B^*$ and $10.21\le log[M_*/M_\odot]\le11.62$ respectively}. The impact parameters found are larger than that predicted by the $W_{2796}$ vs D relationship of the general population of Mg~{\sc ii} absorbers.  At a given D, \usmg\  host galaxies are more luminous and massive compared to typical Mg~{\sc ii} absorbers. However, the measured SFRs are slightly lower than that of main-sequence galaxies with same M$_\star$ at $z\sim0.5$. We report a correlation between $L_{[O~\textsc{ii}]}$ and W$_{2796}$ for the full population of Mg~{\sc ii} absorbers, driven mainly by the host galaxies of weak Mg~{\sc ii} absorbers that tend to have low $L_{[O~\textsc{ii}]}$ and large impact parameters.
{We find at least $\sim$33\% of the \usmg host galaxies (with a limiting magnitude of $m_r<23.6$) are isolated and the large $W_{2796}$ in these cases may originate from 
gas flows (infall/outflow) in single halos of massive but not starburst galaxies. We also find galaxy interactions could be responsible for large velocity widths in at least $\sim$17\% cases.}
\end{abstract}

\begin{keywords}
galaxies: evolution; galaxies: groups: general; galaxies: high-redshift; galaxies: haloes; quasars: absorption lines
\end{keywords}
%\fi

%%%%%%%%%%%%%%%%%%%%%%%%%%%%%%%%%%%%%%%%%%%%%%%%%%
%    $\int = 25$
%%%%%%%%%%%%%%%%% BODY OF PAPER %%%%%%%%%%%%%%%%%%

\section{Introduction}

Our current understanding of galaxy formation and evolution is based on the model known as the ‘cosmic baryon cycle’, according to which, galaxies evolve by means of a slowly varying equilibrium between inflows from the intergalactic medium (IGM), high velocity outflows from the galaxy and the in-situ star-formation taking place within the galaxy \citep{Angles-Alcazar, Peroux2020b}. Biconical galactic scale outflows (with velocities of 100-1000 $km\,s^{-1}$), probed by neutral or singly ionized species like Na~{\sc i}, Mg~{\sc ii} and Fe~{\sc ii} in absorption, are ubiquitous in high redshift (i.e. $0.5\le z\le 1.5$) galaxies \citep[e.g.][]{Tremonti2007,Chen_2010,Martin_2012,Rubin2014,Bordoloi2014}. The probability to detect a wind
is found to depend weakly on the intrinsic properties of the galaxies but strongly on the galaxy orientation.
Notwithstanding this, the importance of star formation rate (SFR) and stellar mass (M$_*$) of the host galaxy is 
reflected by the correlations observed between the maximum wind velocity ($v_{\rm max}$) and M$_*$, and {between} the equivalent width of the flow and SFR. {\it While the presence of a wind is well established in these galaxies, its location (important for deriving wind parameters) with respect to the stellar disk can not be constrained accurately.}

On the other hand, quasar absorption lines in principle allow us to probe the spatial distribution and kinematics of the gas in foreground galaxies at very small impact parameters (D), thereby allowing us to probe the nature of gas flows in these galaxies. Presence of cool circumgalactic medium (CGM) around galaxies out to projected distances of $\sim$200 kpc \citep[for example,][]{Bergeron1991, Steidel1995, Chen_2010, Nielsen_2013, Rubin2018} is well established and an anti-correlation between the rest equivalent width of Mg~{\sc ii}~$\lambda$2796 absorption (i.e. $W_{2796}$) and the impact parameter has been derived using spectra of distant quasars. While early models reproduced this correlation using photoionized halos around galaxies \citep[see for example,][]{Petitjean1992,Srianand_1994}, it was apparent that continuous gas flow is needed to sustain the observed covering fraction of gas over a long period.

This observed anti-correlation  between $W_{2796}$ and D allows us to select quasar and foreground galaxy pairs with small impact parameters using absorption systems with large $\rm{W_{2796}}$ \citep[see,][]{Bouche2007,bouche2012}. The \MgII\ absorption systems with $W_{2796}> 3$\ang\ are known as Ultra-Strong \MgII\ absorbers \citep[referred hereafter as \usmg;][]{Nestor2011}. They constitute only 0.8\% of the \MgII\ absorber population having $W_{2796}>0.02$ \AA. Observationally such large equivalent widths are seen in a very high fraction (i.e. 30-50\%) of (i) outflows detected in $z\sim$0.5 galaxies \citep{Rubin2014}; (ii) Milky Way sightlines that probe disk+halo gas \citep[after applying a factor 2 correction to match with QSO absorbers,][]{Savage2000}; (iii) Galaxy On Top Of Quasars  \citep[GOTOQs;][]{noterdaeme2010a, Straka_2015, Joshi2017, Joshi2018} and (iv) high-z C~{\sc i} absorbers \citep{Ledoux2015,Zou2018}. From figure 2 of \citet{Rao2017} it is also evident that more than 50\% of \usmg\ absorbers are damped Lyman-$\alpha$ systems (DLAs; neutral hydrogen column density, $N$(\HI) $\ge2\times10^{20}$\,cm$^{-2}$).

It is well known that the measured $W_{2796}$ using low dispersion spectra are related to the number of absorbing clouds and velocity dispersion between them, and not directly related to the column density \citep[][]{Petitjean1990}. For a fully saturated \MgII\ line, $W_{2796} \geqslant 3\ang$ would correspond to a minimum velocity width of 320 $\rm{km\,s^{-1}}$. Gas having such velocity spread usually have large metallicities \citep{Ledoux2006}. 
%The origin of 
Large velocity spread could originate from, (i) galactic-scale outflows \citep{bouche2006,weiner2009,Rubin_2012}, (ii) filamentary accretion onto galaxies \citep[][]{steidel2002,Chen_2010}, (iii) dynamical mergers \citep[][]{richter2012} and intra-group gas \citep{Rubin2010,Gauthier2013}. In such cases, measured metallicities and galaxy orientations with respect to the quasar sightlines are used to distinguish between the different possibilities \citep{Peroux2013, Kacprzak_2014, Zabl_2019, Bordoloi2011, Peroux2020a, Zabl2021}.

It is possible that absorption line-based selection of galaxies (unbiased by the galaxy luminosity) may pick a population that is  different from that observed through  galaxy surveys that rely on optical colours and emission line strengths. Large $W_{2796}$ systems, in particular the \usmg systems have been associated with host-galaxies that are going through a rapid star-formation episode or have been through a phase of rapid star-formation in the recent past, i.e, they are either star-burst or post star-burst galaxies \citep{Nestor2011}. By studying the {average} photometric properties of more than 2800 \MgII\ systems, \citet{Zibetti_2007} suggested that the stronger \MgII\ systems (with $W_{2796} \geq 1$\AA) are associated with actively star-forming galaxies. 
Based on the \OII\ emission associated to \MgII-selected systems, \citet{Noterdaeme2010} %\cite{Menard2011} 
found a strong correlation between $W_{2796}$ and \OII\ luminosity. In a similar work, \citet{Menard2011} interpreted this as a correlation between  $W_{2796}$ and SFR, but this interpretation was later shown to be vulnerable to fiber losses effects \citep{Lopez2012,Joshi2017}.

Discussions presented above suggests that \usmg\ absorbers are ideal targets for studying the gas flows at low impact parameters to star forming galaxies and/or interacting groups of galaxies. Studying such systems can provide important insights into the baryonic cycle that governs the galaxy evolution.
Motivated by this, we embarked on a detailed study  of \usmg\ systems at z$\sim$0.5 with the aim to (i) identify their host galaxies and characterize the galaxy environment around these absorbers, (ii) investigate whether we preferentially select a particular galaxy population using the \usmg selection, (iii) study the connection between the galaxy properties and that of the absorption features, and (iv) identify potential quasar-galaxy pairs where galactic outflow can be studied through down-the-barrel absorption towards the galaxy, and absorption along the quasar line-of-sight simultaneously.  This paper is organized as follows. In section \ref{sec:sample}, we discuss the sample. The observational setup, data reduction and calibrations are described in section \ref{sec:observations}. The identification of the \usmg host galaxies and the method of inferring their physical properties based on the available spectro-photometric data are discussed in section~\ref{sec:galaxy_props}.  In section \ref{sec:results}, we present the results of our analysis and the nature of the \usmg host galaxies based on the inferred galaxy properties. As we go along, to investigate whether the \usmg host galaxies are drawn from a specific galaxy population or not, we compare their properties against various low-z Mg~{\sc ii} and DLA absorber samples available in the literature. Wherever possible, we also compare the properties of our sample with those of high-z C~{\sc i} absorbers. In section \ref{sec:discuss}, we provide an overall discussion on the \usmg systems. Our conclusions are summarized in section~\ref{sec:conclusion}. Throughout this paper we assume a flat $\Lambda$CDM cosmology with $H_0 = 70\, \rm{km\, s^{-1}\, Mpc^{-1}}$ and $\Omega_{m,0} = 0.3$.

\begin{figure*}
    \begin{minipage}[t]{0.495\textwidth}
    \centering
    \includegraphics[viewport=5 10 400 400,width=0.9\textwidth,clip=true]{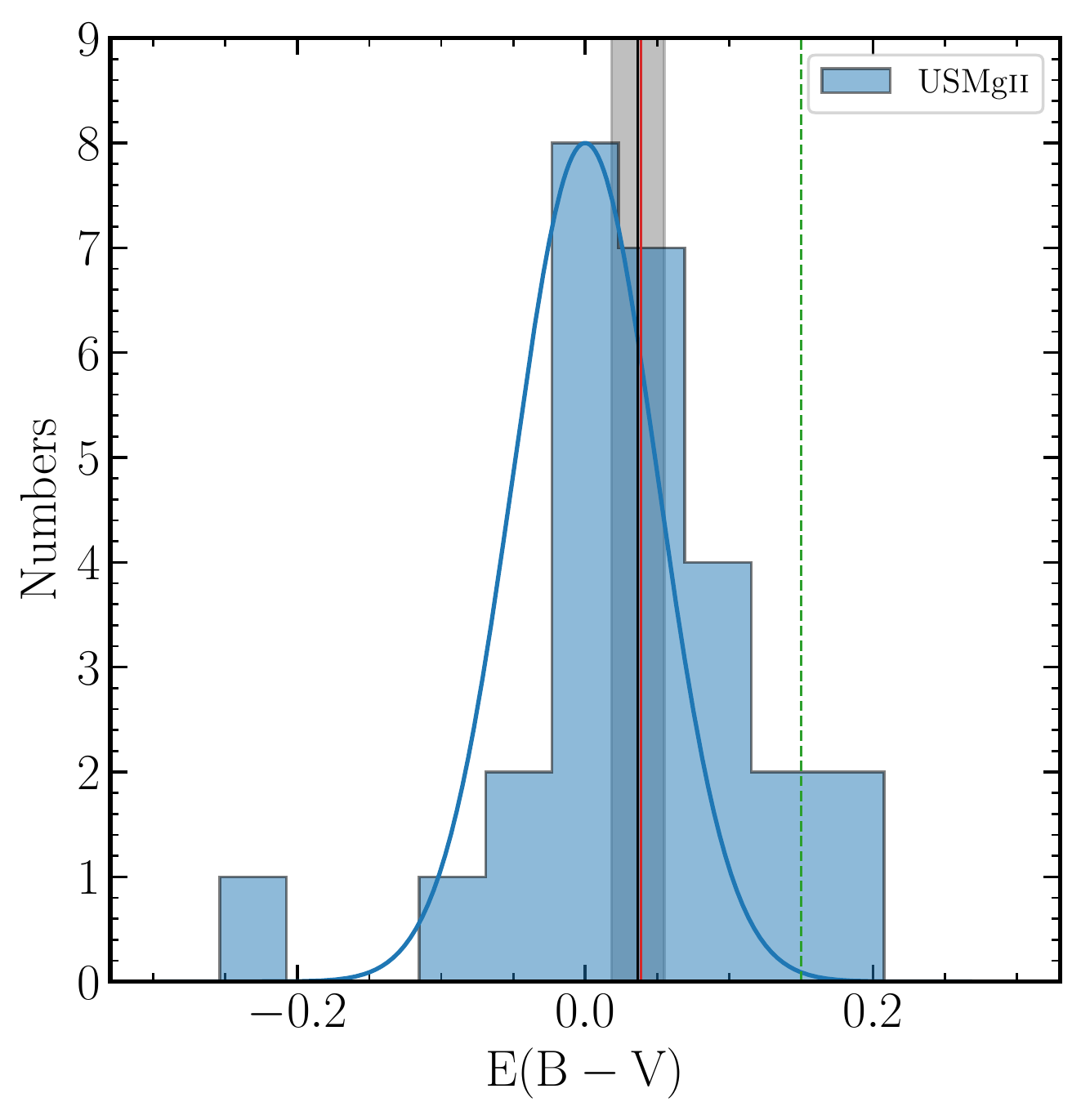}
    \end{minipage}
    %\hfill
    \begin{minipage}[t]{0.495\textwidth}
    \centering
    \includegraphics[viewport= 5 10 460 400, width=1.03\textwidth,clip=true]{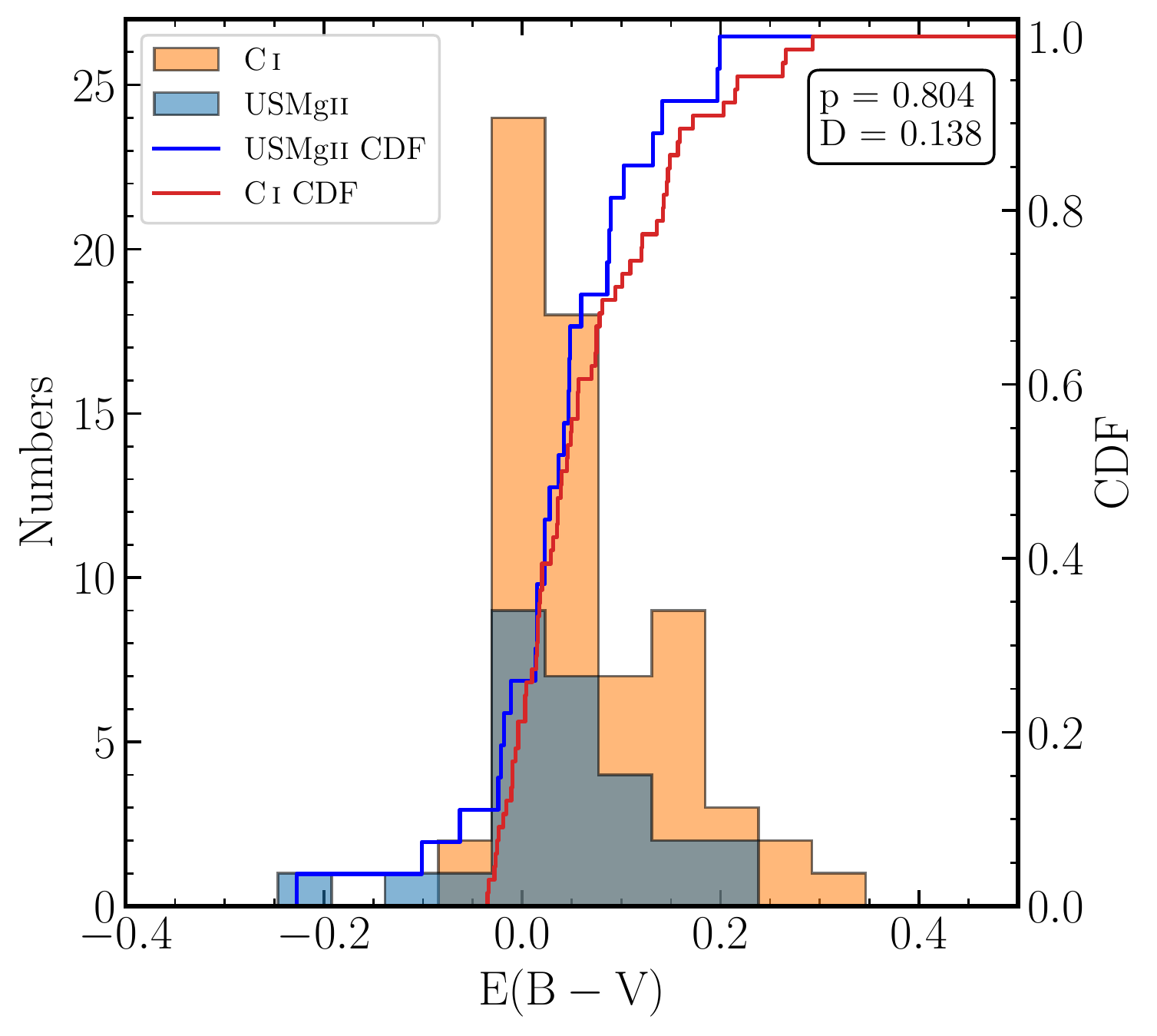}
    \end{minipage}
    \caption{
    {\it Left panel:} The histogram of the V band colour excess, E(B-V), for the \usmg systems in our sample. The deep blue Gaussian ($\mu = 0$, $\sigma = 0.050$) corresponds to the distribution of E(B-V) that is expected to arise out of the template matching procedure. We inferred this from the distribution of negative E(B-V) values in our sample (see text for more details). The vertical dashed green line corresponds to the $3\sigma$ above the mean of the distribution. The solid vertical black line corresponds to the expected colour excess for the $\rm{W_{2796} = 3}$\AA\ based on the relation obtained by \citet{Budzynski2011} with the gray region showing the $1\sigma$ uncertainty range. The vertical red line corresponds to the median color excess for our \usmg sample. {\it Right panel:} { Comparison of the distribution of colour excess}, E(B-V), between the \usmg\ absorbers (blue histogram) and \CI\ absorbers (orange histogram) taken from \citet{Ledoux2015}. The solid blue and red lines correspond to the cumulative distribution of E(B-V) for the \usmg\ and \CI\ absorbers, respectively.} 
    %\RD{suggest to keep same color for USMgII in both panels to avoid confusion}
    \label{fig:dust_fit}
\end{figure*}

\section{Our sample of \usmg\ systems at ${z \sim 0.5}$:}

\label{sec:sample}
We compiled a sample of \usmg systems that are accessible to the South African Large Telescope \citep[SALT,][]{buckley_2005} (i.e., declination, $\delta \leqslant +10^\circ$), using the Sloan Digital Sky Survey \citep[SDSS,][]{York2000} \MgII/\FeII\ absorber catalog of \citet{Zhu2013} in the redshift range $0.4 \leqslant$ \zabs\ $\leqslant 0.6$. The lower limit in \zabs\ was chosen so that the SDSS spectra could cover the Fe~{\sc ii}~$\lambda$2600 absorption line, while the upper limit on \zabs\ is set to be able to cover the  [O\textsc{ii}] $\lambda\lambda3727,3729$ doublet, H$\beta$ and [O\textsc{iii}] $\lambda\lambda4960,5008$ nebular emission lines associated with the \usmg absorption, in the wavelength range over which the Robert-Stobie Spectrograph (RSS) attached to SALT has good sensitivity. Application of these criteria has resulted in a total of 109 \usmg\ systems along the line-of-sight towards 106 different background quasars. A careful visual inspection of the SDSS spectrum of each of these quasars has led to the removal of 34 systems from our sample owing to the false identification of \MgII\ doublets. In 48 cases \CIV\  or Si~{\sc iv} Broad Absorption Lines (BAL) are misidentified as \MgII. After removing these we are left with a sample of 27 \usmg systems.

Details of all the 109 \usmg systems retrieved from the SDSS \MgII\ / \FeII\ absorber catalog of \citet{Zhu2013} are provided in Table~\ref{tab:append_1} of Appendix~\ref{sec:append_a}, where we have explicitly mentioned the reason for rejecting or accepting a \usmg\ system in our sample along with information such as the quasar and absorption redshifts. Among these 27 systems, the galaxy association for 3 \usmg systems have already been studied in the literature in some detail. One is the mutiband photometric study of the environment of the \usmg system J0240-0812 \citep{Nestor_2007}, and another is the spectro-photometric study of the \usmg system J2207-0901 \citep{Gauthier2013}. In addition, using the X-shooter spectrograph at the Very Large Telescope, \citet{Rahmani2016} have studied the galaxy associated with the \zabs = 0.5896 \usmg\ absorber, which is a known DLA. The \OII\ nebular emission from another three of our \usmg systems (i.e. \zabs = 0.5928 towards J0930+0018, \zabs = 0.5561 towards J1025-0046 and \zabs = 0.5519 towards J1216+0350) were detected in the SDSS spectrum and these systems are part of the GOTOQ sample of \citet{Joshi2017}.

In order to reconfirm that all the selected Mg~{\sc ii} systems are bona fide \usmg systems, we measured the rest equivalent widths of \MgII\ $\lambda\lambda 2796, 2803$ absorption lines by Gaussian fitting after approximating the observed continuum spectrum around these absorption lines with smooth polynomials. { Our Gaussian fits to absorption lines are shown in Figure~\ref{fig:abs_profiles} and \ref{fig:caii_absorption}}. During the fit, we kept the redshift and the velocity width same for both the lines of the Mg~{\sc ii} doublet. For the \usmg systems in the spectrum of J0334-0711, J0856+0740, J2045-0704 and, J1114-0023, even at SDSS resolution, the \MgII\ $\lambda\, 2796 $ absorption line itself splits into two Gaussian profiles. For these systems, we fit the \MgII\ $\lambda\lambda2796, 2803$ absorption profile with a pair of Gaussian doublets. For each pair, like before, we imposed the redshift and the velocity width for the individual Gaussian to be same. While the total equivalent width is greater than 3\AA, except for J1114-0023, individual components in these systems are not \usmg\ absorbers. For these systems, we quote the mean redshift of the pair of Gaussian doublets as the absorption redshift. Our analysis confirmed that all the 27 systems selected are indeed \usmg absorbers. Similarly, we also measured the rest equivalent width of the other associated absorption lines like \FeII\ $\lambda\, 2600$ and \MgI\ $\lambda\, 2853$ by fitting a single Gaussian { (also shown in Figure~\ref{fig:abs_profiles})}. The complete sample along with their absorption properties are presented in Table~\ref{tab:sample}. In this table, the quasar name, ${\rm z_{qso}}$ and \zabs\ are summarised in columns 2, 3 and 4 respectively. The next four columns provide rest equivalent widths of \MgII\ $\lambda 2796$, \MgII\ $\lambda2803$, \FeII\ $\lambda2600$ and \MgI\ $\lambda2853$ respectively. For the cases with detection significance level less than $3\sigma$, we provide the $3\sigma$ upper limits. For these cases, the $1\sigma$ limit is computed from the total uncertainty across the FWHM of the \MgII\ absorption line profile around the expected location of the absorption line of our interest. 

Next, we identified additional non-\usmg absorber in the SDSS spectra of quasars in our sample. We find 11 \MgII\ systems but none at 0.4$\le$\zabs$\le$0.6. We then tried to identify if any of the targeted potential host galaxies corresponds to the non-\usmg absorbers identified in the spectra of our sample. We found only one system at \zabs = 0.7981 ($\rm{W_{2796} = 2.14\pm0.18}$\AA) towards J1140-0023. We provide the details of this absorber in the last row of Table~\ref{tab:sample}.

\subsection{Line of sight reddening towards quasars} We calculate the reddening of the background quasar by the absorbing gas assuming the extinction curve of the gas to be similar to the extinction curves of the Small Magellanic Cloud (SMC), Large Magellanic Cloud (LMC), Large Magellanic Cloud Supershell (LMC2) and, our Milky Way (MW) galaxy \citep{Gordon_2003}. For the rest wavelength range considered here, the choice of different extinction curves are found to make no significant difference. We fitted all the quasar spectra in our sample with a standard quasar spectrum template \citep{Selsing2016} redshifted exactly to the quasar redshift and applied the SMC extinction at the redshift of the \usmg absorption, keeping the V band extinction coefficient, $A_V$, as the only free parameter, apart from a multiplicative scaling  \citep[as described in][]{srianand2008a}. The best fit V-band colour excess, E(B-V), for each system is listed in the last column of Table \ref{tab:sample}.

In Figure~\ref{fig:dust_fit}, the distribution of 
%excess, 
E(B-V) is shown in the left panel. For 6 \usmg systems, we find that the color excess, E(B-V) is negative. The system J1146-0237 has an extremely negative E(B-V) of $-0.227$, which is related to the fact that the quasar has weak emission lines and its spectrum  does not exactly follow the standard quasar template. We used the remaining cases with negative values of E(B-V) to compute the expected scatter from the spectral energy distribution (SED) fitting procedure. The deep blue Gaussian shown in Figure~\ref{fig:dust_fit} corresponds to the distribution of E(B-V)  (with a $\sigma = 0.050$) that is expected to arise from our template matching procedure. The vertical dashed green line corresponds to the $3\sigma$ limit of our E(B-V) measurements. We find only two \usmg systems present in our sample to have E(B-V) larger than this value. The highest E(B-V) of $\sim 0.2$ is measured for J0156+0343. As this is a C~{\sc iv} BAL quasar, it is not clear how much of the reddening is due to the \usmg\ absorber.

Using a sample of about 8300 strong \MgII\ absorption systems identified in the SDSS spectroscopic survey (DR6), \citet{Budzynski2011} have shown that $\rm{W_{2796}}$ strongly correlates with reddening of the background quasar spectra \citep[see also, ][]{york2006, Menard2008}  with  $\mathrm{E(B-V)} =[(8.0\pm3.0) \times 10^{-4}]\times(W_{2796})^{(3.48 \pm 0.3)}$. The solid vertical black line in the left panel of the Figure \ref{fig:dust_fit} corresponds to the expected $\rm{E(B-V)}$ for the $\rm{W_{2796}} = 3\AA$ based on this relationship. The solid vertical red line corresponds to the median color excess of our \usmg sample. 
Higher E(B-V) seen in our sample is consistent with their results.

In the right panel of Figure \ref{fig:dust_fit}, the blue and orange histograms correspond to the E(B-V) measurements for \usmg and high-$z$ \CI- selected absorbers from \citet{Ledoux2015} respectively. The solid blue and red lines correspond to the cumulative distribution function for the \usmg and \CI\ absorbers respectively. The median E(B-V) values for the \usmg and C~{\sc i} absorbers are 0.037 and 0.042 respectively. A two sample Kolmogorov-Smirnov (KS) test between these two samples indicates that the probability of these two samples to arise from the same parent population is $\sim$80\%. This is not surprising given the fact that a good fraction (i.e. $\sim$47\%) of C~{\sc i} absorbers have $\mathrm{W_{2796} > 3.0\AA}$  \citep[see table 2 of][]{Zou2018}.

\subsection{Detection of other metal lines}

Out of the 27 \usmg systems, six (along the line-of-sight of J0240$-$0812,  J1016$+$0752, J1025$-$0046, J1030$-$0132, J1216$+$0350 and J2301$-$0212) exhibit associated \MnII\ absorption having more than $\rm{3\sigma}$ significance { (see Figure~\ref{fig:abs_profiles})} with the rest equivalent width in the range, $0.25\le\rm{W_{2576}^{Mn\textsc{ii}}}(\text{\AA})\le1.32$,
while 10 systems (along the line-of-sight of J0218$-$0832, J0334$-$0711,  J1025$-$0046, J1214$+$0804, J1114$-$0023, J1216$+$0350, J2121$+$0039, J2127$+$0827, J2203$-$0022 and  J2301$-$0212) exhibit associated \CaII\ absorption { (see Figure~\ref{fig:caii_absorption})} with the rest equivalent width in the range ($0.18\le \rm{W_{3935}^{Ca\textsc{ii}}}(\text{\AA})\le1.08$)
in the SDSS spectra of the background quasars. In the case of non-detection, the $\rm{3\sigma}$, upper limits are typically in the range 0.5-0.9\ang. We do not find any significant correlation between $\mathrm{W^{CaII}_{3935}}$ and E(B-V). In addition, the distributions of measured E(B-V) of sightlines with and without Ca~{\sc ii} detected are not statistically different (KS test p-value = 0.23).

For the sake of comparison, in Figure~\ref{fig:wca2_vs_dust}, we plot E(B-V) vs Ca~{\sc ii} equivalent width in our sample with those found in low-z DLAs \citep{Wild2005}, high-z C~{\sc i} sample \citep{Zou2018} and our Galaxy \citep{Murga2015}. We do not detect associated Na~{\sc i}  absorption (at $>3\sigma$ level)
for any of the \usmg systems in our sample in the SDSS spectrum. The $\rm{3\sigma}$ upper limit on the rest equivalent width lies between 0.5\ang to 0.9\ang. As our SALT spectra cover limited wavelength range we could not use them for the above discussed absorption line searches.

\begin{table*}
     \begin{threeparttable}
     \caption{Details of our \usmg\ sample.
     Columns 2, 3 and 4 provide the quasar name, its redshift and the \usmg absorption redshift respectively. The next four columns (i.e. 5 to 8) provide the rest equivalent width of \MgII$\lambda\, 2796$, \MgII$\lambda\, 2803$,  \FeII $\lambda\, 2600$ and \MgI $\lambda\, 2853$ absorption respectively, where `$-$' signifies the absorption is outside of the wavelength coverage of the observed spectrum. In the case of non-detections we provide $3\sigma$ upper limits. Last column provides the measured E(B-V) derived using the SMC extinction curve.
     }
     \centering
     \begin{tabular}{lcccccccr}
     \hline
     No. & Quasar & $z_{qso}$  & $z_{abs}$  & $W_{2796}\,(\text{\AA})$  & $W_{2803}\,(\text{\AA})$ & $W_{2600}\,(\text{\AA})$ & $W_{2853}\,(\text{\AA})$ & $\rm{E(B-V)\, (SMC)}$ \\
       (1)&(2)&(3)&(4)&(5)&(6)&(7)&(8)&\multicolumn{1}{c}{(9)}\\
     \hline
     \hline
     \multicolumn{9}{c}{Targets with SALT spectrum}\\
      1& J000413.73-082625.4          & 2.247 & 0.5544  & $3.07\pm 0.16$    & $3.05\pm0.15$ & $2.05\pm0.15$  & $0.82\pm0.15$ & $0.015\pm0.002$\\
      2& J015635.18+034308.1          & 1.371 & 0.5581   & $4.20\pm 0.30$   & $3.76\pm0.26$ & $3.30\pm0.40$  & $1.56\pm0.26$ & $0.199\pm0.001$\\
      3& J021820.10-083259.4\tnote{a} & 1.219 & 0.5896  & $3.07\pm 0.11$    & $2.81\pm0.10$ & $2.15\pm0.14$  & $1.43\pm0.14$ & $0.022\pm0.001$\\  
      4& J024008.21-081223.4\tnote{b} & 2.230 & 0.5311  & $3.88\pm 0.17$    & $3.56\pm0.16$ & $2.01\pm0.15$  & $\leqslant 0.89$ & $0.015\pm 0.005$\\
      5& J033438.28-071149.0          & 0.634 & 0.5977  & $3.55\pm 0.23$    & $3.13\pm0.24$ & $1.89\pm 0.22$ & $0.34\pm0.01$ & $-0.011\pm0.001$\\
      6& J085627.09+074031.7          & 1.890 & 0.5232  & $4.14\pm 0.47$    & $3.19\pm0.61$ & $1.22\pm 0.20$ & $0.34\pm0.05$ & $0.197\pm0.001$ \\
      7& J092222.58+040858.7          & 0.713 & 0.4549  & $3.63\pm 0.21$    & $3.11\pm0.18$ & --             & $0.72\pm0.12$ & $0.049\pm 0.003$\\  
      8& J111400.00-002342.6          & 0.952 & 0.5610  & $5.75\pm 1.36$    & $4.62\pm1.33$ & $3.33\pm 0.95$ & $2.40\pm0.47$ & $0.102\pm0.004$\\
      9& J121453.29+080457.7          & 1.430 & 0.4908  & $3.38\pm 0.21$    & $2.97\pm0.19$ & $2.27\pm0.12$  & $\leqslant0.63$ & $0.089\pm 0.003$\\ 
      10& J121628.03+035031.8\tnote{d} & 0.996 & 0.5519  & $3.51\pm 0.12$    & $3.53\pm0.12$ & $2.89\pm0.18$ & $0.81\pm0.12$ & $-0.063\pm0.001$\\
      11& J155003.71+031325.0          & 1.789 & 0.5694  & $3.11\pm 0.08$    & $2.97\pm0.08$ & $1.82\pm0.15$ & $1.04\pm0.16$ & $0.141\pm0.004$\\
      12& J204501.32-070452.6          & 0.670 & 0.5649  & $3.77\pm 0.28$    & $3.12\pm0.32$ & $2.37\pm0.15$ & $1.21\pm0.38$ & $0.059\pm0.003$\\
      13& J210851.53-074726.5          & 1.486 & 0.5187  & $3.43\pm 0.17$    & $2.69\pm0.14$ & $2.24\pm0.24$ & $1.43\pm0.25$ & $0.088\pm0.003$\\
      14& J212143.98+003954.2          & 1.348 & 0.5509  & $3.44\pm 0.33$    & $3.57\pm0.34$ & $2.85\pm0.48$ & $1.60\pm0.24$ & $0.132\pm 0.003$\\    %          
      15& J212727.20+082724.6          & 0.745 & 0.4392  & $4.26\pm 0.16$    & $3.99\pm0.15$ & $3.75\pm0.16$ & $1.28\pm0.13$ & $0.042\pm0.001$\\  
      16& J220330.04-002211.4          & 1.782 & 0.4381  & $4.49\pm 0.31$    & $3.99\pm0.27$ & - & $0.82\pm0.13$ & $0.022\pm0.002$\\
      17& J220702.53-090127.7\tnote{c} & 1.296 & 0.5623  & $4.08\pm 0.18$    & $3.66\pm0.16$ & $2.93\pm0.32$ & $1.39\pm0.14$ & $0.047\pm0.002$\\  
      18& J230101.28-021200.0          & 0.619 & 0.5367  & $3.16\pm 0.07$    & $3.07\pm0.07$ & $2.20\pm0.12$ & $0.65\pm0.07$ & $0.086\pm0.001$\\  
      19& J232653.15+002142.9          & 2.190 & 0.5624  & $4.37\pm 0.46$    & $5.22\pm0.55$ & $4.29\pm0.76$ & $2.00\pm0.23$ & $0.013\pm0.007$\\  
      20& J233548.62-023734.3          & 1.234 & 0.5081  & $3.76\pm 0.10$    & $3.57\pm0.10$ & $2.54\pm0.11$ & $1.15\pm0.13$ & $0.047\pm0.002$ \\ % 
      21& J233818.25-005610.5          & 0.894 & 0.4801  & $3.05\pm 0.12$    & $2.74\pm0.10$ & $1.95\pm0.31$ & $0.51\pm0.08$ & $-0.018\pm0.003$\\

       %&  &   &   &   &   &   &  \\  
            \multicolumn{9}{c}{Targets without SALT spectrum}\\
        
      22& J093020.60+001828.0\tnote{d} & 2.4300 & 0.5928  & $3.48\pm 0.25$   & $3.58\pm 0.25$ & $2.90\pm0.22$   & $1.75\pm0.36$ &  $-0.101\pm0.004$\\
      23& J101610.82+075209.1 & 2.187  & 0.5961  & $3.09\pm 0.38$   & $3.62\pm 0.45$ & $2.04\pm0.20$   & $1.31\pm0.17$ & $0.028\pm0.006$\\  
      24& J102510.10-004644.9\tnote{d} & 2.212  & 0.5561  & $3.22\pm 0.11$   & $2.80\pm 0.10$ & $2.05\pm 0.08$& $0.97\pm 0.07$ & $0.037\pm0.001$ \\  
      25& J103059.75-013237.7 & 2.155  & 0.5783  & $3.27\pm 0.10$   & $3.20\pm 0.10$ & $2.39\pm 0.11$  & $0.49\pm 0.01$ & $-0.024\pm0.001$ \\  
      26& J110817.93+062833.0 & 1.202  & 0.5721  & $3.02\pm 0.16$  & $2.89\pm0.15$  & $1.72\pm0.33$ &  $0.49\pm0.16$ & $0.023\pm0.004$\\
      27& J114614.24-023716.1 & 2.207  & 0.5295  & $4.13\pm 0.22$   & $3.22\pm 0.17$  & $\leqslant{0.59}$  & $\leqslant{0.91}$  &  $-0.227\pm0.004$\\ 
       \multicolumn{9}{c}{The non-\usmg absorber in our lines of sight}\\     
      28& J111400.00-002342.6 & 0.952 & 0.7981  & $2.15\pm0.18$   & $1.88\pm0.16 $ & $1.37\pm0.16$  & $0.42\pm0.09 $ & $0.084\pm0.006$\\     
      \hline
     \end{tabular}
      \begin{tablenotes}
        \item[a] Host galaxy of this \usmg\ system was studied by \citet{Rahmani2016} using both spectroscopy and multiband photometry
        \item[b] The environment of this system was explored in \citet{Nestor_2007} using multiband photometry.
        \item[c] A group of 4 galaxies associated to this \usmg\
        was studied by \citet{Gauthier2013} using both spectroscopy and multiband photometry.
        \item[d] These systems are part of the GOTOQ sample of \citet{Joshi2017}.
    \end{tablenotes}
     \label{tab:sample}
          \end{threeparttable}
 \end{table*}

\begin{table*}
        \caption{Log of observations for our \usmg\ sample. 
        Systems marked with { *} represent non-detection of any associated host galaxy, while systems marked with { +} represent cases where the host galaxy sits on top of the quasar along the line of sight (see text for more details). {The last two columns correspond to the number of other candidate galaxies within 50 $\leqslant D[kpc] \leqslant$ 100 and D $\leqslant$ 50 kpc with $m_r \leqslant 22.5$ based on SDSS photometry.}
        }
            \centering
    \begin{tabular}{llcccccccc}
\hline
    No. & Quasar   &  Date & Exposure (s)  & Observation &  PA (deg.) & Grating     &  Wavelength   & Other candi-     & Total candi-\\
        &          &       &               &  Mode       &            & Angle (deg) &  Range (\ang) & date galaxies & date galaxies\\
        &          &       &               &             &            &             &               &  ($50\leqslant D[kpc] \leqslant100$)  &  ($D\leqslant 50$ kpc) \\
    (1) & \multicolumn{1}{c}{(2)} & (3) & (4) & (5) & (6) & (7) & (8) & (9) & (10) \\
    \hline
    \hline
     1 & J000413.73-082625.4   & 2017-07-18  & 2400  & Long-slit  & 80  &  17.75  & 5200 -- 8200 & 0 & 1\\
       &                      & 2017-09-17  & 2400 &  Long-slit  & 80  &  17.75  & 5200 -- 8200  &  & \\
     
     2 & J015635.18+034308.1   &  2015-11-07 & 2500  & Long-slit  & 350 &  17.00  & 4900 -- 7900 & 0 & 5\\
       &                      & 2015-11-13  & 2500 & Long-slit   & 350 &  17.00  & 4900 -- 7900 &  & \\
     
     3 & J021820.10-083259.4  &  2016-01-10 & 2500  & Long-slit  & 80  &  18.125 & 5300 -- 8300 & 0 &  1 \\
     
     4 & J024008.21-081223.4  & 2018-10-05  & 2500  & Long-slit  & 80  &  17.75  & 5200 -- 8200 & 0 & 1\\
     
     5 & J033438.28-071149.0  & 2017-11-19  & 2300  & MOS        & 44.8 & 17.75  & 5200 -- 8200 & 0 & 1\\
       &                      & 2017-12-11  & 2300  & MOS        & 44.8 & 17.75  & 5200 -- 8200 &   & \\
       
     6 & J085627.09+074031.7   & 2018-02-09  & 2200  & Long-slit  & 315 &  17.00  & 4900 -- 7900 & 0 & 1\\
       &                      & 2018-02-10  & 2200  & Long-slit  & 315 &  17.00  & 4900 -- 7900 &    &\\
       
     7 & J092222.58+040858.7 & 2016-02-08  & 2400  & Long-slit  & 0   &  16.625  & 4800 -- 7800 & 0 & 1\\
       &                      & 2016-03-01  & 2400  & Long-slit  & 0   &  16.625  & 4800 -- 7800 &  &   \\
     8 & J111400.00-002342.6     &  2021-04-15 & 2580  & Long-slit  & 0   &  18.125  & 5300 -- 8300 & 1  &  1\\
       
     9 & J121453.29+080457.7 & 2016-04-09  & 2300  & Long-slit  & 100 &  17.375  & 5050 -- 8050 & 1 & 1\\
       &                      & 2016-03-01  & 2160  & Long-slit  & 100 &  17.375  & 5050 -- 8050 &  & \\
       
     10& J121628.03+035031.8{ +}   & 2016-02-05  & 2400  &  Long-slit &   0 &  17.75  & 5200 -- 8200 & 0 & 1\\
       &                      & 2016-02-06  & 2400  &  Long-slit &  90 &  17.75  & 5200 -- 8200 &   &   \\
       
     11& J155003.71+031325.0  & 2015-06-19  & 2500  &  Long-slit & 96  &  16.625 & 4800 -- 7800 & 0   & 1\\
       &                      & 2015-07-14  & 2500  &  Long-slit & 96  &  18.125 & 5300 -- 8300 &  &\\
       
     12& J204501.32-070452.6{ *}  & 2017-08-18   & 2500   & Long-slit  & 77  &  18.125  & 5300 -- 8300 & 0 & 1 \\
     
     13& J210851.53-074726.5{ +}    & 2017-06-03   & 2400   & Long-slit  &  70 &  17.75  & 5200 -- 5300 & 0 & 1\\
       &                      & 2017-06-18   & 2400   &  Long-slit &  70 &  17.75  & 5200 -- 5300 &  & \\
       
     14& J212143.98+003954.2 &  2015-06-20  & 2340   &  Long-slit & 90  &  16.625 & 4800 -- 7800 & 0 & 1\\
       &                      & 2020-10-18   & 2200   &  Long-slit & 240 &  18.125  & 5300 -- 8300 &  & \\
       
     15& J212727.20+082724.6   & 2015-06-18   & 2500   & Long-slit  & 95  &  16.625  & 4800 -- 7800 & 0 & 1\\
       &                      & 2016-06-20   & 1844   &  Long-slit & 95  &  18.125  & 5300 -- 8300 &  & \\
       
     16& J220330.04-002211.4  & 2017-09-13   & 2500   & Long-slit  & 45  &  17.00   & 4900 -- 7900 & 0 & 1\\
     
     17& J220702.53-090127.7  & 2017-09-16   & 2400   &  Long-slit & 77  &  18.125  & 5300 -- 8300 & 1 & 1\\
     
     18& J230101.28-021200.0{ *}  & 2017-10-11   & 2500   &  Long-slit & 308 &  18.125  & 5300 -- 8300 & 0 & 0\\
       &                      & 2017-10-17   & 2500   &  Long-slit & 283 &  17.375  & 5050 -- 8050 &  & \\
       
     19& J232653.15+002142.9   & 2017-10-17   & 2500   &  Long-slit &  5  &  18.125  & 5300 -- 8300 & 0 & 1\\

     20& J233548.62-023734.3  & 2015-06-19   & 2271   & Long-slit  &  100 & 16.625  & 4800 -- 7800 & 0 & 1\\
       &                      & 2015-06-21   & 2500   &  Long-slit &  100 & 18.125  & 5300 -- 8300&  &\\
       
     21& J233818.25-005610.5  &  2018-08-21  & 2200   &  MOS       &  90  & 17.00   & 4900 -- 7900& 1 & 1\\
       &                      & 2018-09-11   & 2324   &  MOS       &  90  & 17.00   & 4900 -- 7900& \\
     \hline
    \end{tabular}
    \label{tab:observation_log}
\end{table*}

\section{SALT observations and data reduction}
\label{sec:observations}

We first identified all galaxies seen within an impact parameter of $ 50$ kpc from the quasar line of sight using the SDSS photometry (typically complete for  $m_r \leqslant 22.5$) and having consistent photometric redshifts  within $1\sigma$. Note that based on the known anti-correlation between D and $\mathrm{W_{2796}}$, we expect the isolated host galaxies to lie within D$\le$10 kpc. { Even with the scatter in the relation the observed D values are $<50$ kpc in the available literature data (see discussions in Section 5.4). Thus we mainly focus on observing candidate galaxies within this impact parameter.} We have completed observations for 21 absorbers out of 27 discussed in the previous section.

\begin{figure*}
    \begin{subfigure}{0.245\textwidth}
    \centering\includegraphics[width=\textwidth]{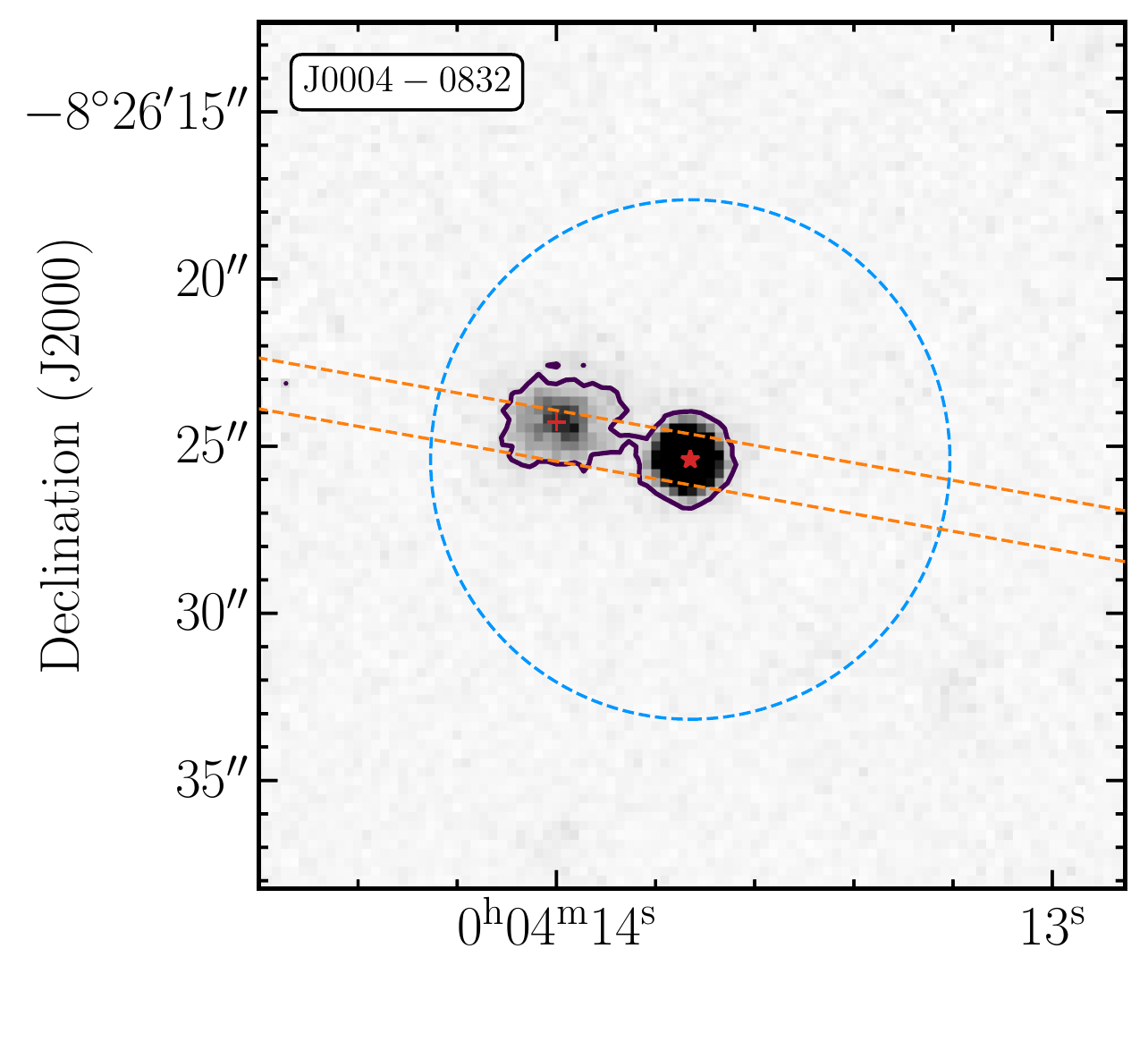}
    %\caption{Caption text 1}
  \end{subfigure}
  \begin{subfigure}{0.245\textwidth}
    \centering\includegraphics[width=\textwidth]{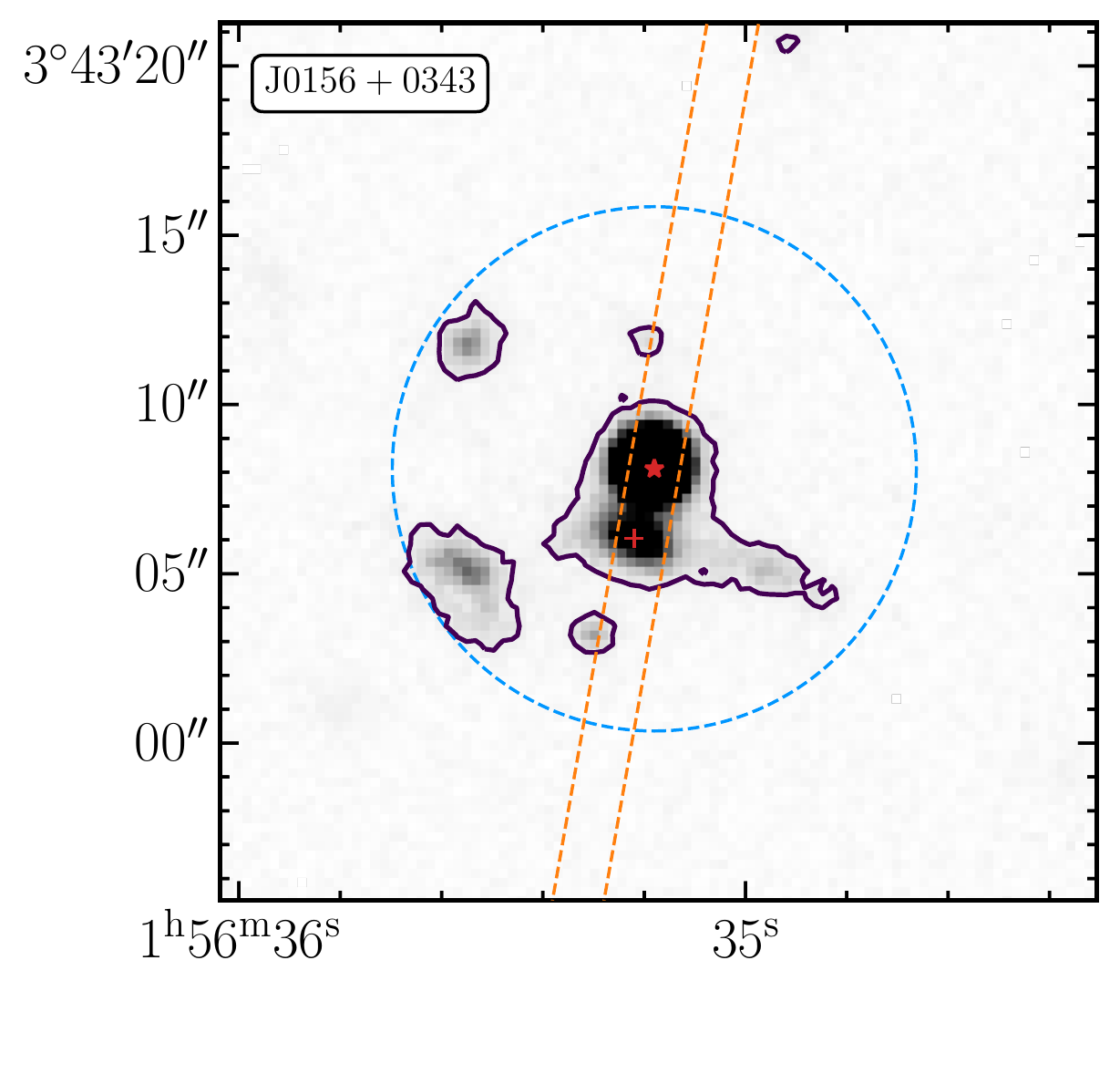}
    %\caption{Caption text 1}
  \end{subfigure}
  \begin{subfigure}{0.245\textwidth}
    \centering\includegraphics[width=\textwidth]{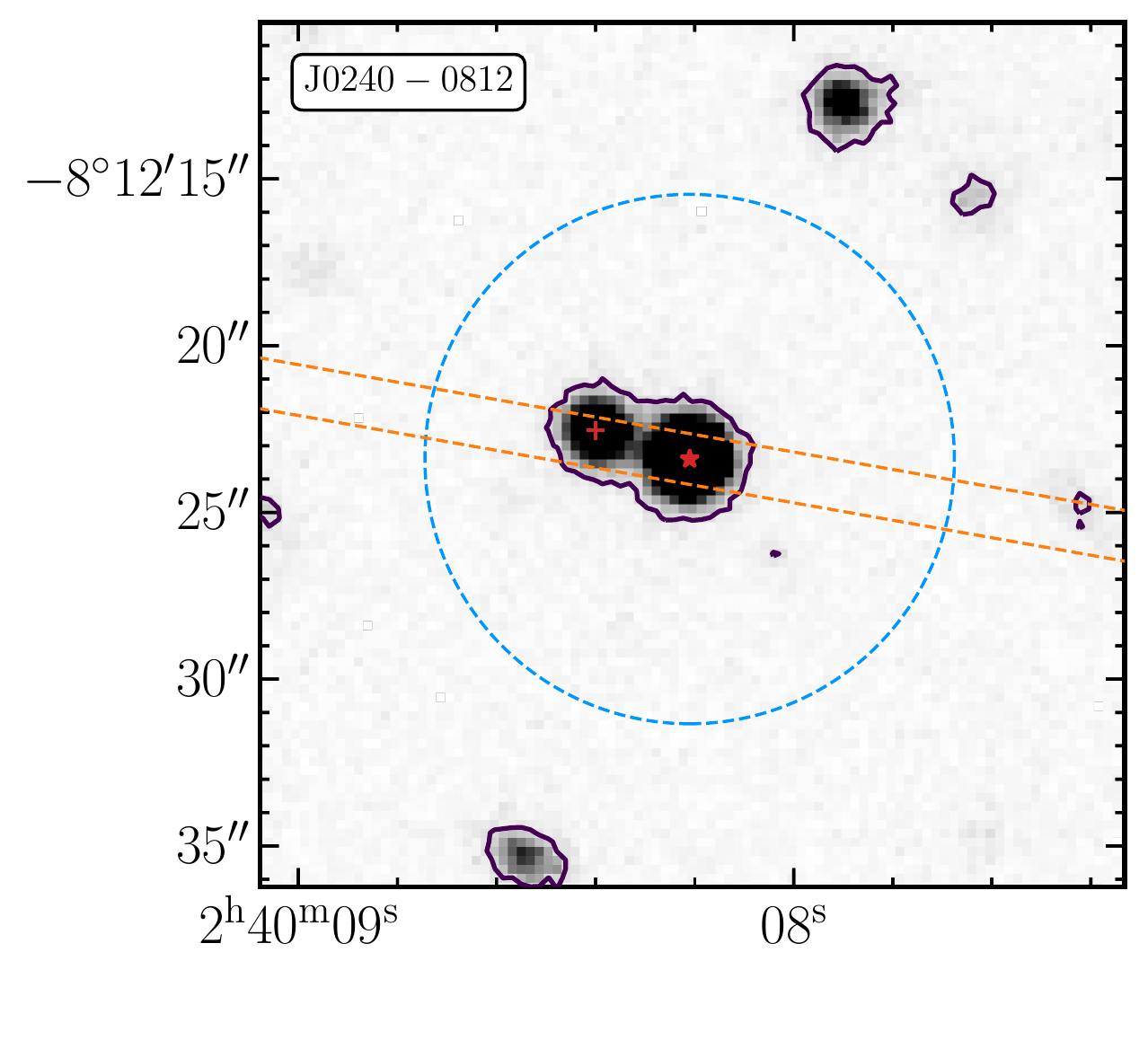}
    %\caption{Caption text 2}
  \end{subfigure}
  \begin{subfigure}{0.245\textwidth}
    \centering\includegraphics[width=\textwidth]{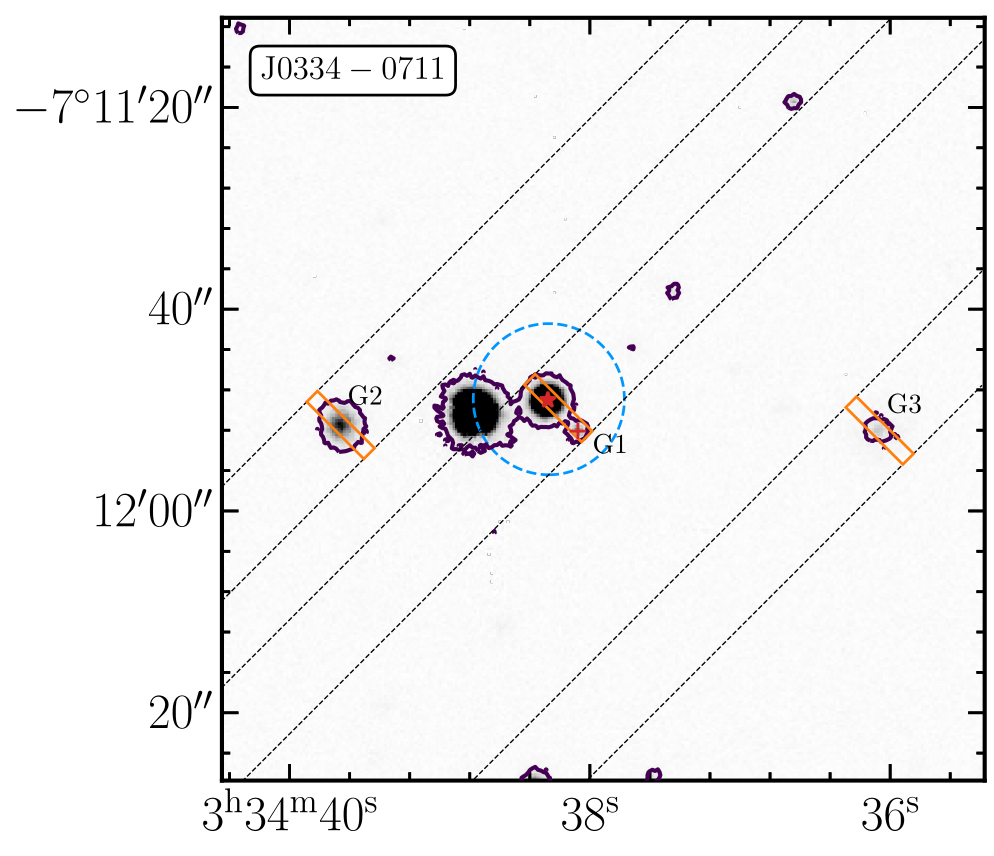}
    %\caption{Caption text 1}
  \end{subfigure}
  \newline
  
  \begin{subfigure}{0.245\textwidth}
    \centering\includegraphics[width=\textwidth]{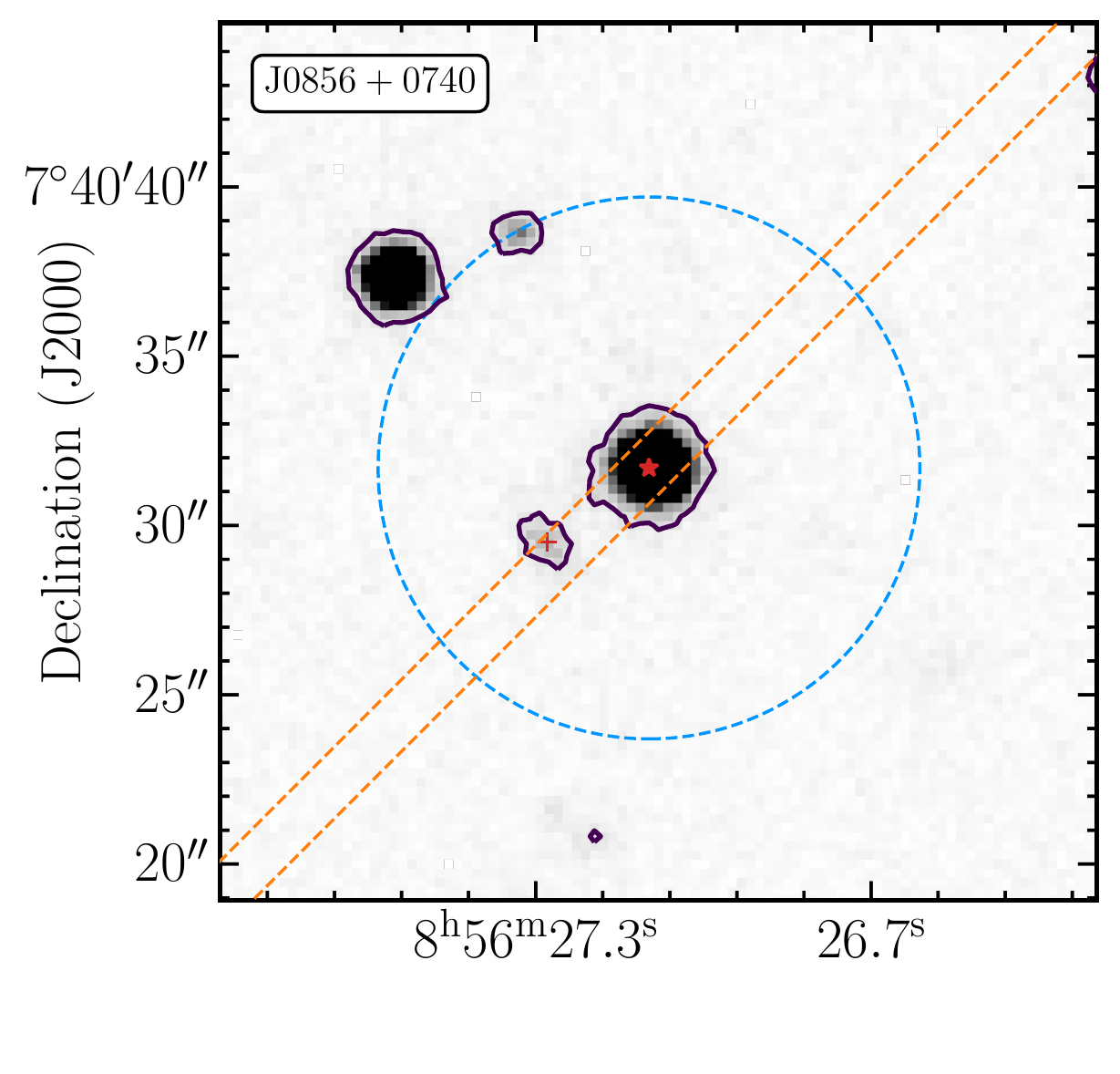}
    %\caption{Caption text 1}
  \end{subfigure}
  \begin{subfigure}{0.245\textwidth}
    \centering\includegraphics[width=\textwidth]{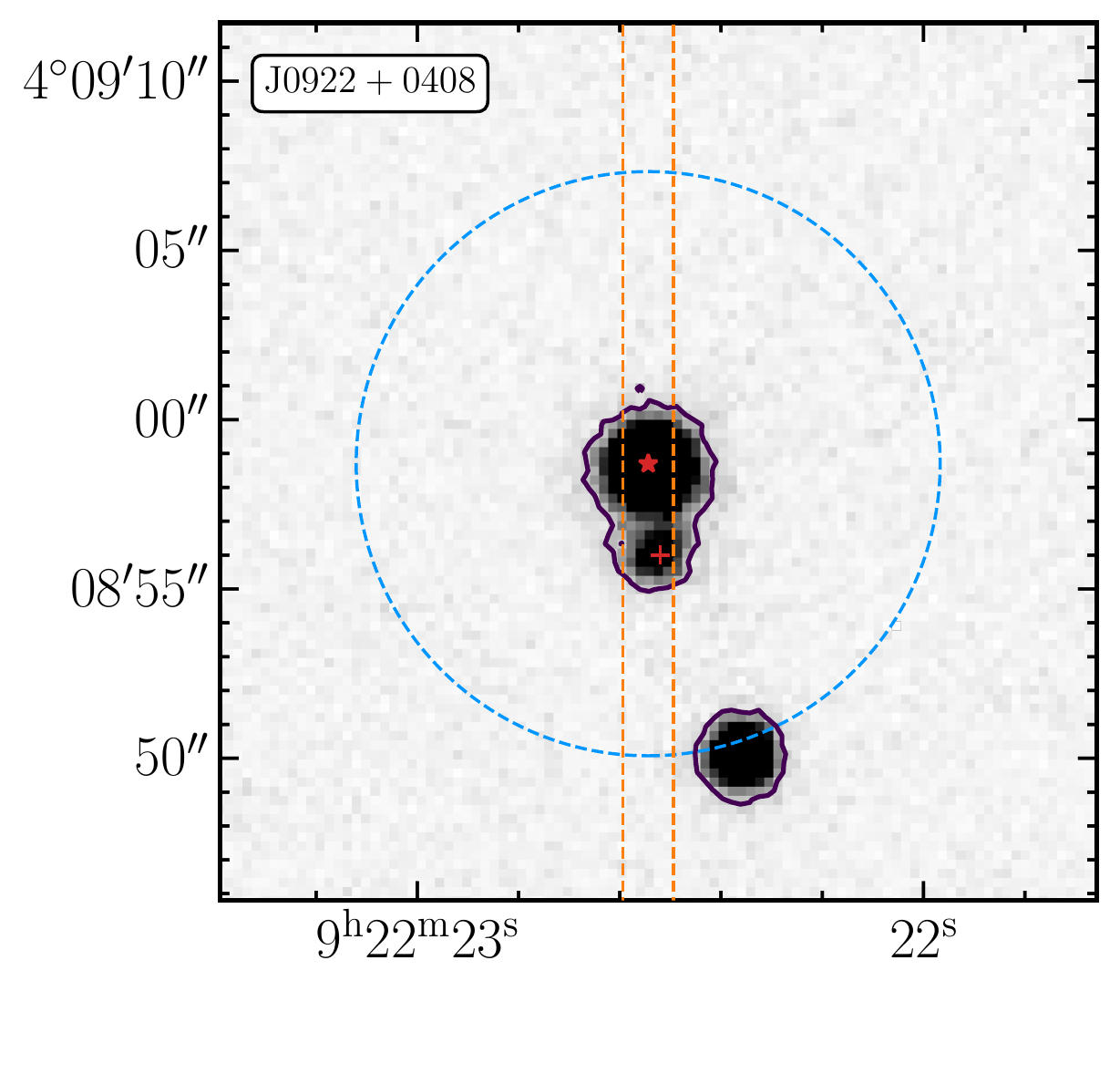}
    %\caption{Caption text 2}
  \end{subfigure}
  \begin{subfigure}{0.245\textwidth}
    \centering\includegraphics[width=\textwidth]{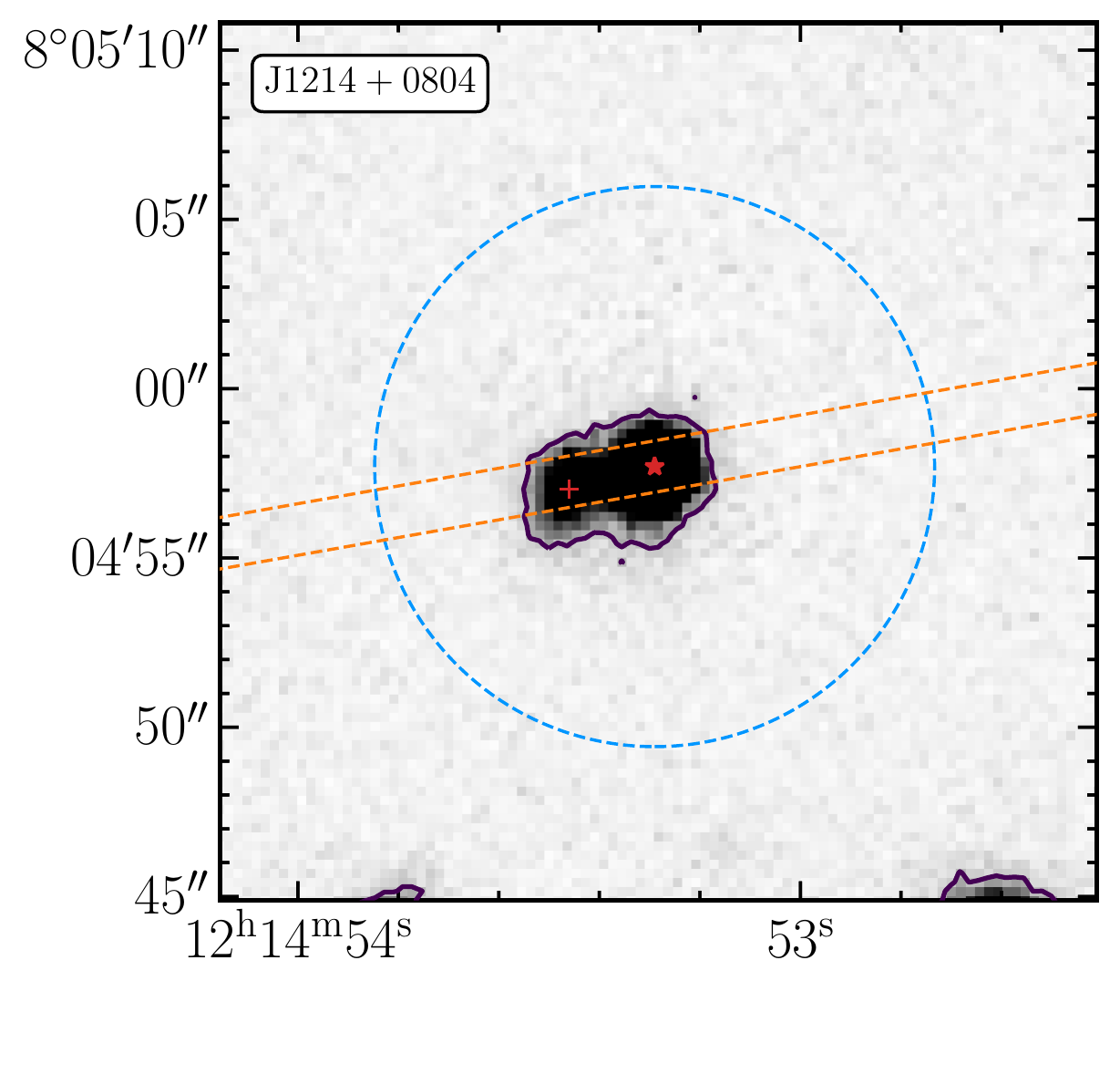}
    %\caption{Caption text 3}
  \end{subfigure}
  \begin{subfigure}{0.245\textwidth}
    \centering\includegraphics[width=\textwidth]{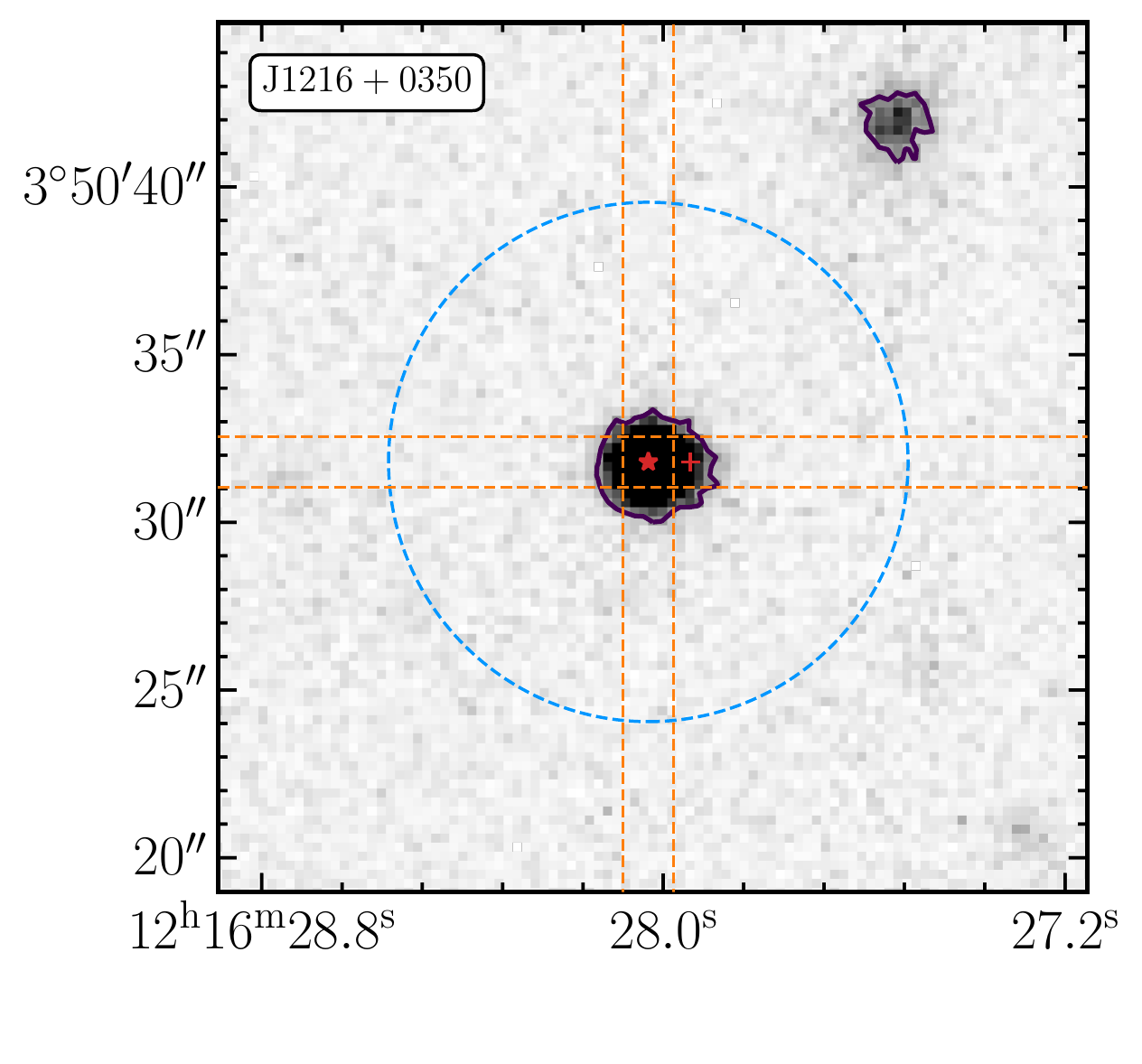}
    %\caption{Caption text 4}
  \end{subfigure}
  \newline
  
  \begin{subfigure}{0.245\textwidth}
    \centering\includegraphics[width=\textwidth]{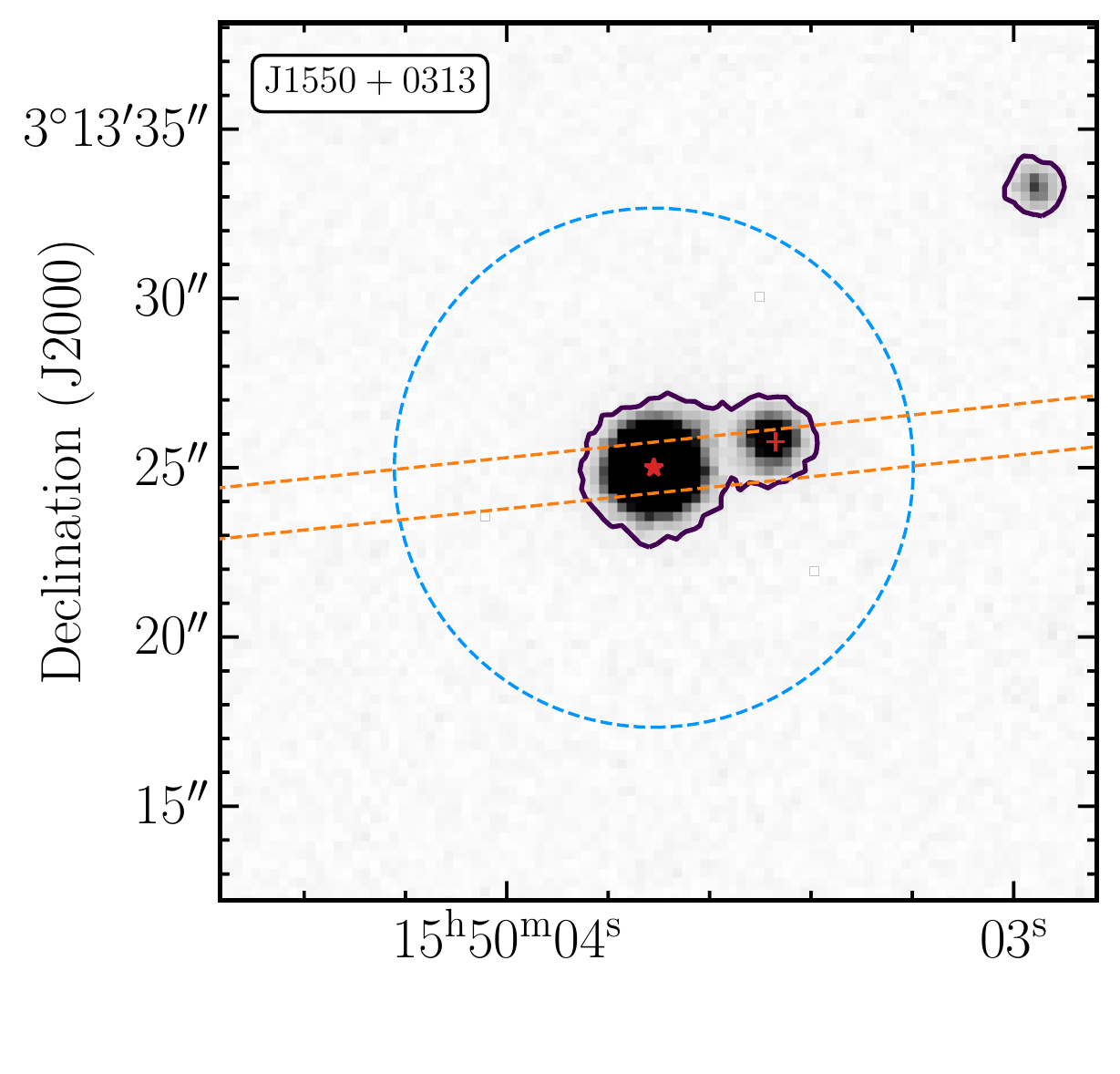}
    %\caption{Caption text 1}
  \end{subfigure}
  \begin{subfigure}{0.245\textwidth}
    \centering\includegraphics[width=\textwidth]{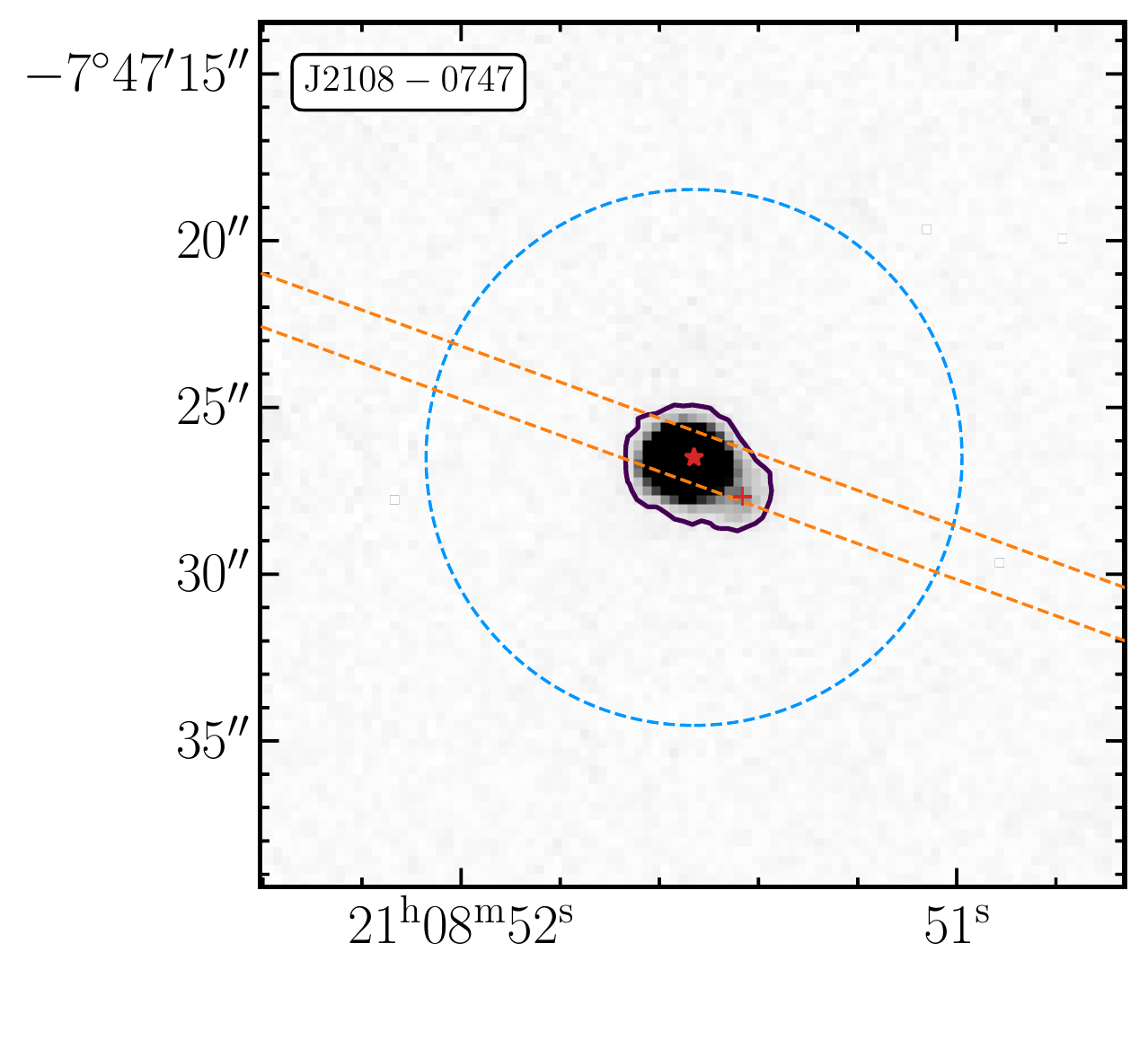}
    %\caption{Caption text 4}
  \end{subfigure}
  \begin{subfigure}{0.245\textwidth}
    \centering\includegraphics[width=\textwidth]{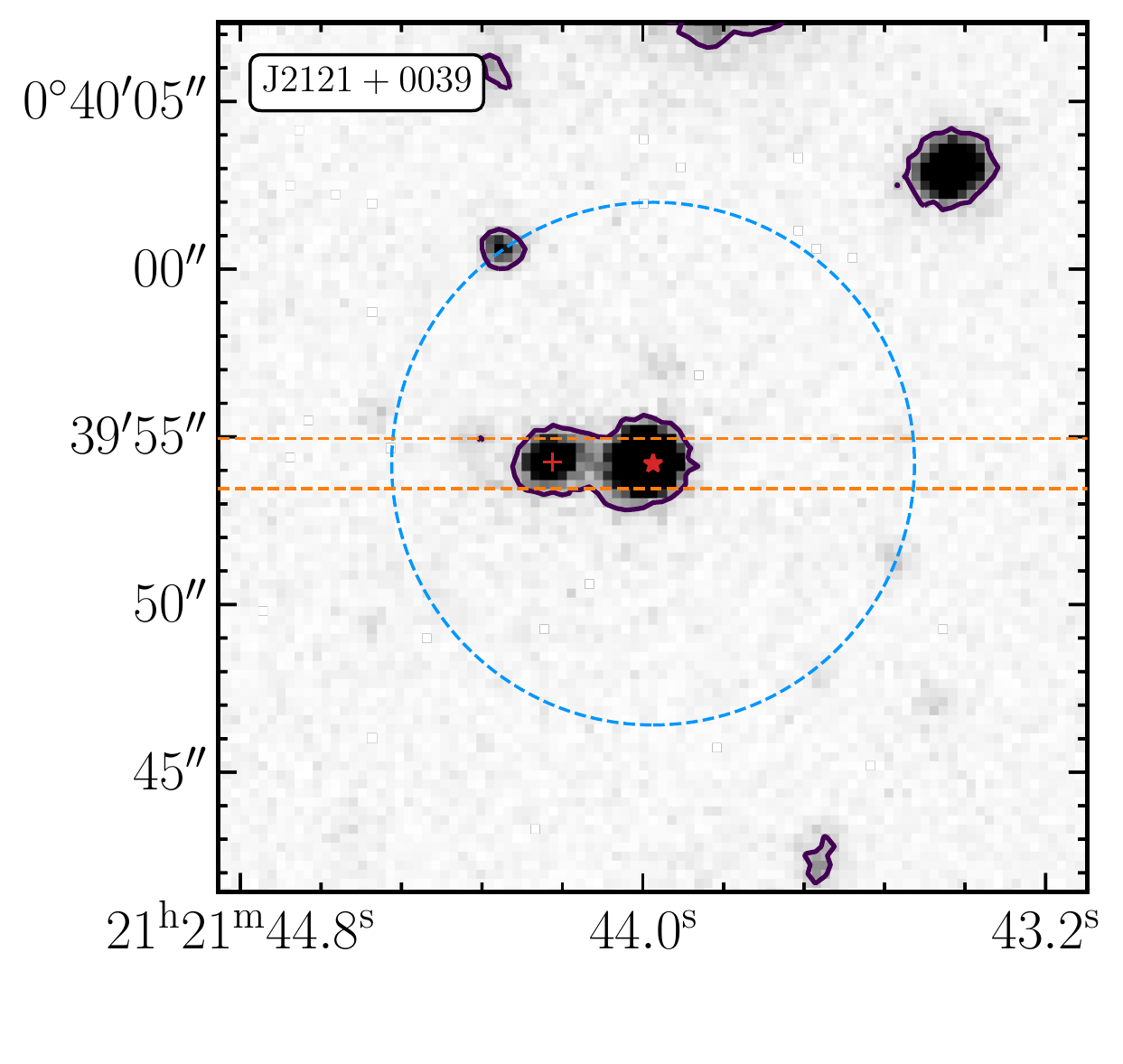}
    %\caption{Caption text 3}
  \end{subfigure}
  \begin{subfigure}{0.245\textwidth}
    \centering\includegraphics[width=\textwidth]{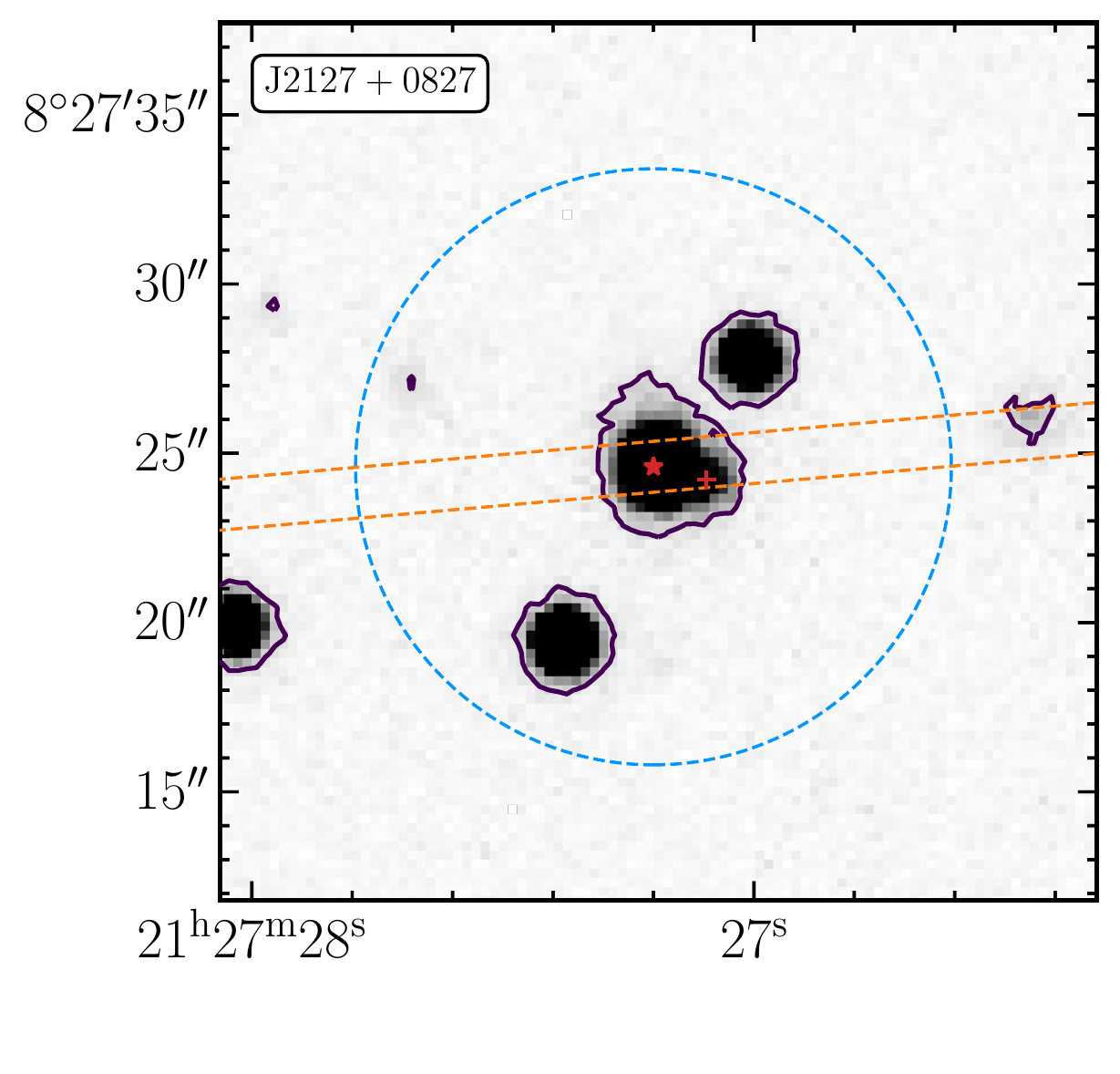}
    %\caption{Caption text 3}
  \end{subfigure}

  \begin{subfigure}{0.245\textwidth}
    \centering\includegraphics[width=\textwidth]{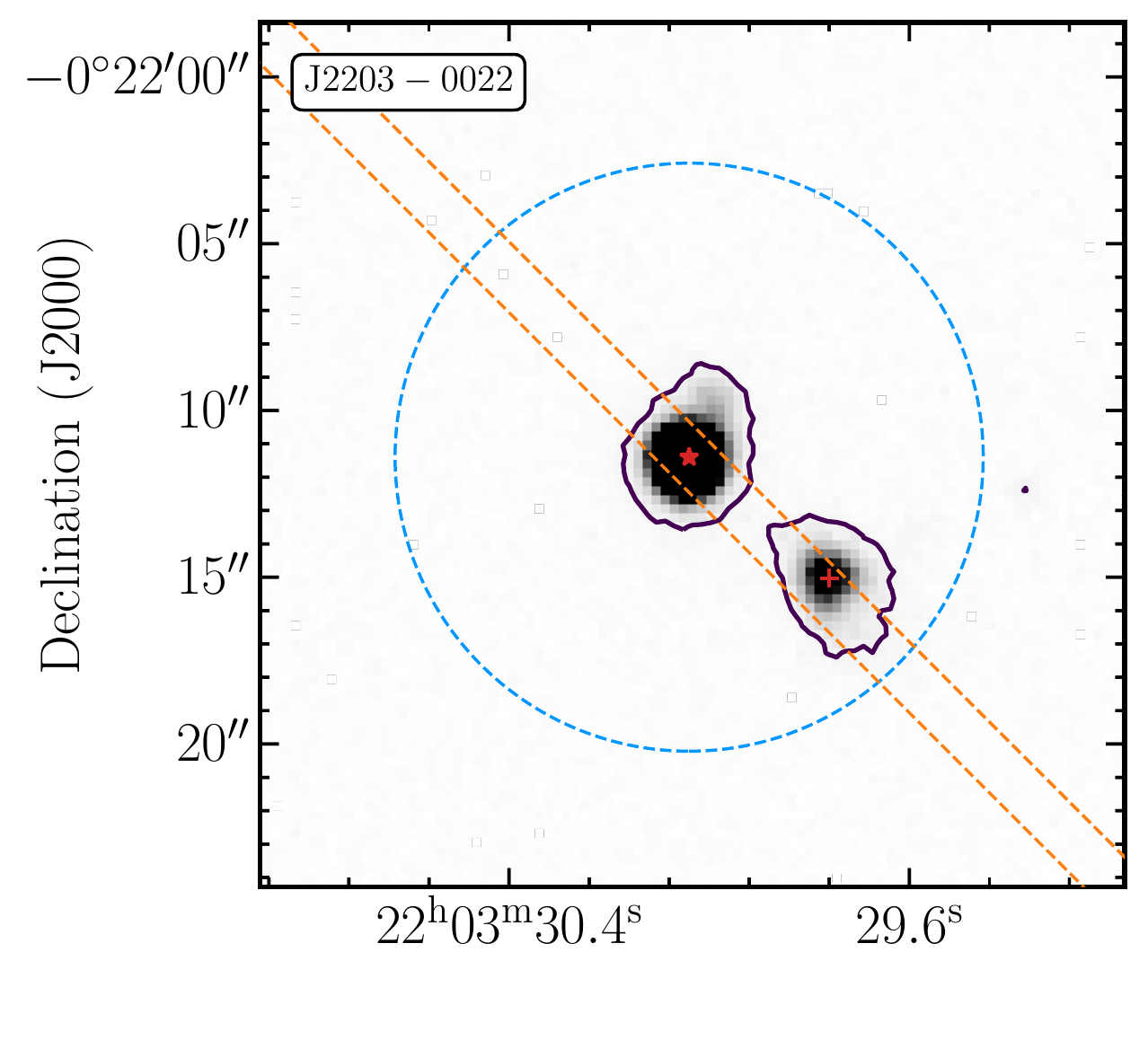}
    %\caption{Caption text 4}
  \end{subfigure}
  \begin{subfigure}{0.245\textwidth}
    \centering\includegraphics[width=\textwidth]{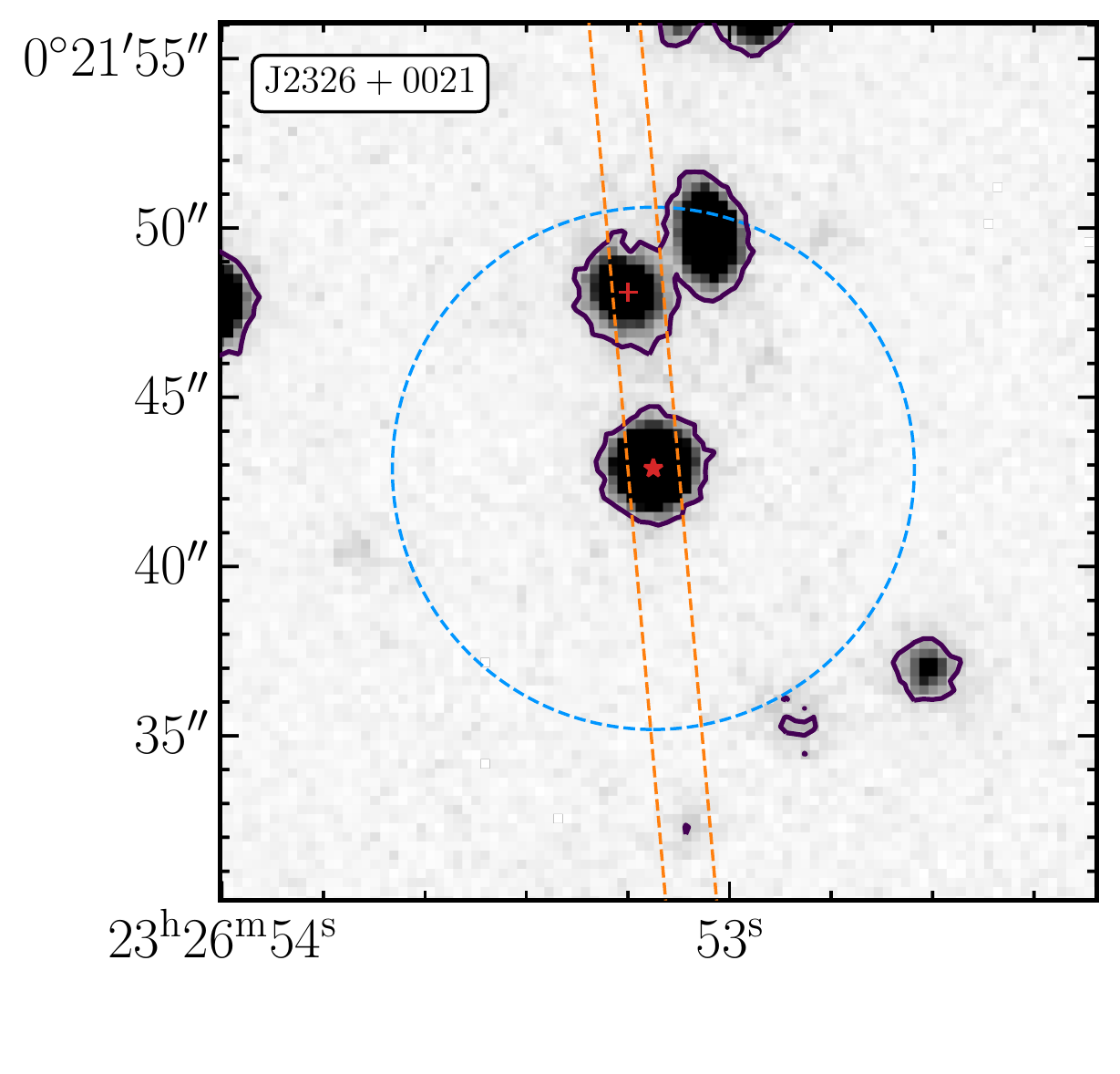}
    %\caption{Caption text 1}
  \end{subfigure}
  \begin{subfigure}{0.245\textwidth}
    \centering\includegraphics[width=\textwidth]{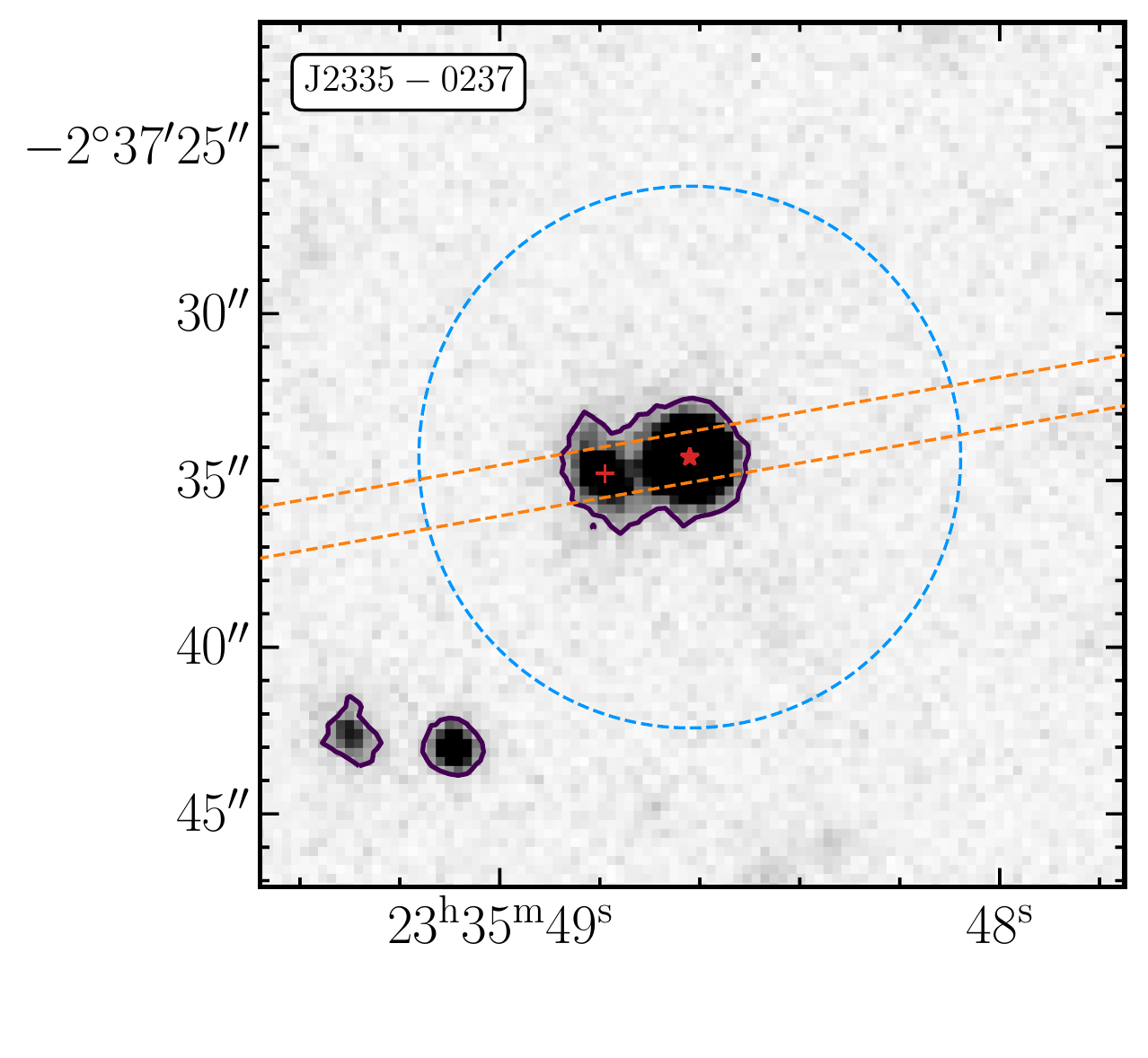}
    %\caption{Caption text 4}
  \end{subfigure}
  \begin{subfigure}{0.245\textwidth}
    \centering\includegraphics[width=\textwidth]{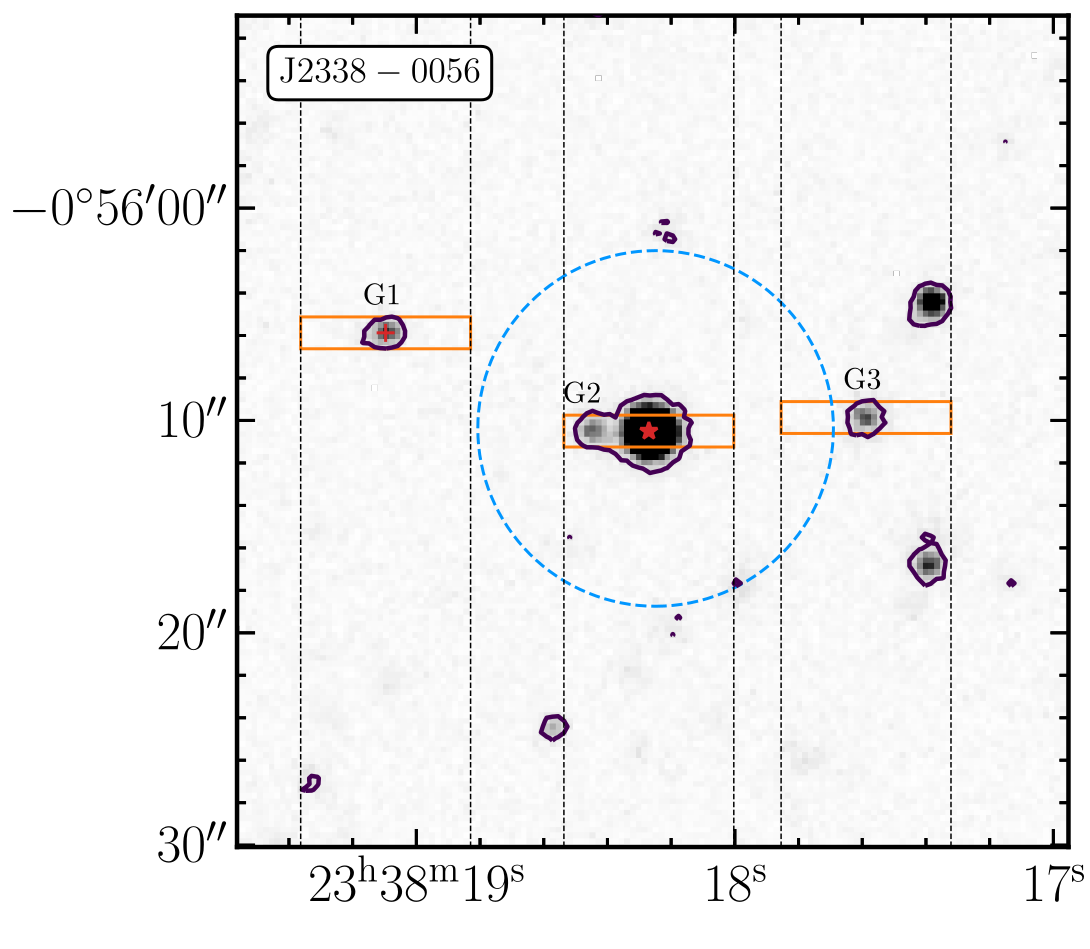}
    %\caption{Caption text 4}
  \end{subfigure}

    \begin{subfigure}{0.245\textwidth}
    \centering\includegraphics[width=\textwidth]{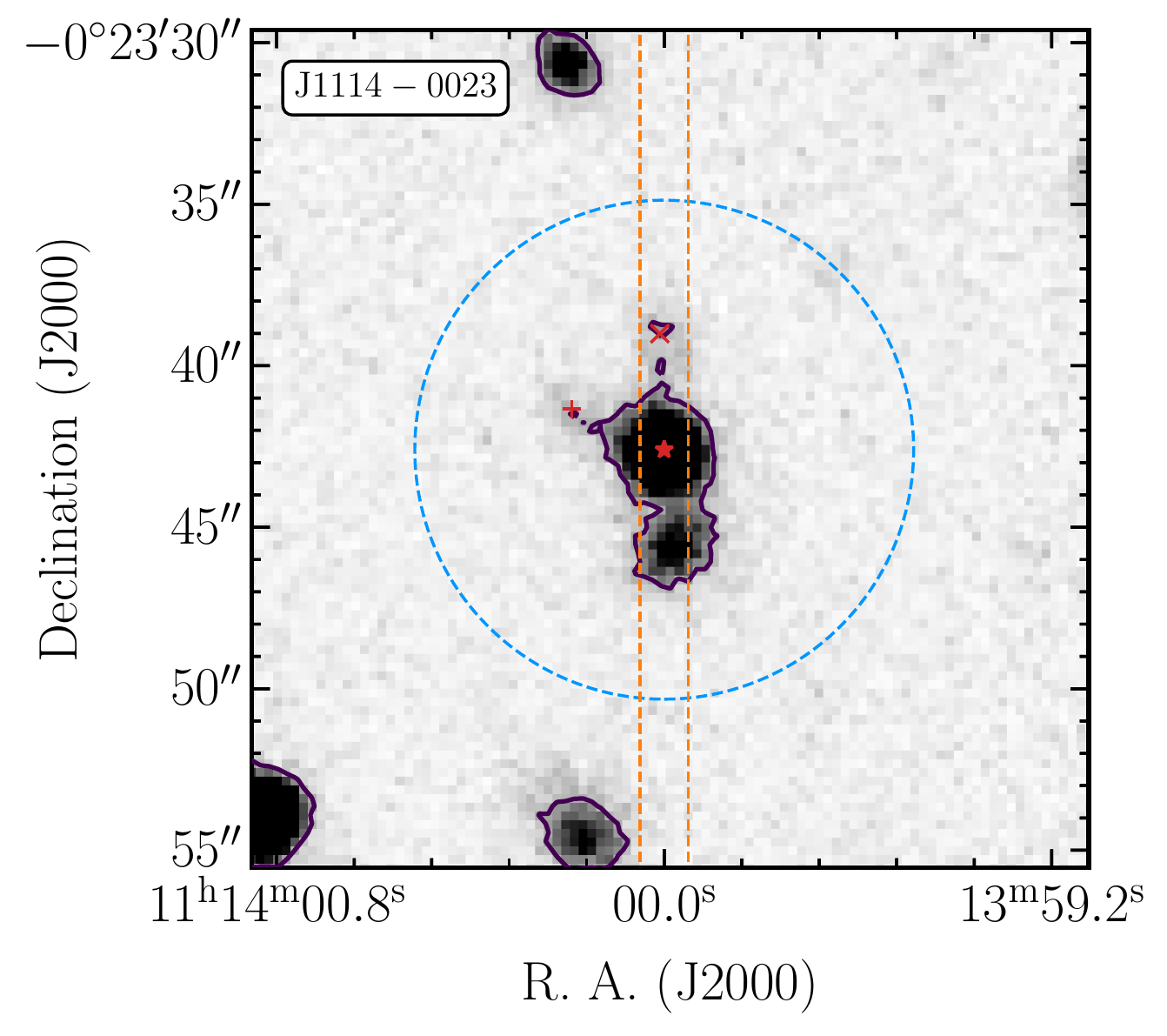}
    %\caption{Caption text 4}
  \end{subfigure}
  %\begin{subfigure}{0.245\textwidth}
  %  \centering\includegraphics[width=\textwidth]{J0218_slit.pdf}
    %\caption{Caption text 1}
  %\end{subfigure}
  \begin{subfigure}{0.245\textwidth}
    \centering\includegraphics[width=\textwidth]{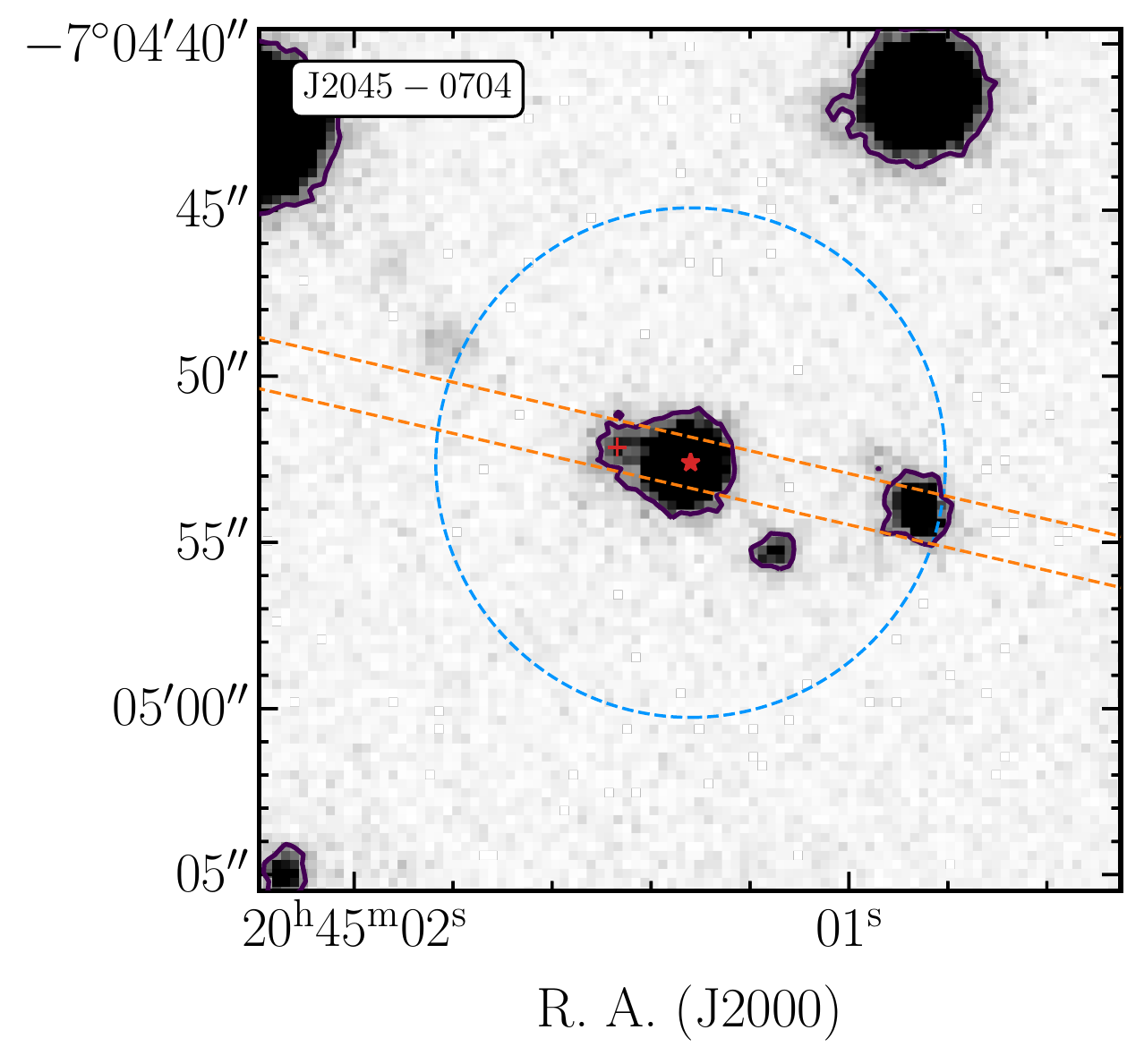}
    %\caption{Caption text 4}
  \end{subfigure}
  \begin{subfigure}{0.245\textwidth}
    \centering\includegraphics[width=\textwidth]{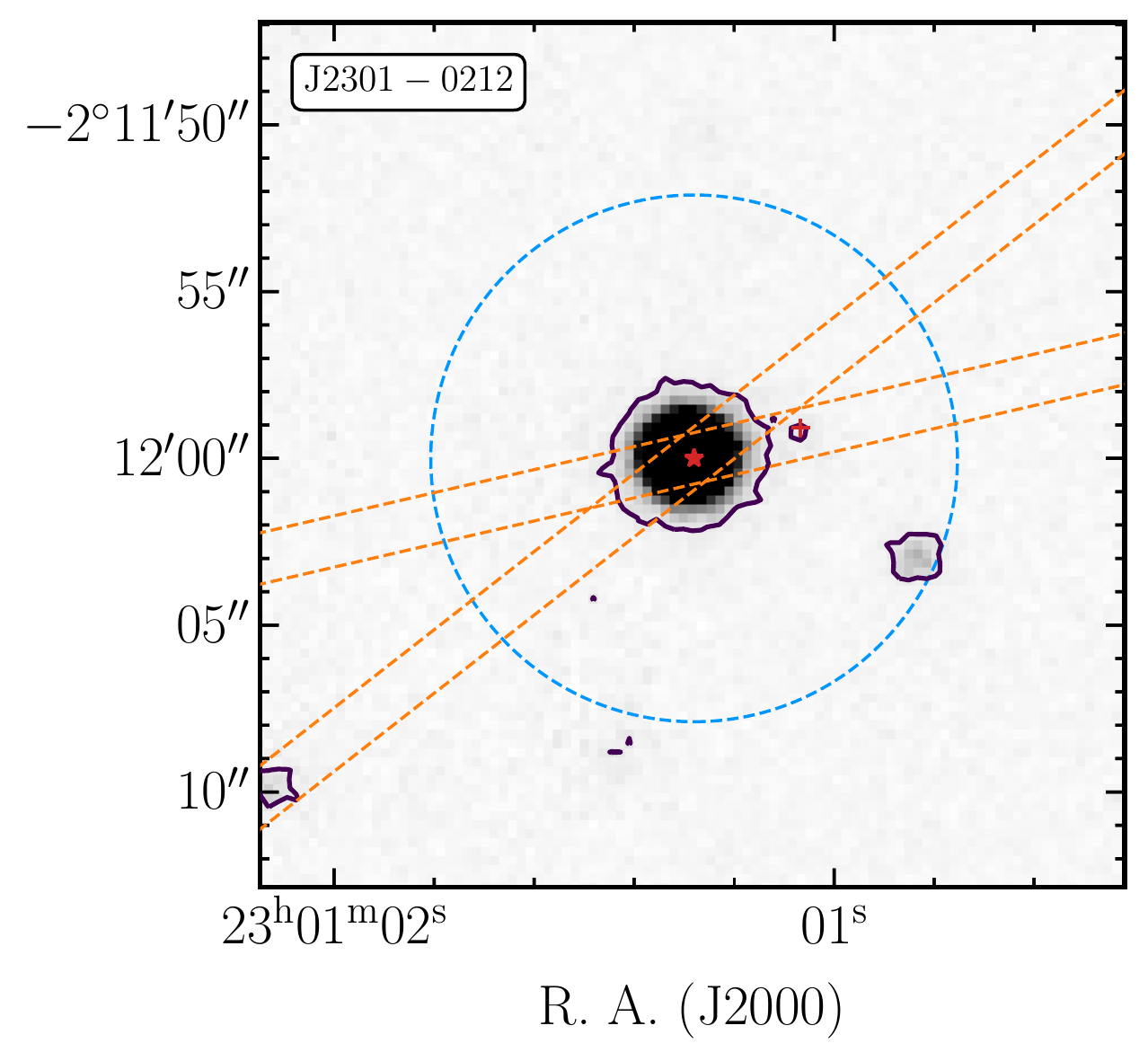}
    %\caption{Caption text 4}
  \end{subfigure}

  \caption{The DECaLS r-band images of the observed \usmg fields, except for the systems J2207-0901 and J0218-0832 that have been studied in detail by \citet{Gauthier2013} and \citet{Rahmani2016} respectively. In each of these fields, the quasar sits at the centre and is marked with a red star while in the case of detection, the centre of the associated \usmg host galaxy is marked with a red plus sign. The contours correspond to the 3$\sigma$ noise level on top of the mean background counts. The blue dashed circle corresponds to the circle with a radius 50 kpc around the quasar. Our SALT programme aims to obtain redshifts of potential galaxies within 50 kpc. The parallel red dashed line corresponds to the slit of width $1.5^{\prime\prime}$ used during the SALT observations. For two fields J0334-0711 and J2338-0056 multiple potential galaxies are seen and were targeted using MOS. The GOTOQ system, J1216+0350, has been observed with two different slit position angles to constrain the impact parameter by the method of triangulation using the \OII\ emission line. The 16 systems in the first 4 rows correspond to detections while the 3 systems presented in the last row are non-detections. The potential host galaxies based on the photo redshifts in the cases of non-detections are marked with red `+' sign. For the \usmg system, J1114-0023, we have marked the \MgII\ absorbing galaxy at \zabs $\sim$ 0.7981 ($W_{2796} = 2.15\ang$) with red `x'.
  }
  \label{fig:qso_fields}
\end{figure*}

In Table~\ref{tab:observation_log}, we provide the number of such galaxies (last column) in addition to our observational log. It is clear from this table that in only one case we have more than one host-galaxy candidate within 50 kpc. These are
5 galaxies in the case of \zabs = 0.5581 towards J0156+0343 within 50 kpc. 
In 6 cases the potential galaxy image is merged with that of the quasar (see Figure~\ref{fig:qso_fields}). Thus we have 26 host galaxy candidates and we obtained spectra of 23 of them (including from the literature as mentioned above).
%Thus the observational completeness of our sample is 88\% { (for galaxies brighter than $m_r$=22.5 mag or luminosity greater than 0.27 L$_*$)}  and within 50 kpc to the quasar sightline at $z_{abs}$. 
The SALT spectroscopic observations were carriedout using the Robert Stobie Spectrograph \citep[RSS,][]{RSS_1, RSS_2}  either in the long-slit  or the Multi Object Spectroscopy (MOS) mode from June, 2015 to April, 2021 (Program IDs: 2015-1-SCI-021, 2015-2-SCI-031, 2017-1-SCI-011, 2017-2-SCI-009 and 2018-1-SCI-019, 2020-2-SCI-019). 
For all our observations, we have used the PG0900 grating along with a slit-width of $1.5^{\prime\prime}$ and a suitable grating angle depending on the system observed to get the required wavelength coverage of [O~{\sc ii}], [O~{\sc iii}] and H$\beta$ emission lines. Such observational settings resulted in a velocity resolution of $\sim300\, km\,s^{-1}$. We have used MOS mode only for two cases (J0344-0711 and J2338-0056) and additional galaxies observed in these cases are beyond 50 kpc. In the case of the GOTOQs, there are no potential \usmg host galaxies seen in SDSS images within 50 kpc and the host galaxy lies on top of the quasar image.  We therefore considered two slit position angles to separate the \OII\ emission from the quasar continuum trace in the 2D spectra using the triangulation method. Our method is similar to what is discussed in \citet{fynbo2010, srianand2016}.

The details of observations along with the observational settings are provided in Table~\ref{tab:observation_log}. 
The second column provides the quasar identification. The date of observations, exposure time and mode of observation are presented in columns 3, 4 and 5, respectively. 
 Sixth, seventh and eight columns of this table present the slit position angle, grating angle and the wavelength coverage of the setting used respectively. 
The number of additional galaxies having consistent photometric redshifts within 100 kpc to the quasar sightlines 
with photometric redshift consistent with the absorption redshift of the \usmg\ systems is indicated in column 9 of Table~\ref{tab:observation_log}. 

Our sample is, by far, the largest sample of \usmg\ systems for which the host galaxy properties are probed over the redshift range $0.4 \leqslant z \leqslant 0.6$. 
The field of these targets and the slit orientations are shown in Figure~\ref{fig:qso_fields}. 
Details of these \usmg systems that  were observed with SALT are presented in the top part of Table~\ref{tab:sample}, and the remaining 5 are presented in the bottom part of this table. This unobserved target list also includes two GOTOQ absorbers (\zabs = 0.5928 towards J0932020.60+001828.0 and \zabs = 0.5561 towards J1025$-$0046), for which, based on the \OII\ emission detected on top the SDSS quasar spectrum, we have some information on the host galaxy from the SDSS quasar spectrum itself.

The raw CCD frames were initially subjected to the preliminary data processing with the SALT data reduction and analysis pipeline \citep{crawford2010}. Next, we use standard \texttt{IRAF} \citep{tody1986}
routines to obtain the wavelength calibrated 2D spectra and the flux calibrated 1D spectra of the quasars as well as the host galaxies. In summary, each of the science frames were first  flat-field corrected, cosmic ray zapped, and wavelength calibrated against a standard lamp. Next we applied the extinction correction due to the Earth's atmosphere and then the 1D spectra of the quasar and the galaxy were extracted. The long-slit spectra were all individually flux calibrated against standard stars observed with the same settings as of the quasar within a couple of nights of our observations. Once the flux calibration is done, we apply the air to vacuum wavelength transformation and also correct for the heliocentric velocity.

Since we could only observe 21 out of 27 systems, we first establish that the observed sub-sample represents an unbiased population of the original sample. We performed the two-sample Kolmogorov-Smirnov (KS) test between the parent and the observed sub-sample based on the absorption line properties [such as the rest equivalent width of \MgII$\lambda$ 2796 ($W_{2796}$), the doublet ratio of \MgII\ doublet ($W_{2796}$ / $W_{2803}$), $W_{2600}$ / $W_{2796}$ (referred to as $\mathcal{R}$), and the absorption redshift ($z_{\rm abs}$)]. Results of the KS test are summarised in Table~ \ref{tab:ks-test}. For each of the absorption properties, we find that the $p$-value of the KS test, i.e., the probability of both the samples (parent and the observed) being drawn from the same underlying population is very close to 1, which implies that purely based on absorption line properties the observed sub-sample essentially represents an unbiased population of the parent sample.
%}}
\begin{table}
    \centering
     \caption{Results of the KS test between the observed sub-sample and the parent sample} 
    \begin{tabular}{lcccc}
        \hline
        Properties & $W_{2796}$  & $ W_{2796}/W_{2803}$ & $W_{2600}/W_{2796}$ & $z_{abs}$ \\
         \hline
         \hline
         
        Statistics (D) &  0.101 & 0.095 & 0.057 & 0.127\\
        p-value & 0.998 & 1.000 & 1.000 & 0.975\\
        \hline
    \end{tabular}
    \label{tab:ks-test}
\end{table}
%

\iffalse
\begin{table}
    \centering
     \caption{The result of the KS test}
    \begin{tabular}{lcccc}
        \hline
        Properties & $W_{2796}$  & $ W_{2796}/W_{2803}$ & $W_{2600}/W_{2796}$ & $z_{abs}$ \\
         \hline
         \hline
        Statistics (D) &  0.048 & 0.032 & 0.089 & 0.071\\
        p-value & 1.000 & 1.000 & 1.000 & 1.000\\
        \hline
         
    \end{tabular}
    \label{tab:ks-test_2}
\end{table}
\fi

{
\subsection{Survey completeness}
\label{sec:completeness}
We consider deeper images and photometric redshifts available in DECaLS galaxy catalogs \citep{Dey2019} to quantify completeness of our sample at the faintest magnitude levels (i.e. $m_r<23.6$).
In Table~\ref{tab:decals} in the appendix, we summarize details of all the galaxies in DECaLS, within an impact parameter of 100 kpc and photometric redshift consistent with the \zabs\ for the \usmg\ in our sample.  First we consider the 36  (26 if $m_r<22.5$) galaxy candidates within $D\le50$ kpc. For 12 \usmg\ absorbers there is only a single identified galaxy. Our observational completeness is 92\% (one galaxy not observed) and spectroscopic measurement completeness is 83\% (i.e. 10/12). For six \usmg\ absorbers we have two galaxy candidates available within 50 kpc. In 3 cases both the galaxies are observed and in the remaining 3 cases only the brightest galaxy is targeted.  But the galaxies unobserved were fainter than 22.5 mag and not visible in the SDSS. Therefore, the observational completeness is 75\% (100\%) for the limiting magnitude $m_r$= 23.6 (22.5) mag. We measure the redshifts of 7 of the 9 targeted galaxies. Therefore, the redshift measurement completeness is 58\% (78\%) for the limiting magnitude $m_r$= 23.6 (22.5) mag. So for 18 \usmg\ absorbers discussed above our observational completeness is 79\% (90\%) and redshift measurement completeness of 71\%(81\%) for the limiting magnitude $m_r$= 23.6 (22.5) mag. Our survey completeness for D$\le$50 kpc are summarised in the Table \ref{tab:completeness}.

\begin{table}
    \centering
    \caption{Summary of the observational and the redshift completeness of our \usmg survey for D$\le$50 kpc.}
    \begin{tabular}{lcc}
         \hline
          Survey &\multicolumn{2}{c}{ Limiting magnitude ($m_r$) } \\
          Completeness &  23.6    & 22.5 \\
        \hline
        \hline
         Observations & 79\%   & 90\% \\
         Redshift measurement & 71\% & 81\% \\
         \hline
    \end{tabular}
 
    \label{tab:completeness}
\end{table}

Next we consider galaxies in the impact parameter $50\le$D[kpc]$\le$100, which is important to identify whether the \usmg\ absorber is part of a galaxy group. It is evident from Table~\ref{tab:decals} that for 7 \usmg (i.e. 33\% of the sample) systems only one galaxy candidate is present. In the remaining 11 cases there are candidate galaxies with impact parameter in the range 50-100 kpc. But good fraction of them have $m_r<22.5$ mag. In this impact parameter range our search is very much incomplete as we do not target most of these faint sources that also have large errors in the photometric redshifs.

In the case of J0156+0343, there are 6 galaxy candidates present within D=50 kpc, we have spectra of two of these galaxies which have consistent redshifts. Description of this system is presented in detail in section~\ref{sec:gp_spec}.  In the case of J2203-0022 there are three potential galaxies. We targeted two of them (with the lowest D) and could measure redshift in only one case. No nebular emission was detected in the other case. Note the source that was not observed has $m_r=23.5$ mag.
In the case of J1114-0023, there are three galaxies within D of 50 kpc. We have not targeted two of these galaxies which are fainter than $m_r = 22.5$ mag. The galaxy observed by us did not have correct redshift. 
}

\section{Identification \& properties of the \usmg\ host galaxies}
\label{sec:galaxy_props}
We searched for emission and absorption line signatures
of the host galaxies of \usmg\ systems
in the extracted spectra. 
%We have 21 useful spectra (out of 24 obtained) for redshift measurements i.e. $\sim$88\% completeness for the redshift identification.
Apart for 3 cases (shown in the bottom row in Figure~\ref{fig:qso_fields}) we were able to identify at least one galaxy with redshift consistent within 300 $\rm{km\, s^{-1}}$ 
with \zabs\ of the \usmg systems.
For the \zabs = 0.5623 \usmg system towards J2207-0901, \citet{Gauthier2013} have identified 4 potential associated galaxies, the closest galaxy being at an impact parameter of 38 kpc and the farthest being at an impact parameter of 246 kpc. For the \usmg system at \zabs = 0.5896 towards the quasar J0218-0832, \citet{Rahmani2016} have identified the host galaxy at an impact parameter of 16 kpc. For these two cases, we will use their identified galaxies in our work. In the case of the \zabs = 0.5311 \usmg absorber towards J0240-0812 our spectrum confirms the potential host galaxy identified by \citet{Nestor_2007} based on photometric redshifts. For the remaining 15 cases, we detect host galaxy(ies) showing at least one emission line among \OII\ $\lambda\lambda\, 3727, 3729$, \OIII\ $\lambda\lambda\, 4960, 5008$ and $H\beta\, \lambda\, 4862$ having consistent redshift with the \usmg absorption.

\begin{figure*}
    \begin{minipage}[t]{.35\textwidth}
        \centering
        \includegraphics[height=6.25cm, width=0.985\textwidth]{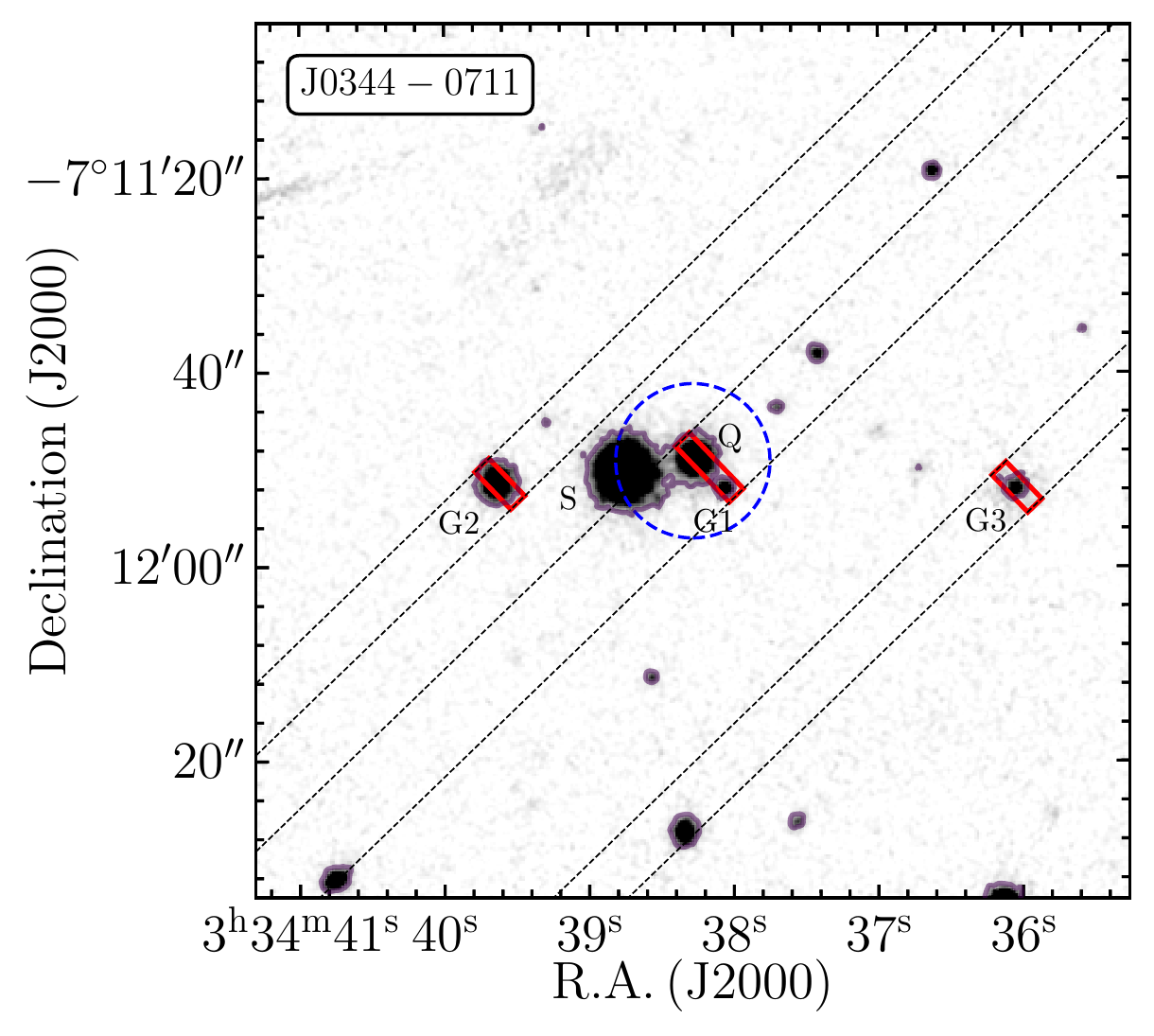}
    \end{minipage}
    %\hfill
    \begin{minipage}[t]{.645\textwidth}
        \centering
        \includegraphics[height=6.2cm, width=0.985\textwidth]{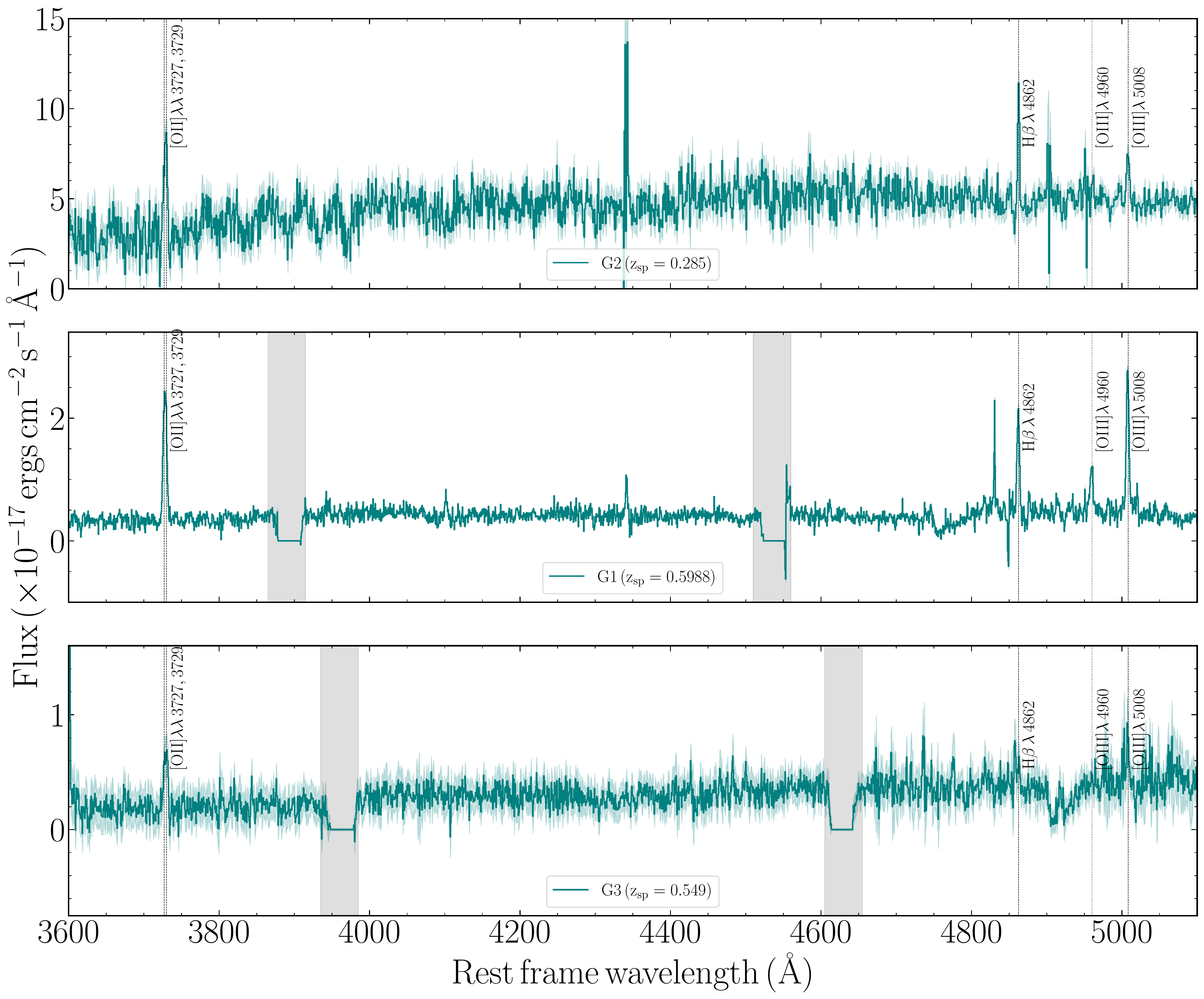}
    \end{minipage}  
    \caption{Left panel: the observational configuration for Multi Object Spectroscopy (MOS) mode observations of the \usmg system J0334-0826. The blue dashed circle represents a projected distance of 50 kpc around the quasar at the absorption redshift. Three nearby galaxies (named `G1', `G2' and `G3') having photometric redshifts  consistent with the redshift of the \usmg systems were targeted.   Rest frame galaxy spectra of these galaxies are shown in the right panel. Since the galaxy `G2' was also observed with SDSS, here we display the SDSS spectrum owing to a superior SNR. The black dashed vertical lines mark the locations of nebular emission lines confirming the redshift of these galaxies. The measured redshifts of the galaxies `G1', `G2' and, `G3' are  0.5989, 0.285 and 0.549 respectively. Only galaxy `G1' is consistent with the absorption redshift and hence, is assumed to be associated with the \usmg absorption.}
    \label{fig:J0334MOS}
\end{figure*}

\begin{figure*}
    \begin{minipage}[t]{.31\textwidth}
        \centering
        \includegraphics[width=\textwidth, height=\textwidth]{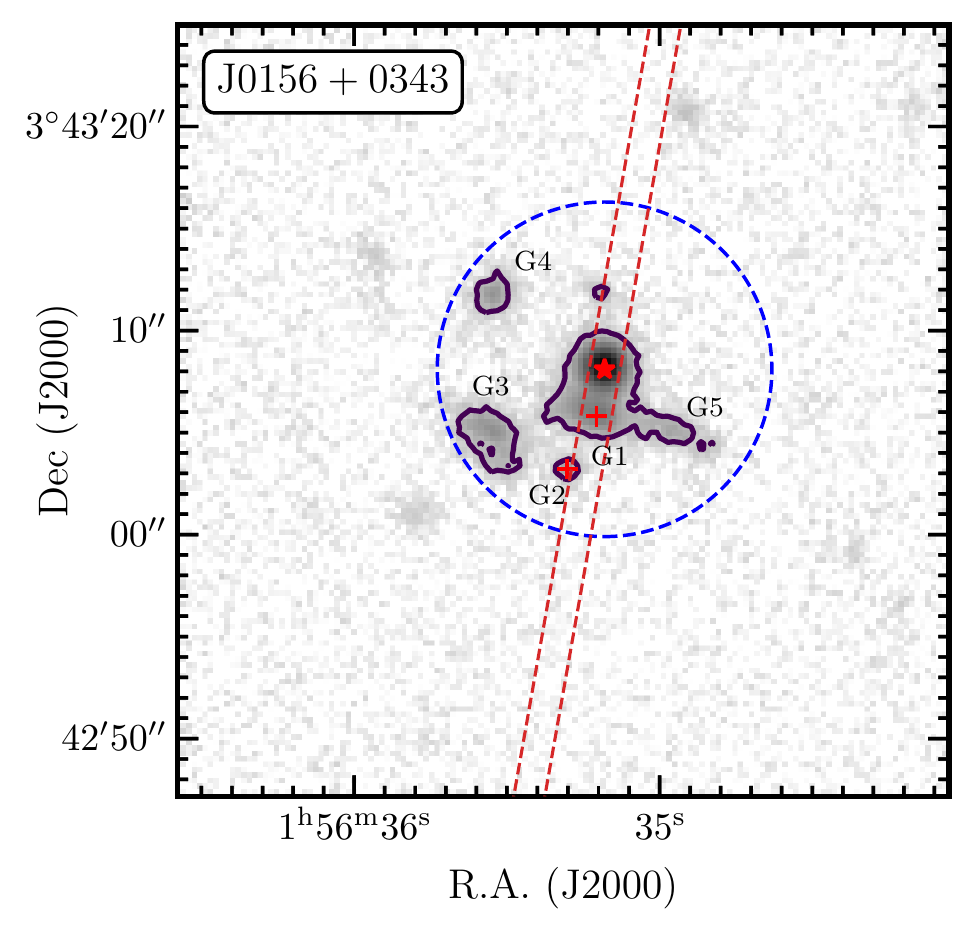}
    \end{minipage}
    \hfill
    \begin{minipage}[t]{.32\textwidth}
        \centering
        \includegraphics[width=\textwidth, height=\textwidth]{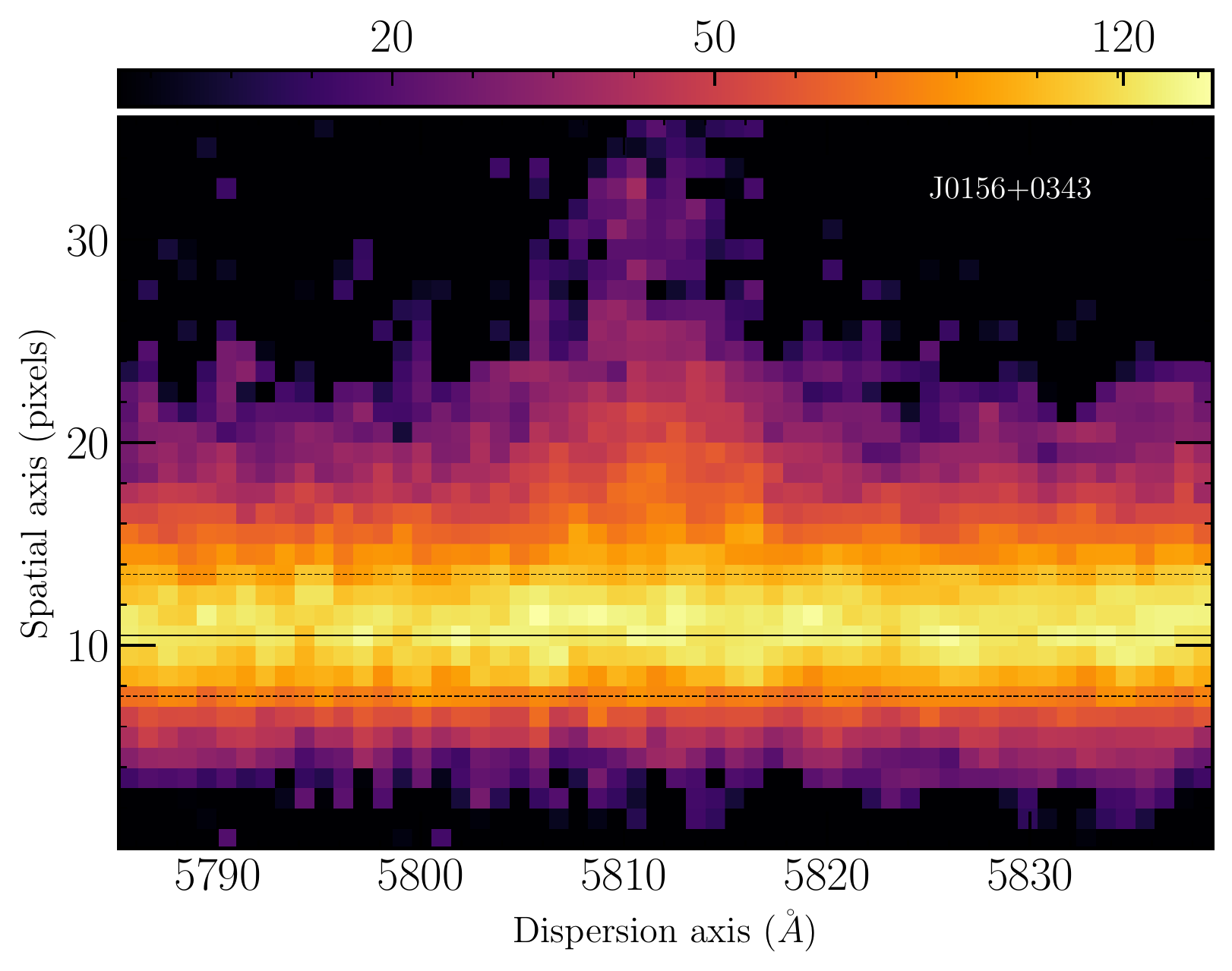}
    \end{minipage}  
    \hfill
    \begin{minipage}[t]{.32\textwidth}
        \centering
        \includegraphics[width=\textwidth, height=\textwidth]{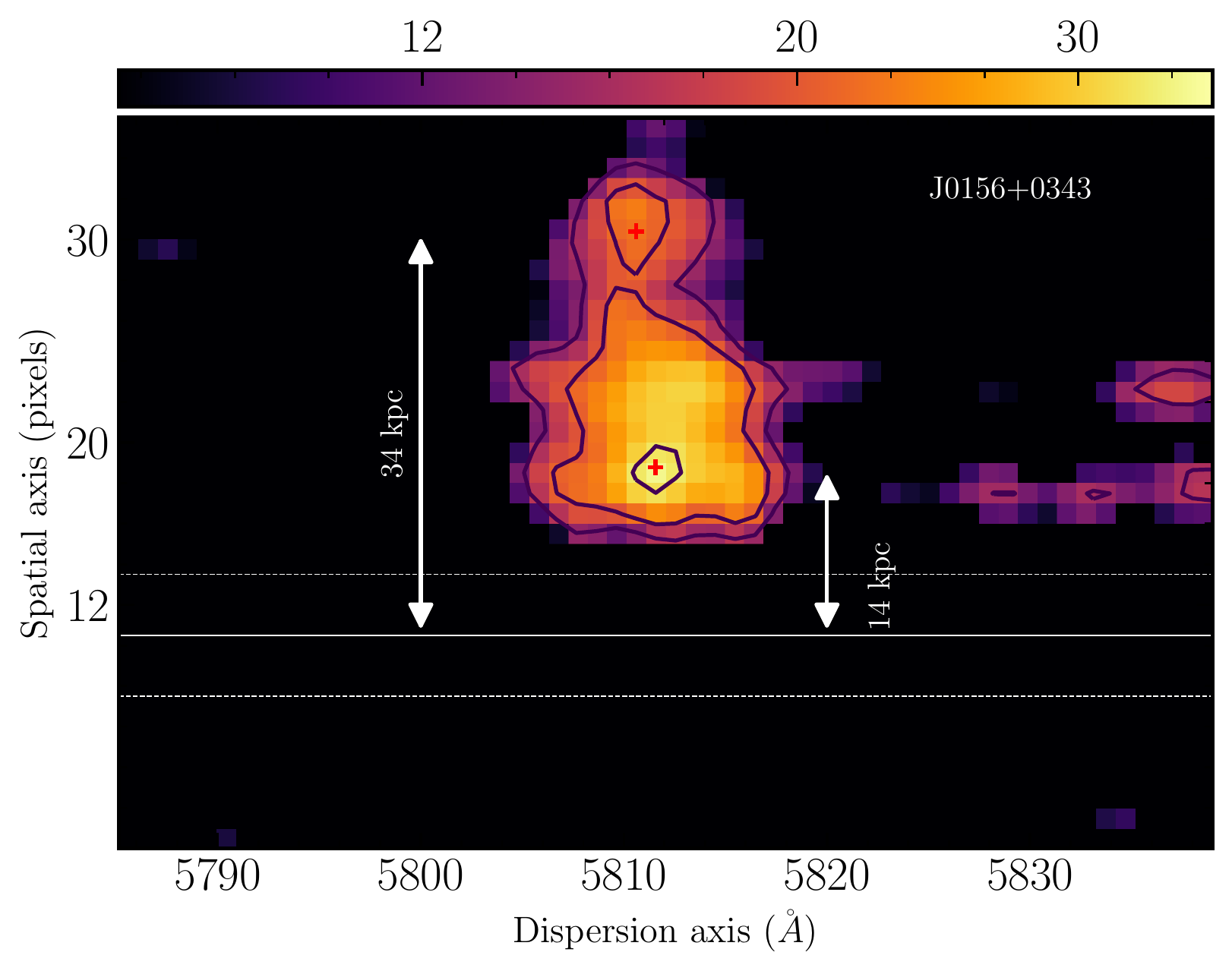}
    \end{minipage} 
    \caption{{\it Left panel:} DECaLS r-band image of J0156+0343 with the slit position shown with red dashed lines. We identify 5 possible galaxies within 50 kpc to the quasar sightline (blue circle). 
    {\it Middle} and {\it Right} panels show the 2D spectrum before and after the continuum subtraction. We detect \OII\ emission associated with G1 (D = 14 kpc) and G2 (D = 34 kpc)  indicated by "+" in the figure. G5 could be interacting strongly with G1. We also see strong \OII\ emission in between G1 and G2 that could be related to the gas expelled from the interaction between G1 and G5. Photometric redshifts of G3 and G4 are $0.455 \pm 0.106$ and $0.524\pm0.112$ respectively, that are also consistent with that of the \usmg absorption.}
   
    \label{fig:J0156Merger}
\end{figure*}

We have obtained MOS mode observations for two objects (see Figure~\ref{fig:qso_fields}). In the case of \zabs = 0.5977 \usmg systems towards J0334$-$0711, we could get good spectra of all the three galaxies close to the quasar sightline (see Figure~\ref{fig:J0334MOS}). The closest galaxy G1 has a redshift of $z = 0.5989$ consistent with being the host galaxy of the \usmg\ system.  
The measured redshifts of galaxies G2 and G3 are 0.285 and 0.549, respectively, suggesting that they are not associated with the \usmg absorber. The impact parameter of galaxy G3 with respect to the quasar sightline is 211 kpc and we do not detect any \MgII\ absorption with a 3$\sigma$ rest equivalent with limit of  0.28\AA.
We also used MOS observations for \zabs = 0.480 \usmg absorber towards J2338-0056 using three slits. The galaxy G1 (see Figure~\ref{fig:qso_fields}) has a redshift of 0.4798 and is at an impact parameter of 79 kpc from the quasar sightline. Although both G2 and G3 have photometric redshifts consistent with that of the \usmg absorption, we did not get spectra of sufficient SNR (SNR $<$ 3) to be able to measure their spectroscopic redshifts.
While redshift of G1 is consistent with $z_{abs}$, we do not consider this galaxy in our correlation analysis as G2 could be the real host galaxy at an impact parameter of 18 kpc. 

As seen in Figure~\ref{fig:qso_fields}, 
for 13 cases (i.e., 12 objects listed in Table~\ref{tab:photometric_props} plus the host galaxy identified for \zabs = 0.4801 absorber towards J233818.25-005610.5), the host galaxies are photometrically separated from the background quasars in the SDSS photometry. 
In the SALT observations as well, we were able to get the galaxy spectra without any contamination from the quasar emission in these cases. We matched our spectroscopic measurements with the r- and i-band fluxes of the host galaxies to correct for the fact that our slit might not have covered the whole galaxy. For four galaxies (associated with absorbers along the lines of sight towards J0156+0343, J1216+0350, J2108-0747 and J2127+0827), for which we can not directly compare the photometric flux of the \usmg host galaxies with its spectroscopic flux due to the quasar contamination, we compare the quasar spectrum obtained from the SALT and SDSS to estimate the flux-loss of the quasar. For these four systems, we assumed that the slit-loss suffered by the galaxy light is the same as that of the quasar.

In addition, we have detected  the stellar/interstellar \CaII $\lambda\lambda\, 3935,3970$ absorption and  strong H~{\sc i} Balmer absorption lines in the spectra of 5 \usmg host galaxies. These galaxies are also identified in Table~\ref{tab:spectroscopic_properties}. 
Recently, it has been found that in about 88\% of all the red galaxies that shows \OII\ emission lines, the excitations are due to various types of AGN, mostly the low ionization nuclear emission-line regions (LINERS, 63\%) \citep{Yan2006, Kocevski2011}. However, \citet{Yan2012} based on the spatial extention of the emitting regions rule out AGN contribution in LINER like emission in passive galaxies.
Observing [N\textsc{ii}] and H$\alpha$ emission lines is important to quantify any contribution of AGN in cases where we have indications of old stellar populations.

As mentioned above, there are three GOTOQs in our sample. To check whether there is any faint emission from galaxies close to the QSO sightline in other cases as well, we searched for the average \OII\ emission from the remaining 24 \usmg systems within the SDSS fibre using a median stacked spectrum. We detect a signal at the expected position of the \OII\ emission at the 2$\rm{\sigma}$ level.  Upon careful investigation, we notice that this stacked emission is mainly dominated by four systems (\zabs = 0.5509 towards J2121+0039, \zabs = 0.4392 towards J2127+0827, \zabs = 0.5961 towards J1016+0752 and \zabs = 0.5610 towards J1114-0023). Note none of them are in the GOTOQ sample of \citet{Joshi2017} as the emission lines are seen at a lesser significance level in individual spectrum. In the case of J2121+0039, our observations confirm the presence of the host galaxy close to the QSO sightline. Similarly, also for J2127+0827, we identify the host galaxy close to the QSO sightline. In the case of J1114-0023, available images show the presence of a galaxy ($m_r$ = 21.2 mag) along the slit very close to the quasar (identified with + in Figure~\ref{fig:qso_fields}). Our long-slit spectra did not show any nebular emission from this galaxy. However, we detect an emission line galaxy with \zem = 0.7983 at an impact parameter of 22.4 kpc (marked with a cross in Figure~\ref{fig:qso_fields}) associated with the \zabs = 0.7981 \MgII\ system. While this is not an \usmg\ system, we measure all its properties also.

The fourth system, \zabs = 0.5961 towards J1016+0752, has not been observed with SALT. However, upon careful inspection, we identify clear signatures of \OII\ $\rm{\lambda\lambda\, 3727, 3729}$, \OIII\ $\rm{\lambda\lambda\, 4960, 5008}$ nebular emission lines in the SDSS spectrum. Hence, this is essentially a GOTOQ system but in the Pan-STARRS photometry, we did not identify any detectable extension around the background quasar. When we repeated the stacking exercise for  the remaining systems, we did not find any detectable \OII\ emission with at least 2$\sigma$ significance.

\subsection{Galaxy parameters from spectroscopy}  \label{sec:gp_spec}
 
 In Table~\ref{tab:spectroscopic_properties}, we summarize properties of the identified host galaxies derived from our flux calibrated spectra after applying the corrections as explained above. In column 3 of this table, we provide the host galaxy redshift measured using \OII\ lines. In this case, we fitted the line with double gaussians having the same redshift and velocity width. The flux ratio of the two lines is kept as a free parameters and allowed to vary between 0.34 and 1.50. Columns 6, 7 and 8 provide the measured fluxes of \OII, H$\beta$ and \OIII\ emission lines. The 3$\sigma$ upper limits  are provided in the case of non-detections, and `--' indicates that the measurements are not possible with our spectra because of poor background subtraction due to the skylines present.
 
 The measured impact parameter (D, the projected separation of the host galaxy with respect to the quasar sightline) for host galaxies of \usmg\ are listed in the fifth column of Table~\ref{tab:spectroscopic_properties}. In cases where the host galaxy is well-separated from the quasar sightline, we measure the impact parameter using the coordinates of the centroids of the galaxy and the quasar.

\begin{figure*}
    \begin{minipage}[t]{.31\textwidth}
        \centering
        \includegraphics[width=\textwidth, height=5.5cm]{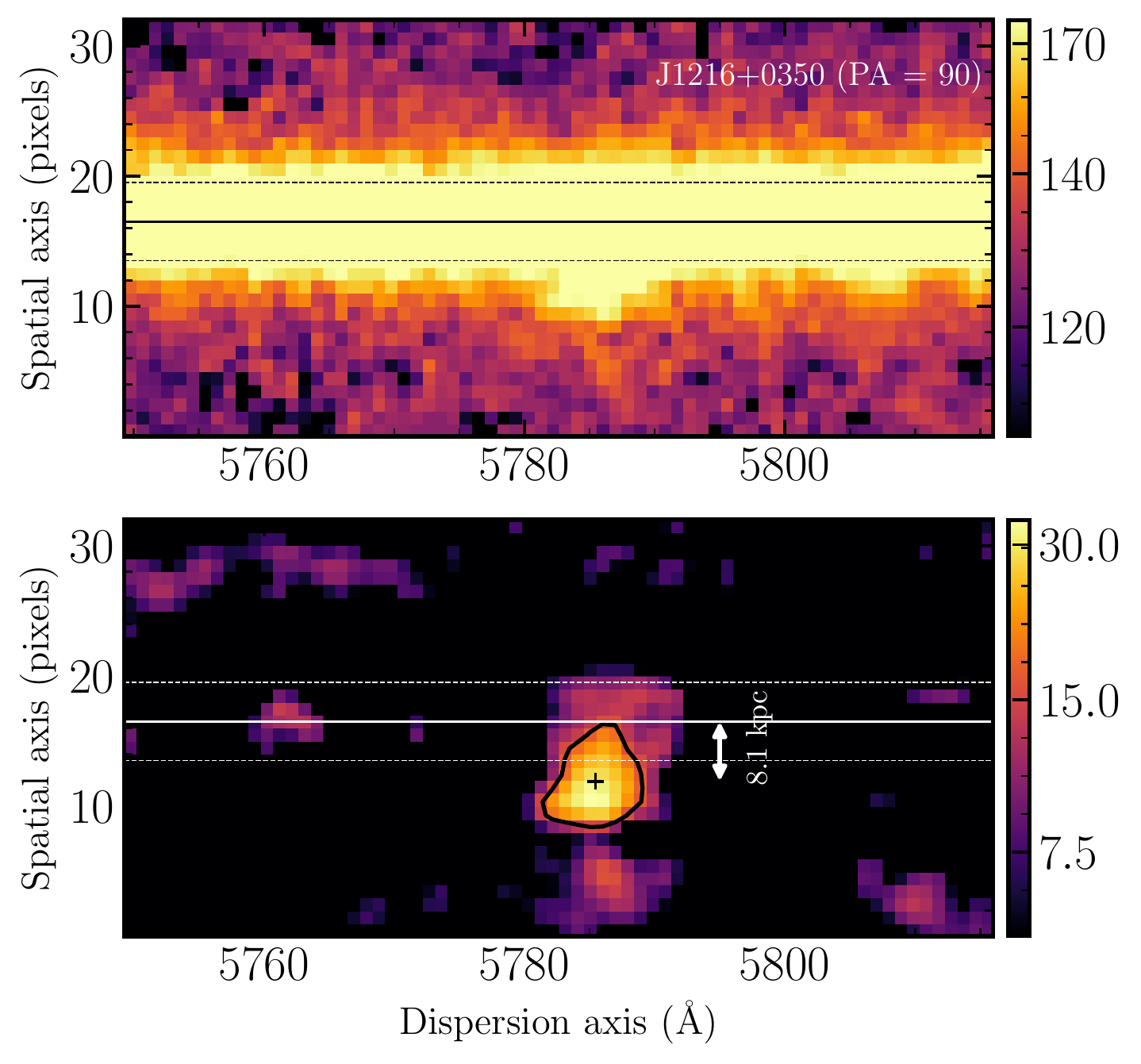}
    \end{minipage}
    \hfill
    \begin{minipage}[t]{.31\textwidth}
        \centering
        \includegraphics[width=\textwidth, height=5.5cm]{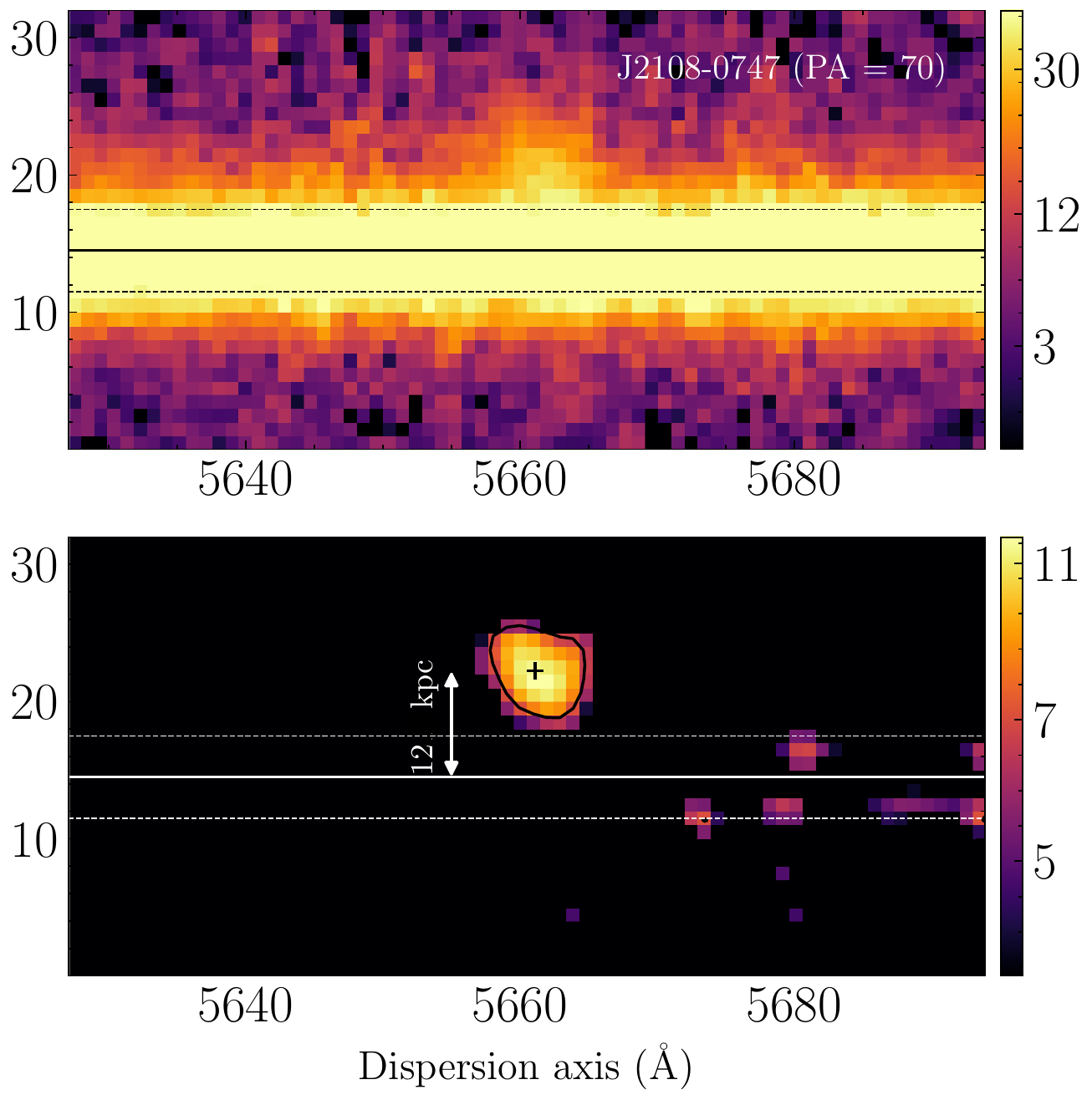}
   \end{minipage}
   \hfill
   \begin{minipage}[t]{.31\textwidth}
        \centering
        \includegraphics[width=\textwidth, height=5.5cm]{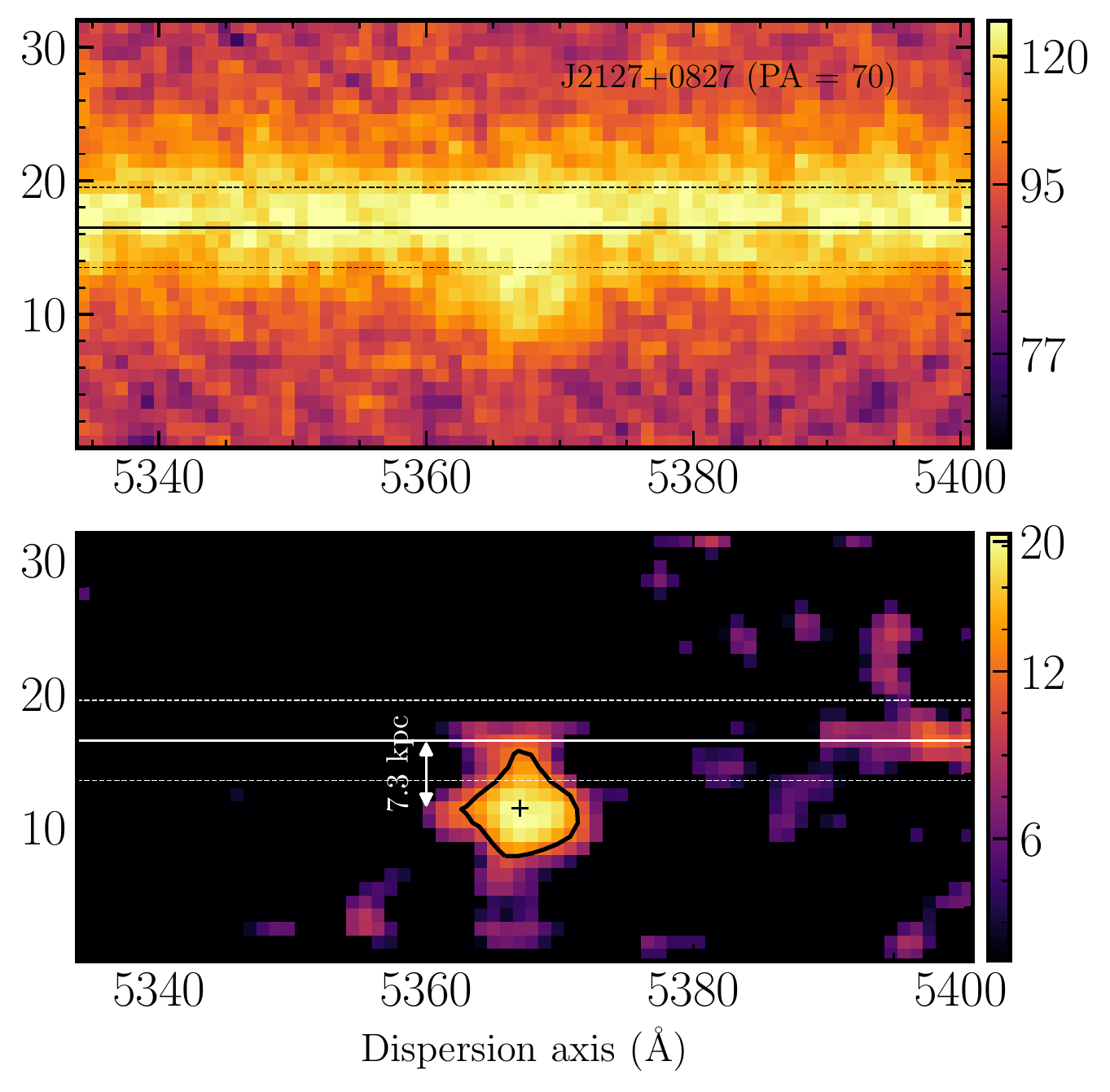}
   \end{minipage}
    \caption{An illustration of the procedure for determining the impact parameter for \usmg systems with the host galaxies at low impact parameter to the background quasar. Observed  and continuum  subtracted 2D spectra  are shown in the top and bottom panels respectively. In the left most panel,
    we show the continuum subtracted 2D spectrum of J1216+0350. \OII\ emission is clearly detected with an off-set of 1.27 arc sec from the quasar trace (solid horizontal line). This corresponds to an impact parameter of 8.1 kpc (see section~\ref{sec:gp_spec} for details). The solid black contour corresponds to the $3\sigma$ noise level on top of the mean background. Similar plots for J2108-0747 and J2127+0827 are shown in the middle and right panels respectively.
    }
    \label{fig:subtraction}
    \end{figure*}

The \usmg system  at $z_{abs} = 0.5519$ towards J1216+0350 is one of the GOTOQs in our sample with the SDSS spectrum of the quasar showing \OII, \OIII\ and $\rm{H\beta}$ emission lines, at the absorption redshift,
detected at $>3\sigma$ confidence. In the Pan-STARRS images, we do not find any significant extension of the background quasar continuum due to the foreground galaxy. Hence, to constrain the impact parameter, we obtained
the SALT spectrum by keeping the slit at two position angles (i.e. PA = 0$^\circ$ and 90$^\circ$). 
In the left most panels of Figure~\ref{fig:subtraction}, we present the 2D spectrum of this quasar obtained along PA = 90$^\circ$ before (top panel) and after (bottom panel) subtracting the quasar continuum. We removed the quasar flux from the 2D spectra by linearly interpolating the quasar flux measured over $\rm{35 \AA}$ on both sides of the \OII\ emission. It is evident that the \OII\ emission is well-separated from the QSO along PA = 90$^\circ$.  The bottom panel also displays the $\rm{3\sigma}$ contour about the median and the centroid is marked with a `+'. Based on this we, constrain the impact parameter to be $\rm{8.1\, kpc}$. Due to poor spectral point spread function, we were unable to use the spectra obtained at PA = 0$^\circ$ to further refine the position of the \OII\ emission.
Interestingly,  we find a galaxy at an impact parameter of 302 kpc with redshift z= 0.5527 within 490 \kms\ of this \usmg\ absorber in the SDSS catalog. However, it is most likely that this galaxy is unrelated to the \usmg\ absorber as it is at a larger impact parameter and has a  much higher velocity separation relative to the \usmg absorption, compared to the galaxy at an impact paramter of 8.1 kpc. 

In the remaining two cases, \zabs = 0.5187 towards J2108-0747 and \zabs = 0.4392 towards J2127+0827, while the host galaxies are not distinctly visible, their quasar images show extensions (see Figure~\ref{fig:qso_fields}). We observed these targets keeping the slit aligned to these extended features and detected emission lines well-detached from the quasar trace in both cases. After subtracting the quasar contribution in the 2D image, we estimate the impact parameters to be 12 and 7.3 kpc, respectively, for J2108-0747 and J2127+0827 (see  Figure~\ref{fig:subtraction}). 

For the case of \usmg absorption at \zabs = 0.5581 along the line of sight to J0156+0343, we identify a group of 5 galaxies (among which a merger is possibly present) close to the line of sight (see Figure~\ref{fig:J0156Merger}). The nearest galaxy G1 could be part of a tidally interacting pair with G5. In our slit spectrum we detect \OII\ emission from the galaxy G1 (D=14 kpc) and G2 (D=34 kpc). We also detect \OII\ emission between these two galaxies.
The galaxies G3, G4 and, G5 have photometric redshifts $0.455\pm0.106$, $0.524\pm0.112$ and $0.488\pm0.104$ respectively. Thus, this absorber is most likely to originate from gas associated with a group of interacting galaxies.

\subsection {Galaxy parameters from SED fitting:}

\begin{figure*}
    \begin{minipage}[t]{.27\textwidth}
        \centering
        %\flushleft
        \includegraphics[height=5cm, width=\textwidth]{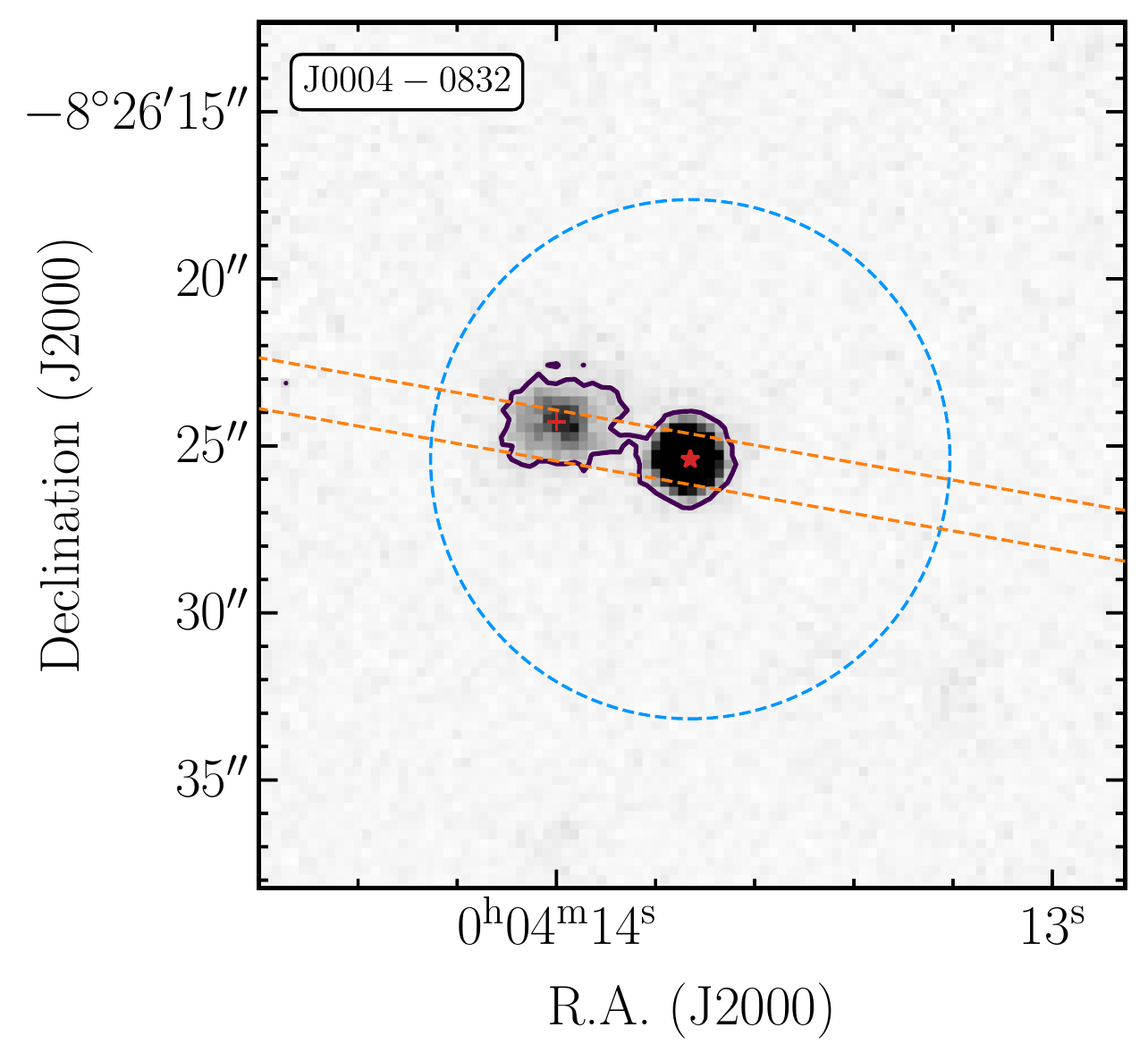}
    \end{minipage}
    \hfill
    \begin{minipage}[t]{.725\textwidth}
        %\centering
        \flushright
        \includegraphics[height=5cm, width=\textwidth]{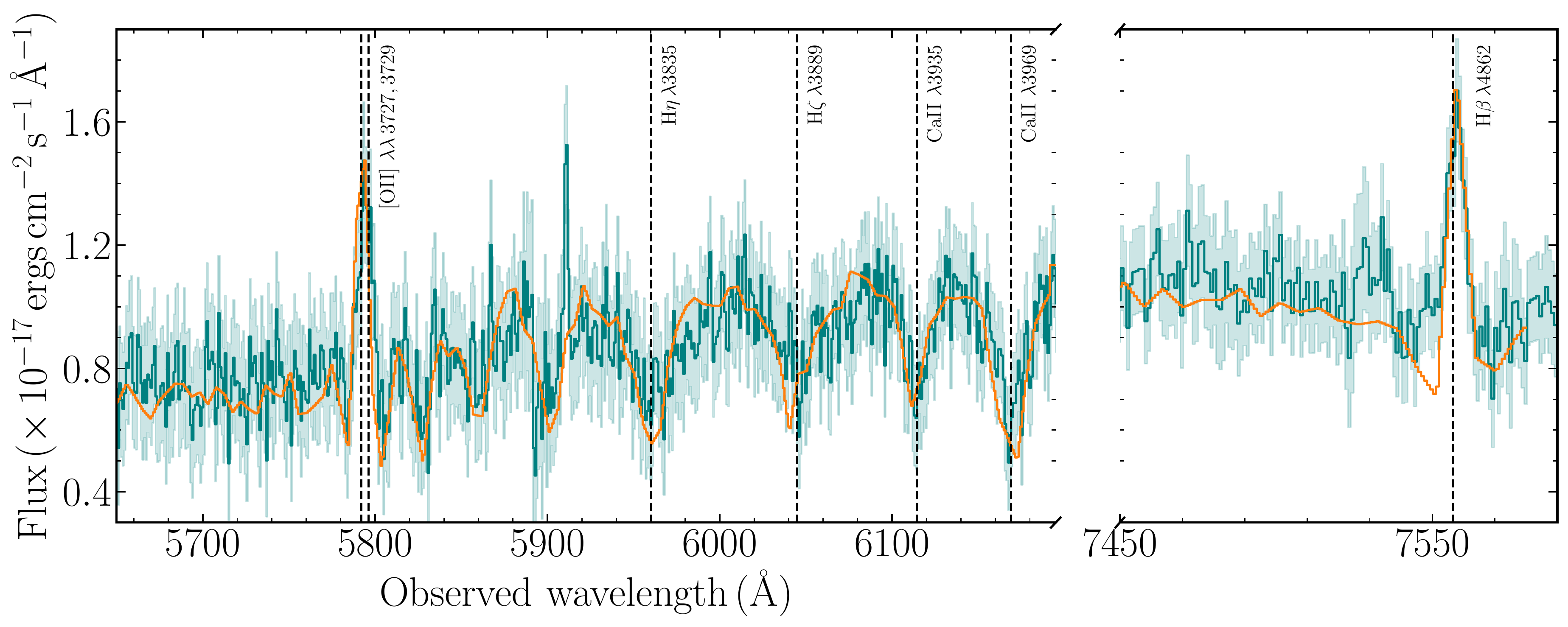}
    \end{minipage}
    
    \centering
    \includegraphics[width=\textwidth, height=6cm]{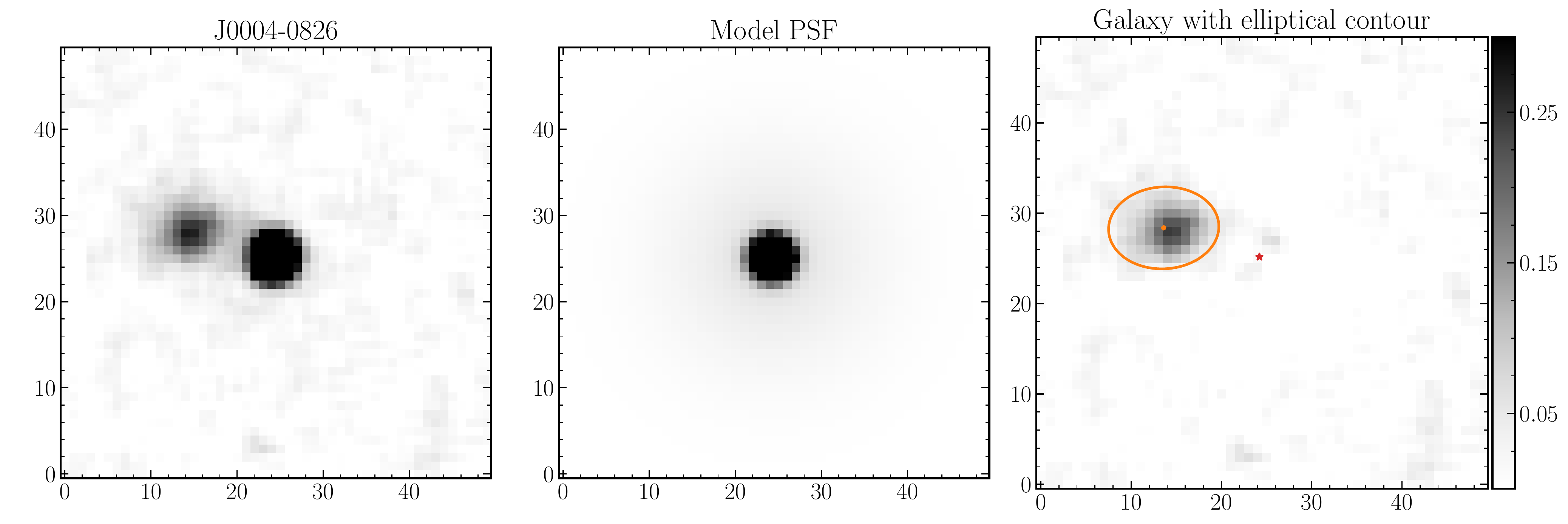}
    \caption{ {\it Top left:}  The DECaLS r-band image of the observational configuration, with the background quasar and the targeted galaxy marked with symbols, `$\star$' and `+', respectively, for the \usmg system at \zabs = 0.5544 towards J0004-0826. The blue dashed circle corresponds to a projected radius of 50 kpc at the redshift of the host galaxy around the quasar. {\bf Red dashed lines show the slit orientation used in our spectroscopic observations.} {\it Top right:}
    observed spectrum of the galaxy  in cyan-green colour with different emission and absorption lines marked using vertical dashed lines.  The synthetic spectrum obtained from the SED fitting exercise, resampled at the observed wavelengths, is overplotted on top of the observed galaxy spectrum (orange). The bottom panels demonstrate the process of modeling and subtracting the quasar PSF from the SDSS photometry to obtain the galaxy fluxes.
    {\it Bottom left:}
    The actual $50"\times50"$ SDSS i-band image of the quasar J0004-0826. {\it Bottom middle:} the model QSO image obtained by the fitting the quasar image with the SDSS PSF. {\it bottom right:} Residual image after the pixel-by-pixel subtraction of the middle panel from the left panel. The orange ellipse presents the inclination and the orientation of the galaxy obtained from fitting the galaxy image.
    The red `$\bullet$' and `$\star$' mark the center of the galaxy and the quasar respectively.    } 
    
    \label{fig:J0004SED}
\end{figure*}

Next we measure the  host galaxy properties such as B-band absolute magnitude and luminosity (${\rm M_B}$ and ${\rm L_B}$), stellar mass ($M_\star$), star formation rate (SFR) and specific star formation rate (sSFR) using available spectroscopic and photometric data and template fitting methods. Since the \usmg host galaxies are at  low impact parameters to the quasar sightline, we model the quasar image using the appropriate SDSS point spread function \citep{Xin2018}, and then subtract it from the original image before extracting the parameters of the host galaxies. The procedure for modeling and subtracting the PSF for the \usmg system J0004-0826 is shown in Figure \ref{fig:J0004SED}. We measure the flux of the host galaxy in different photometric passbands and correct for finite telescope gain using standard approaches presented in the \texttt{SExtractor} software \citep{Bertin1996} and the python fork of the software, Source Extraction and Photometry \citep[][]{Barbary2016}.

We make use of the publicly available python package, Bayesian Analysis of Galaxies for Physical  Inference  and  Parameter  EStimation \citep[\BAGPIPES;][]{Carnall2018}, to extract the galaxy parameters. 
We consider the stellar population models by \citet{Bruzual}, constructed assuming the \cite{Kroupa} IMF, and recently updated by \citet{Charlot2016} to include the MILES stellar spectral library \citep{Barroso} and an updated stellar evolutionary track \citep{Marigo}. We used the default implementations of nebular emission, dust and IGM attenuation in \BAGPIPES. More details of modeling and fitting the galaxy SED to the spectroscopic/photometric data using \BAGPIPES\ can be found in section 3.1 of \citet{Carnall2018}.

To start with, we fix the redshift of the galaxies to their spectroscopic redshifts. We assume that all stars within the galaxy have the same metallicity, and vary it with a uniform prior in the range $0.01 - 2.5$ of the solar metallicity. We parameterize the star formation histories (SFHs) using the delayed exponential model. We also chose a flat prior for the logarithm of the total stellar mass formed, in the units of solar mass in the range $0-13$. 
It is known that derived parameters depend on our choice of SFH. Using  \BAGPIPES, \citet{Carnall2019} have obtained the $M_\star$, SFR, dust attenuation and metallicity posteriors for the galaxies to lie within $\rm{1-2\sigma}$ of the original values in the mock catalog with four different kinds of SFHs (an exponentially declining SFH, delayed exponentially declining SFH, a log-normal SFH and, a double power law SFH) with a bias of about $\rm{0.1}$ dex in $M_\star$ and a bias of $\rm{0.2}$ dex in SFR.  We shall keep this in mind while discussing these parameters.
The whole process is demonstrated in Figure~\ref{fig:J0004SED} for the \usmg system J0004-0826. In the top panel, we show the QSO field in the left and the observed spectrum in the right panels. In the bottom panels, we show the QSO image (left), model PSF (middle) and the galaxy image after subtracting the model QSO PSF (right). In the top left panel, we overlay the best fitted SED on the observed spectrum. This clearly reproduces all the absorption and emission lines and the stellar continuum very well.

The model parameters for the best fitted case are summarized for 12 \usmg host galaxies, where the host galaxies are photometrically separable from the background quasars in Table.~\ref{tab:photometric_props}. For each \usmg system, this table provides, coordinates of the background quasars and the \usmg host galaxies (columns 1 and 2), inferred stellar mass, rest frame B-band absolute magnitude, rest frame B-band luminosity and SFR in columns 2, 4, 5 and 6 respectively. { The stellar mass of the \usmg host galaxies is in the range, 10.21 $\rm{\leqslant\, \log\,(M_\star/ M_\odot) \, \leqslant}$ 11.62, with a median value of $\rm{\log\, (M_\star / M_\odot)}$ = 10.64.}
We convert the stellar mass to the dark matter halo mass ($M_h$) using the stellar-to-halo mass relation provided by \citet{Girelli2020}. Halo masses and the corresponding virial radius, $R_{h}$ (we use $R_{200}$ for this purpose), 
associated with the \usmg host galaxies are given in column 8 and 9 respectively.

Next we estimate the B-band luminosities using  the synthetic spectrum obtained from the SED fitting of the \usmg host galaxies. First, we de-redshift the synthetic spectrum to its rest frame by conserving the total energy and then we calculated the rest frame B-band flux by appropriately weighting the B-band filter transmission curve. Following the prescription of \citet{Fukugita}, we convert this flux to the absolute magnitude of the galaxy (given in column 4 of Table~\ref{tab:photometric_props}). Then we determined the B-band luminosity of these galaxies by comparing their magnitude with the solar magnitude and luminosity. These values are summarized in column 5 of Table.~\ref{tab:photometric_props}.
 The \usmg host galaxies have absolute B-band magnitude between -19.05 and -21.68, with a median value of -20.49. Only two of the galaxies are brighter than the characteristic B-band luminosity of the field galaxies at redshift $z = 0.5$ \citep[$M_B^\star = -21.15$;][]{Faber2007}.

Further, for a set of isolated host galaxies of  \MgII\  systems,
(defined as host galaxies without any nearest neighbor within 100 kpc and velocity separation within 500 \kms) from the \Magiicat\ survey \citep{Nielsen_2013}, we calculated the rest-frame $\rm{B}$-band absolute magnitudes from the synthetic spectrum obtained from the SED fitting, and compared them against the values provided in the \Magiicat survey and found them to be consistent. We also calculated the galaxy parameters for the host galaxies identified in the \Magiicat\ survey. Although in the \Magiicat\ survey, the impact factor goes up to 200 kpc, we are particularly interested in the galaxies with a smaller impact parameter, as isolated \usmg host galaxies identified in the combined sample of our work and from the literature have a maximum impact parameter of 53.8 kpc. Therefore, we consider the isolated \MgII\ absorption systems from the \Magiicat\ survey up to 60 kpc.  { For these galaxies, we find 8.65 $\rm{\leqslant\, \log\,(M_\star / M_\odot) \, \leqslant}$ 12.12, with a median value of $\rm{\log\, (M_\star / M_\odot)}$ = 10.51.}
In the \Magiicat\ survey, there are a total of 98 isolated \MgII\ host galaxies within the impact parameter of 60 kpc, out of which 76 are observed and identified as photometric sources in the SDSS photometric survey.
In the following sections, we will be comparing the host galaxy parameters of our \usmg\ absorbers with those obtained for these galaxies.

\section{Results}
\label{sec:results}

In this section, we discuss the properties of identified \usmg host galaxies and compare  them with that of GOTOQs \citep{Joshi2017}, Mg~{\sc ii} host galaxies identified in surveys like \Magiicat\ \citep{Nielsen_2013}, 
\textsc{MAGG} \citep{Dutta2020}, \textsc{MEGAFLOW} \citep{Schroetter2019} and  SIMPLE \citep{Bouche2007}.
We also explore possible correlations between the observed properties of galaxies and absorption lines. However, one has to keep in mind that our survey of \usmg host galaxies, GOTOQs, \textsc{MEGAFLOW} and SIMPLE are `absorber-centric' surveys, i.e., aimed at studying the host galaxies of known absorption systems. \textsc{MAGG} survey, on the other hand, is a `galaxy-centric' survey, i.e., aimed at understanding the gas distribution around a population of galaxies while \Magiicat is a combination of both types from the literature.

\begin{table*}
     \begin{threeparttable}
 \caption{Galaxy properties inferred from the observed spectra. Emission line fluxes are given in the units of $(\times 10^{-17} ergs\,cm^{-2}\,s^{-1})$.} 
% \Anand{Give the redshifts and emission line flux for the GOTOQ J1025-0046}{\color{DarkOrchid}{(Done)}}
    \centering
    \begin{tabular}{lccccccccr}
        \hline
         No. & Quasar  & $z_{abs}$ & $z_{gal}$ & D & $f_{\OII\ 3728}$ & $f_{H\beta}$ & $f_{\OIII\ 5008}$ & Z & $\log q$\\
 %        \rule{0pt}{4ex}
           &                               &        &        & (kpc) &                 &                 &                 &     & \\
           (1) & (2) & (3)&(4) & (5) & (6)&(7)&(8) &(9) &(10)\\
          \hline
          \hline
         1 & J000413.73-082625.4\tnote{a}  & 0.5544 & 0.5539 & 26  & 4.96$\pm$0.71   & 3.76$\pm$0.47   & 1.80$\pm$0.20   & $ 8.83\pm0.04 $  &  $ 7.61\pm0.20 $  \\
         2 & J015635.18+034308.1\tnote{a}  & 0.5581 & 0.5590 & 14  & 8.73$\pm$1.22   & $\le$ 0.60        & $\le$ 0.62        & --   &  --   \\
         3 & J021820.10-083259.4\tnote{c} & 0.5896 & 0.5895 & 16   &$\le2.1$         & $\le 1.30$      & $\le 1.70$      & --   &  --   \\
         4 & J024008.21-081223.4\tnote{a}  & 0.5311 & 0.5311 & 18  & 6.93$\pm$0.96   & 3.65$\pm$0.48   & 1.98$\pm$0.60   & $ 8.71\pm0.08 $  & $ 7.54\pm0.19 $ \\
         5 & J033438.28-071149.0           & 0.5977 & 0.5988 & 28  & 23.72$\pm$ 0.56 & 13.22 $\pm$0.43 & 12.08$\pm$ 0.40 & $ 8.69\pm0.07 $  & $ 7.66\pm0.16 $   \\
         6 & J085627.09+074031.7           & 0.5232 & 0.5235 & 24  & 1.72$\pm$0.24   & 0.48$\pm$0.05   & $\le$ 0.62        & --   &  --   \\
         7 & J092222.58+040858.7           & 0.4549 & 0.4548 & 15  & 19.24$\pm$2.80  & 6.92$\pm$0.85   & 14.51$\pm$1.78  & $ 8.45\pm0.09 $  & $ 7.75\pm0.13 $  \\
         8 & J121453.29+080457.7\tnote{a}  & 0.4908 & 0.4911 & 16  & 17.35$\pm$2.47  & 6.45$\pm$1.29        & 4.63$\pm$0.52   & $8.60\pm0.11$   & $7.41\pm0.14$   \\
         9 & J121628.03+035031.8\tnote{b}  & 0.5519 & 0.5514 & 8.1  & 9.39$\pm$1.37   & 2.09$\pm$0.23   & 3.71$\pm$0.45   & $ 8.47\pm0.07 $  & $ 7.60\pm0.10 $   \\
         10 & J155003.71+031325.0\tnote{a}  & 0.5694 & 0.5695 & 23  & 5.34$\pm$ 0.73  & $\le$ 0.30       & --              &  --  & --    \\
         11 & J210851.53-074726.5\tnote{b} & 0.5187 & 0.5184 & 12  & 2.67$\pm$ 0.36  & 0.47 $\pm$0.07  & --              &  --  & --    \\
         12 & J212143.98+003954.2          & 0.5509 & 0.5509 & 19  & 9.80$\pm$ 1.10  & 3.60 $\pm$0.74  & 5.30 $\pm$ 1.75 & $ 8.50\pm0.13 $  & $ 7.51\pm0.21$   \\
         13 & J212727.20+082724.6\tnote{b} & 0.4392 & 0.4397 & 7.3  & 10.30$\pm$ 0.72 & 3.19 $\pm$0.40  & 2.17 $\pm$ 0.34 & $ 8.53\pm0.07 $  & $ 7.22\pm0.10 $  \\
         14 & J220330.04-002211.4          & 0.4381 & 0.4375 & 31  & 47.69$\pm$ 3.21 & 16.51$\pm$0.77  & 23.60 $\pm$ 1.74& $ 8.48\pm0.10 $  & $ 7.58\pm0.11 $   \\
         15 & J220702.53-090127.7\tnote{d} & 0.5623 & 0.5621 & 38  & 2.82$\pm$0.31   & --              & --              & --   &  --   \\
            &                              &        & 0.5623 & 55  & 14.14$\pm$0.78  & --              & --              & --   &  --   \\
            &                              &        & 0.5623 & 209   & 14.13$\pm$1.57  & --              & --              & --   &  --   \\
            &                              &        & 0.5604 & 246   & $\le 7.12$      & --              & --              & --   &  --   \\
         16 & J232653.15+002142.9          & 0.5624 & 0.5621 & 33  & 4.61 $\pm$ 0.49 & $\le$ 0.88        & $\le$ 1.69        &  --  &  --   \\
         17 & J233548.62-023734.3          & 0.5081 & 0.5080 & 16  & 4.47$\pm$ 0.66  & $\le$ 0.53        & $\le$ 0.33        &  --  &  --   \\
         18 & J233818.25-005610.5\tnote{e} & 0.4801 & 0.4798 & 79  & --              & --              & --              & --   & --    \\
          \multicolumn{10}{c}{GOTOQ without SALT observations}\\
          19 & J102510.10-004644.9         & 0.5561 & 0.5562 &$\le$ 6.45 & 6.14$\pm$1.68 & $\le$ 3.29        & $\le$ 1.98        & --   & --    \\
          20 & J093020.60+001828.0         & 0.5928 & 0.5928 &$\le$ 6.5  & 12.19$\pm$2.45 & 4.23 $\pm$ 1.03  & 8.79$\pm$0.97 &  --  &  -- \\
         \multicolumn{10}{c}{The non-\usmg\ absorber in our lines of sight}\\
         21 & J111400.00$-$002342.6        & 0.7981 & 0.7983 & 22.4  & $1.98\pm0.25$   & --              & --              & --   & --    \\
         \hline
         \end{tabular}
         \begin{tablenotes}
           \item[a] Galaxy showing strong stellar absorption from Ca~{\sc ii} and H~{\sc i} in its spectrum.
           \item[b] The host galaxy is at small angular separation from the QSO and its image merges with that of the galaxy. Although only J1216+0350 is a GOTOQ. 
           
           \item[c] Galaxy information is from \citet{Rahmani2016}.
           \item[d] Galaxy information is from \citet{Gauthier2013}. We consider the galaxy with the lowest D for correlation analysis involving impact parameters.
           \item[e] There are two potential host galaxies at smaller impact parameters for which we could not get reliable spectra 
           [see Figure~\ref{fig:qso_fields}].
         \end{tablenotes}
    \label{tab:spectroscopic_properties}
              \end{threeparttable}
\end{table*}

\begin{table*}
     \caption{Properties of the \usmg\ host galaxies not sitting on top of the background quasars inferred from the SED fitting.
     }
     \centering{
     \begin{tabular}{ccccccccr}
     \hline
     Quasar &  Galaxy   &$\rm{log\, (M_\star / M_\odot)}$ & $M_B$ & $[L_B/10^{43}]$  & SFR (SED)  &  SFR$_{\rm [O\textsc{ii}]}$ & $\rm{\log\, (M_h / M_\odot)}$ & $\rm{R_h}$\\
     &  &   &   & (ergs s$^{-1}$) & ($\rm{M_\odot\, yr^{-1}}$)  & ($\rm{M_\odot\, yr^{-1}}$) &    &  (kpc) \\
    (1)  &  (2) &   (3) &   (4) &  (5)  & (6)  & (7)  & (8) & (9)  \\
     \hline
     \hline
       J000413.73-082625.4  &  J000413.99-082624.27 & $10.99^{+0.06}_{-0.06}$ & $-21.68\pm0.09$ &  $15.36\pm0.09$  &   $10.25^{+0.96}_{-0.87}$ & $1.37\pm0.44$ & 12.24  & 228\\
      J015635.18+034308.1  & J015635.21+034306.03 & $10.21^{+0.33}_{-0.33}$ & $-20.06\pm0.25$ &  $3.45\pm0.79$  &   $7.39^{+49.52}_{-6.34}$ & $2.31\pm0.73$  & 11.65    & 145\\
       J024008.21-081223.4  &  J024008.39-081222.53 & $10.57^{+0.07}_{-0.08}$ & $-20.68\pm0.13$ &  $6.13\pm0.76$  &   $5.87^{+0.74}_{-0.87}$ & $1.55\pm0.49$  & 11.87    & 174\\
      J033438.28-071149.0  & J033438.08 -071152.08 & $10.57^{+0.12}_{-0.14}$ & $-20.81\pm0.15$ &  $6.87\pm0.94$  &   $4.55^{+1.16}_{-1.12}$ & $6.50\pm1.86$   & 11.88    & 169\\
      J085627.09+074031.7  & J085627.31+074029.51 & $11.02^{+0.21}_{-0.42}$ & $-19.05\pm0.16$ &  $1.37\pm0.20$  &   $17.90^{+10.29}_{-7.60}$ & $0.26\pm0.08$   & 12.27    & 237\\ 
      J092222.58+040858.7  & J092222.52+040856.00  & $10.26^{+0.09}_{-0.07}$ & $-20.06\pm0.12$ &  $3.46\pm0.39$  &   $2.59^{+0.23}_{-0.34}$ & $2.06\pm0.66$   & 11.66    & 154\\
      J121453.29+080457.7  & J121453.46+080457.03   & $10.71^{+0.13}_{-0.17}$ & $-20.28\pm0.12$ &  $4.22\pm0.46$  &   $4.95^{+0.78}_{-0.82}$ & $2.23\pm0.71$  & 11.97    & 192\\
      J155003.71+031325.0  & J155003.47+031325.77 & $11.62^{+0.19}_{-0.20}$ & $-21.21\pm0.10$  &  $9.99\pm0.96$  &   $33.51^{+22.0}_{-14.13}$ & $1.44\pm0.46$ & 13.56   & 624\\
      J212143.98+003954.2  & J212144.18+003954.25  & $10.37^{+0.15}_{-0.15}$ & $-20.33\pm0.14$ &  $4.42\pm0.56$  &   $7.93^{+1.70}_{-1.89}$ & $1.66\pm0.51$ & 11.75   & 157\\
      J220330.04-002211.4 & J220329.76-002215.03   & $10.22^{+0.11}_{-0.12}$ & $-20.39\pm0.10$ &  $4.68\pm0.43$  &  $3.74^{+0.61}_{-0.70}$ & $4.68\pm1.34$    & 11.64   & 153\\
      J232653.15+002142.9 & J232653.20+002148.10  & $11.07^{+0.24}_{-0.26}$ & $-20.68\pm0.21$ &  $6.13\pm0.21$ &   $20.14^{+10.14}_{-7.74}$ & $1.03\pm0.32$  & 12.34   & 245\\
      J233548.62-023734.3 & J233548.79-023734.79   & $10.81^{+0.15}_{-0.13}$ & $-20.58\pm0.10$ &  $5.58\pm0.53$ &   $6.65^{+2.59}_{-1.49}$ & $0.62\pm0.19$   & 12.06   & 204\\
     \hline
     \end{tabular}
     }
     \label{tab:photometric_props}
 \end{table*}

\subsection{\usmg systems and star forming galaxies:}

 We have identified host galaxies in the case of 18 out of 21 (i.e. $\sim$86\%) \usmg\ absorbers listed in Table~\ref{tab:observation_log}.
 Only for one (i.e. \zabs = 0.5896 towards J0218-0832) of these 18 \usmg\ absorbers (listed in Table~\ref{tab:spectroscopic_properties})
 the host galaxy does not show \OII\ emission. The galaxy redshift in this case was measured based on absorption lines and a weak H$\alpha$ emission \citep{Rahmani2016} corresponding to a SFR of 0.42 M$_\odot$ yr$^{-1}$. 
  Moreover, at least for 16 (i.e., excluding z = 0.4801 absorber towards J2338-0056) out of 21 \usmg absorbers (i.e., all the \usmg\ systems for which we have host galaxy observations listed in Table~\ref{tab:observation_log}) the identified host galaxies clearly show detectable nebular emission lines indicating an on-going star formation. The \OII\ besed SFR are higher than 1 M$_\odot$ yr$^{-1}$ in 14 galaxies.
  From the 21 systems discussed above and the two GOTOQs J1025-0046 and J0930+0018  that were not observed with SALT, we can conclude at least { $\sim$70\%} of the \usmg\ absorbers (i.e. 16/23) are associated with star forming galaxies (i.e., SFR $>1$ M$_\odot$ yr$^{-1}$) and between 4.8--30\% of the absorbers are hosted by passive galaxies. 

Among the three non-detected cases (\usmg absorption towards J2045-0704 at \zabs = 0.5649, J2301-0212 at \zabs = 0.5367, and, J1114-0023 at \zabs = 0.5610), for the first two cases we obtained the upper limits on $L_{[O\textsc{ii}]}$  (or SFR) assuming the targeted potential \usmg host galaxies are indeed the \usmg host galaxies (see last row of Figure \ref{fig:qso_fields}). To calculate the upper limit on the SFR, we first remove the quasar trace from the 2D spectrum around the region of the expected \OII\ emission as discussed in \ref{sec:gp_spec}. Then we selected a 2D rectangular patch centered on the expected \OII\ emission having width equal to the typical width of the \OII\ emission line (FWHM = 350 $km\,s^{-1}$) in our sample and height equal to the FWHM of spectral PSF obtained by fitting the quasar trace. Next, we calculated the total counts within this rectangular patch by summing over all the pixel counts that happens to fall within this patch.  We convert this to \OII\ flux using our flux calibration solutions. The $3\sigma$ upper limits on $L_{[O\textsc{ii}]}$ for the host galaxies of the \usmg\ absorbers along the line of sight to J2045-0704 and J2301-0212 are 0.76 and 1.12$\times10^{41}$ erg s$^{-1}$ respectively. For the \usmg system J1114-0023, the targeted galaxy turns out to be a high redshift galaxy and from the DECaLS photometry, we find that the galaxy with the consistent photometric redshift of the \usmg absorption sits outside our slit.

In their SINFONI Mg~{\sc ii} Program for Line Emitters (SIMPLE), \citet{Bouche2007} have identified host galaxies of 6 \usmg systems at $0.8<z<1.0$ out of 8 systems searched for H$\alpha$ emission.
\citet{Nestor_2011} also have identified the host galaxies associated with two \usmg\ absorption systems at \zabs $\sim$ 0.7 (i.e. \zabs = 0.7646 system towards 0747+305 and \zabs = 0.669 towards Q1417+011). Both the \usmg\ absorption were associated with a pair of starburst galaxies in the impact parameter  range of 29-61 kpc. In the MAGIICAT full sample there is only one \usmg\ system \citep[i.e. \zabs = 0.836627 towards J000448.11$-$415728.8 from][]{Guillemin1997}. The measured impact parameter in this case is 53.8 kpc, and the galaxy does not show detectable \OII\ emission.

\citet{Dutta2020} have identified a total of 53 galaxies either in an isolated or in a group environment, associated with 21 \MgII\ absorbers out of a total sample of 27 \MgII\ absorbers spanning a  range in $\mathrm{W_{2796}}$ of $0.02-3.2$\,\AA\ at  $0.8\le z\le1.5$. This gives a host galaxy detection  rate of $78$ \%  that is 90\% complete down to a \OII\ flux limit of $3\times10^{-18}$\,erg\,s$^{-1}$\,cm$^{-2}$. Only one \usmg system is present in this sample associated with an isolated galaxy with an impact parameter of 18 kpc. \citet{Schroetter2019} have identified at least one host galaxy using the \OII\ emission line associated with 59 strong \MgII\  absorbers ($\rm{W_{2796} \geqslant 0.3\AA }$), with a detection rate of about 75\% out of the 79 strong \MgII\ absorption systems present along the line of sight of 22 quasars they studied as a part of their MEGAFLOW survey using the MUSE IFU. Their  5$\sigma$ detection limit is  $\sim 4\times10^{-18}$\,erg\,s$^{-1}$\,cm$^{-2}$.
There are three \usmg absorbers in this sample, with all of them associated with isolated \OII\ emitting galaxies with  impact parameters in the range 9.1 -24.5 kpc.

We summarize the details of \usmg systems from the literature in Table~\ref{append :usmg_litterature} of the appendix. Out of these 14 \usmg\ literature systems, host galaxies of  13 (i.e. $\sim 93$\%) show nebular emission lines with SFR in excess of 2 M$_\odot$ yr$^{-1}$ (see last column in Table~\ref{append :usmg_litterature}).

In the same redshift range as our \usmg\ sample, \citet{Rahmani2016} have detected host galaxies of DLAs and sub-DLAs with a success rate of 78\% at an impact parameter of 10-30 kpc, with 71\% of the host galaxies showing multiple nebular emission lines [typical limiting flux is in the range $(1.5-6.0)\times 10^{-17} erg~s ^{-1}$ cm$^{-2}$].Note that different literature studies have different selection criteria (pre-selection/blind), observing strategy (long-slit/MOS/IFU), sensitivity limit of observations, definition of absorber-galaxy association (i.e. velocity and impact parameter range considered).  { Despite this, the discussions presented above confirm that the \OII\ emission is frequently detected irrespective of  ${\rm W_{2796}}$. Motivated by this, we check whether there is any correlation between ${\rm W_{2796}}$ and \OII\ luminosity in section~\ref{sec:loii}.}

 \subsection{Velocity difference between the host galaxy and the \usmg\ absorber}

\begin{figure}
    \centering
    \includegraphics[viewport=7 7 550 400, width=0.475\textwidth,clip=true]{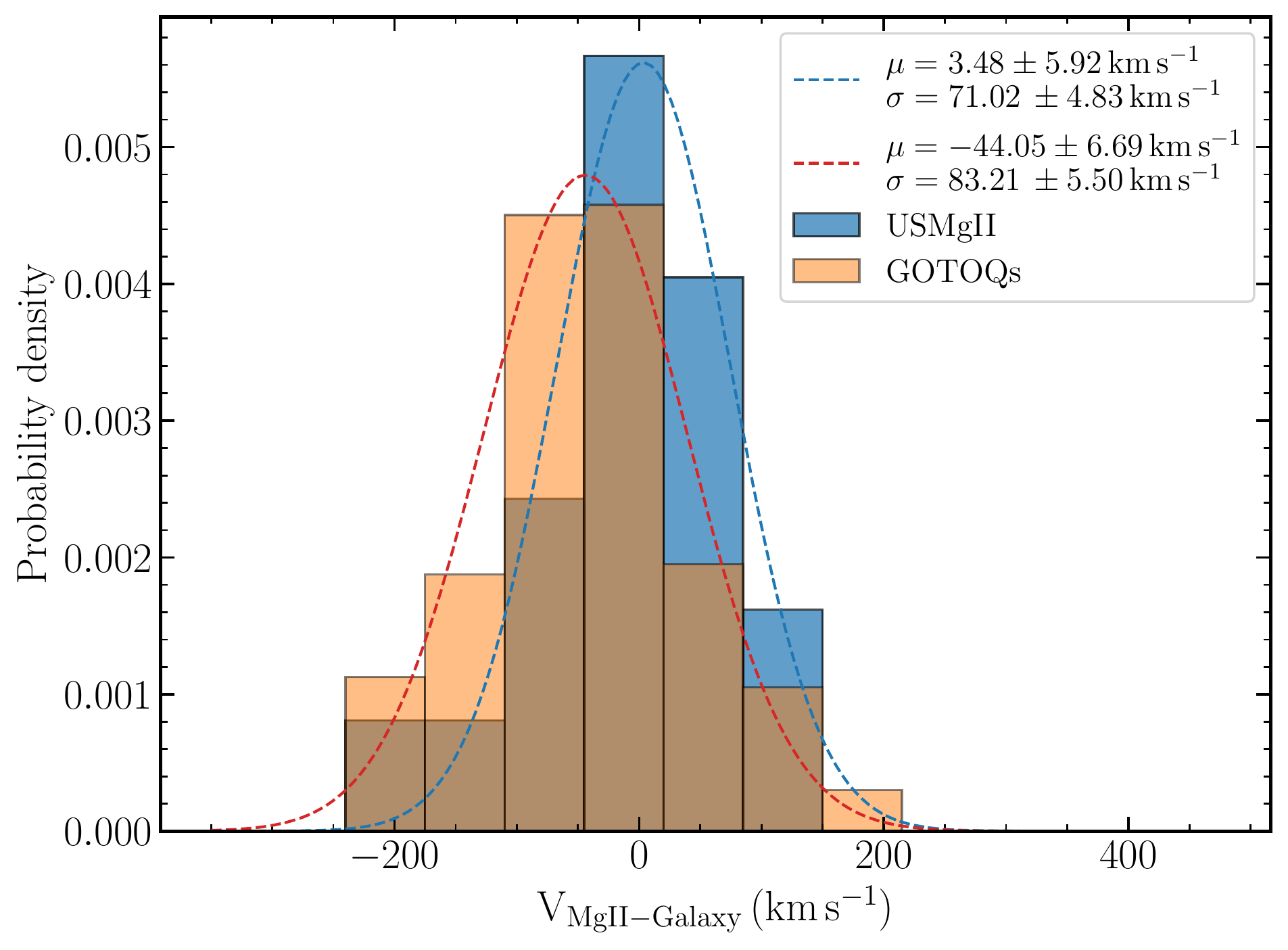}
    \caption{The distribution of the line of sight velocity separation between the \MgII\ absorption and the \OII\ emission detected from the associated host galaxies. The blue and orange histograms represent the \usmg host galaxies from our sample and the GOTOQs from \citet{Joshi2017}, respectively. The dashed blue and orange line corresponds to the Gaussian fits to the histograms associated with the \usmg and the GOTOQs, respectively. The Gaussian parameters for these fits are also provided in the figure.}
    \label{fig:veldiff}
\end{figure}

In Figure~\ref{fig:veldiff}, we plot the histogram of the rest frame velocity offset between the host galaxy and the \MgII\ absorption. The velocity difference is measured using $z_{abs}$ of \MgII\ absorption seen in the quasar spectrum with respect to the galaxy redshift measured using the \OII\ line. The distribution is well fitted with a Gaussian having a mean velocity offset of, $\mu = 3.48\pm 5.92$ \kms with a dispersion of $\sigma = 71.02\pm 4.83$ \kms. All the identified host-galaxies have velocity offset well within 250 \kms\ to the absorption redshift.  Note that the uncertainty in measured galaxy redshift from the \OII\ is typically 60 \kms. This figure also shows the distribution of velocity difference found for GOTOQs from \citet{Joshi2017} in the same redshift range. In this case we find $\mu = -44.05\pm6.69$ \kms with a dispersion of $\sigma = 73.21\pm 5.50$ \kms. Significant albeit small non-zero $\mu$ value found here is interesting. We will investigate this in more detail using long-slit spectroscopy of bright GOTOQs in our upcoming paper.

\citet{Huang2021} have found $\mu \sim 0$ \kms with a dispersion of $\sigma = 84$ \kms
for their full sample of galaxies associated with Mg~{\sc ii} absorbers at $z=0.10-0.48$.
They also found the dispersion to be larger (i.e. $\sigma\sim235$ \kms) for the non-isolated galaxies. 
In the case of the MAGG sample with absorbers at $0.8<z<1.5$, the measured median velocity difference is $\mu=-8.0$ \kms and $\sigma\approx62$\,\kms\ for the single galaxies and $\approx208$\,\kms\ for the group galaxies \citep[][]{Dutta2020}. Seventy five percent of these galaxies are within $\pm200$ \kms\ of the \MgII\ absorption.
Interestingly in our sample, the only system with clearly identified non-isolated host galaxies ({ beyond D=100 kpc}) is \usmg\ absorber towards J2207-0901 at \zabs = 0.5623  studied by \citet[][] {Gauthier2013}. If we combine this with the non-isolated host galaxies studied by \citet{Nestor2011}, we find the mean velocity offset to be $\mu = 106.4$ \kms\ and $\sigma = 201.3$ \kms. This trend of larger velocity dispersion for non-isolated galaxies, though based on only 3 \usmg\ absorbers, is consistent with what has been found for literature samples discussed above.

\subsection{Impact parameter: distribution and correlations}
\label{sec:IP}

\begin{figure}
    \centering
        \includegraphics[width=0.45\textwidth]{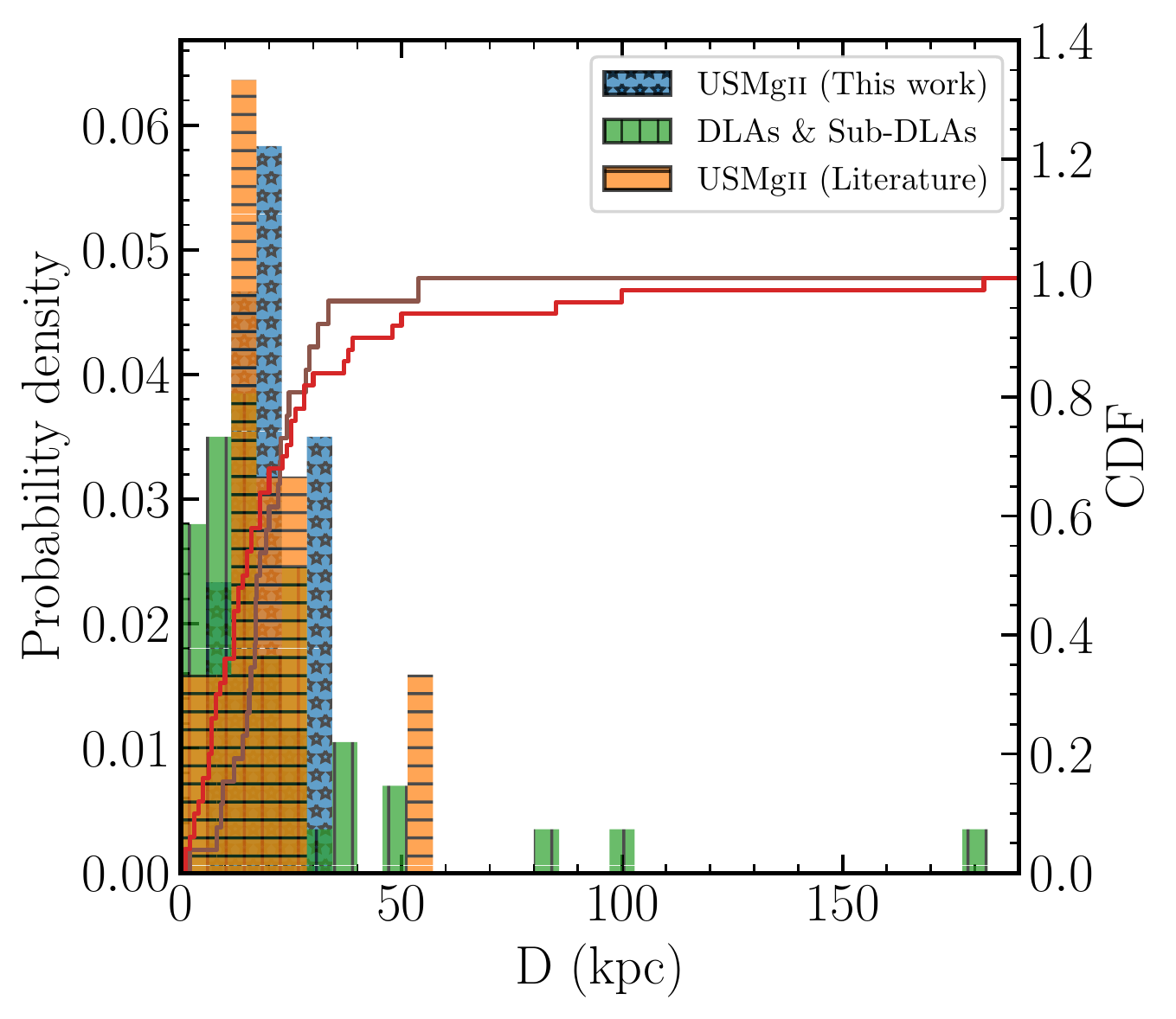}
    \caption{The impact parameter distribution of the \usmg host galaxies identified in our survey (blue), from the literature (orange)  and the host galaxies of  DLAs and sub-DLAs \citep[green,][]{Rahmani2016}. {{The cumulative distribution of D for the \usmg absorbers (ours + literature sample) and that for the DLAs and sub-DLAs are shown in brown and red respectively.}}
}
    \label{fig:location}
\end{figure}

The measured impact parameter (D) of the \usmg\ host galaxies are listed in column 5 of Table~\ref{tab:spectroscopic_properties} and its distribution is shown in Figure~\ref{fig:location}.
In the case of multiple host galaxy identifications we consider the  nearest one (i.e. smallest D values).
Recall, in the case of  \usmg\ system at \zabs = 0.4801 towards J2338-0056, the galaxy with the consistent redshift is at D = 79 kpc, while there are two nearby galaxies for which we could not measure the redshifts.
When we consider the nearest identified galaxy to the \usmg\ absorber (excluding the case of J2338-0056), the D values are in the range 7.3 to 38 kpc with a median D of 18 kpc. If we include two GOTOQs, \zabs = 0.5561 system towards J1025-0046 and \zabs = 0.5928 towards J0930+0018, then the median D is 16 kpc.

{ As mentioned before DLAs tend to have large ${ W_{2796}}$, therefore it is important to compare the D-distributions of \usmg\ and DLAs in order to understand differences in their origin.}
In Figure~\ref{fig:location} we also show the distributions of D for the \usmg host galaxies found in the literature (see Table~\ref{append :usmg_litterature} in the appendix), as well as the host galaxies of DLAs and the sub-DLAs (at $0.01\le$ \zabs$\le 3.4$) given in \citet{Rahmani2016}.
In the sample of \usmg systems from the literature, the impact parameter for the nearest galaxy ranges from 2 to 53.8 kpc with a median D of 17 kpc.
The impact parameter of the DLAs and sub-DLAs ranges from 1 to 182 kpc with a median D of 14 kpc.
We performed a KS test between the  distributions of D for the two samples (sub-DLAs + DLA sample vs. the combined \usmg sample, i.e., ours + literature sample).
We find the maximum difference between the cumulative distributions to be 0.30 and the probability of this difference to occur by chance being 0.08. This suggests that the two samples of absorbers belong to different parent populations at the $\sim2\sigma$ level.
The difference mainly comes from few DLA host galaxies with large impact parameters.

{ Next we explore the correlation between ${\rm W_{2796}}$ and other parameters such as E(B-V), different equivalent width ratios and velocity offset. We do not find any clear trend between E(B-V) and D in our sample. Spearman rank correlation analysis between the two gives a correlation coefficient of, $\rm{r_S}$ = $-$0.05, confirming the lack of a correlation.  We also do not find any significant correlation (or anti-correlation) between D  and $\mathrm{W_{2853}^{MgI}/W_{2796}^{MgII}}$. We do see some trend of large $\mathcal{R} = \mathrm{W_{2600}^{FeII}/W_{2796}^{MgII}}$ having smaller D (when we exclude the \zabs = 0.5624 system towards J232653.15+002142 that has Mg~{\sc ii} doublet ratio less than 1 and $\mathrm{W_{2600}^{FeII} = 4.29}$\AA). We find a Spearman correlation coefficient of $r_S$ = -0.48 between $\rm{D}$ and $\mathcal{R}$. It will be interesting to  confirm this trend with a larger sample.

\begin{figure}
    \centering
    \includegraphics[width=0.49\textwidth]{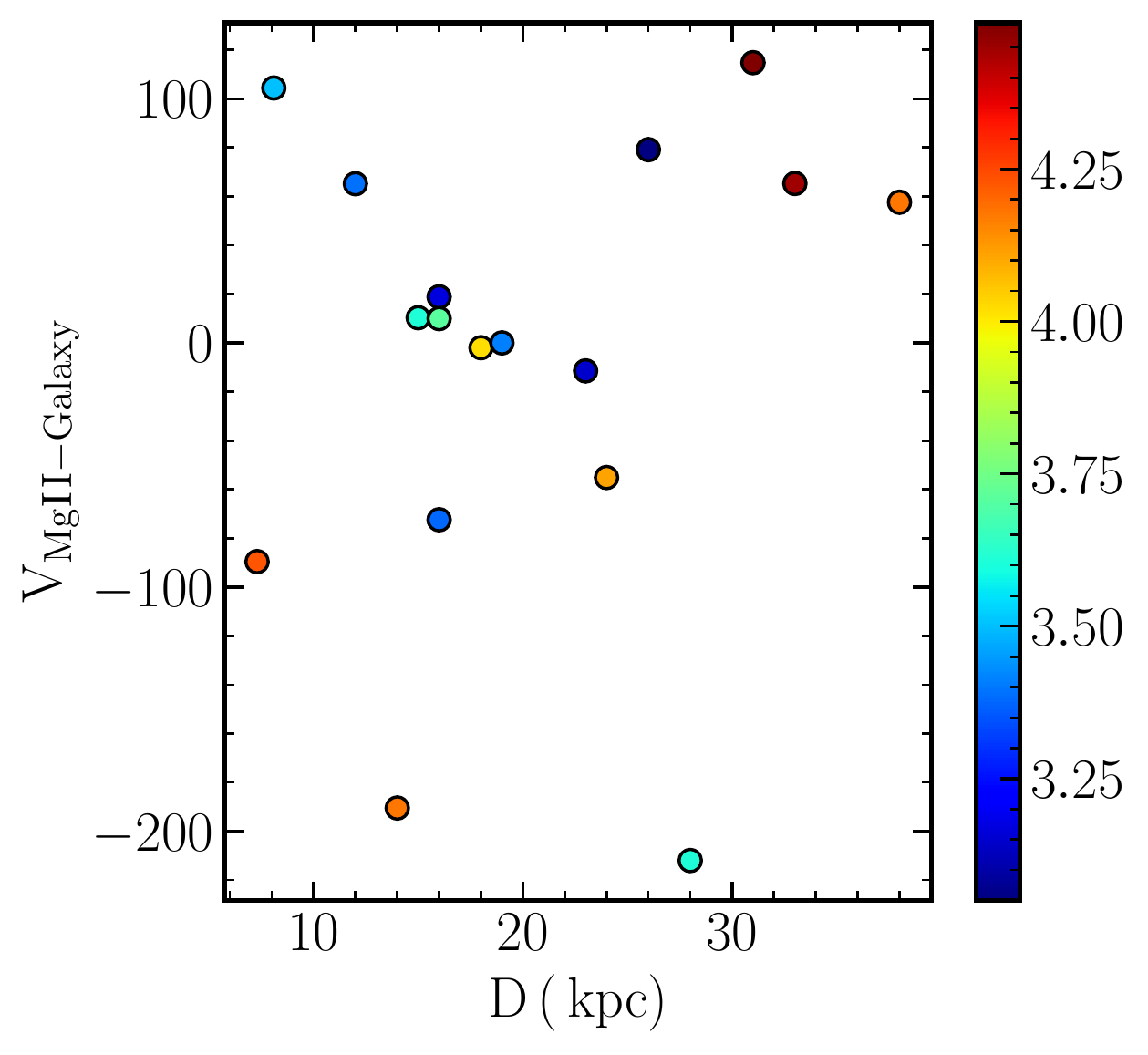}
    \caption{Velocity difference between the host galaxies and \usmg\ absorbers is plotted against the impact parameter. The points are colour-coded by $\mathrm{W_{2796}}$ as shown in the colour bar. No correlation is evident in this figure.}
    \label{fig:voff}
\end{figure}

In Figure~\ref{fig:voff}, we plot the velocity offset of the host galaxy with respect to the absorption redshift and the impact parameter for our \usmg\ sample. There is no clear correlation evident between these two quantities. The figure also suggests that there is no correlation between $W_{2796}$ and either D or the velocity offset. As $W_{2796}$ is related to the line-of-sight velocity spread, lack of strong velocity offset is consistent with the velocity field in the absorbing gas being symmetric with respect to the systemic redshift of the galaxies. 

} 

\subsection{Impact parameter vs $\mathrm{W_{2796}}$}
\label{sec:wvsd}
\begin{figure}
    \centering

         \includegraphics[viewport=5 7 410 390, width=0.48\textwidth,clip=true]{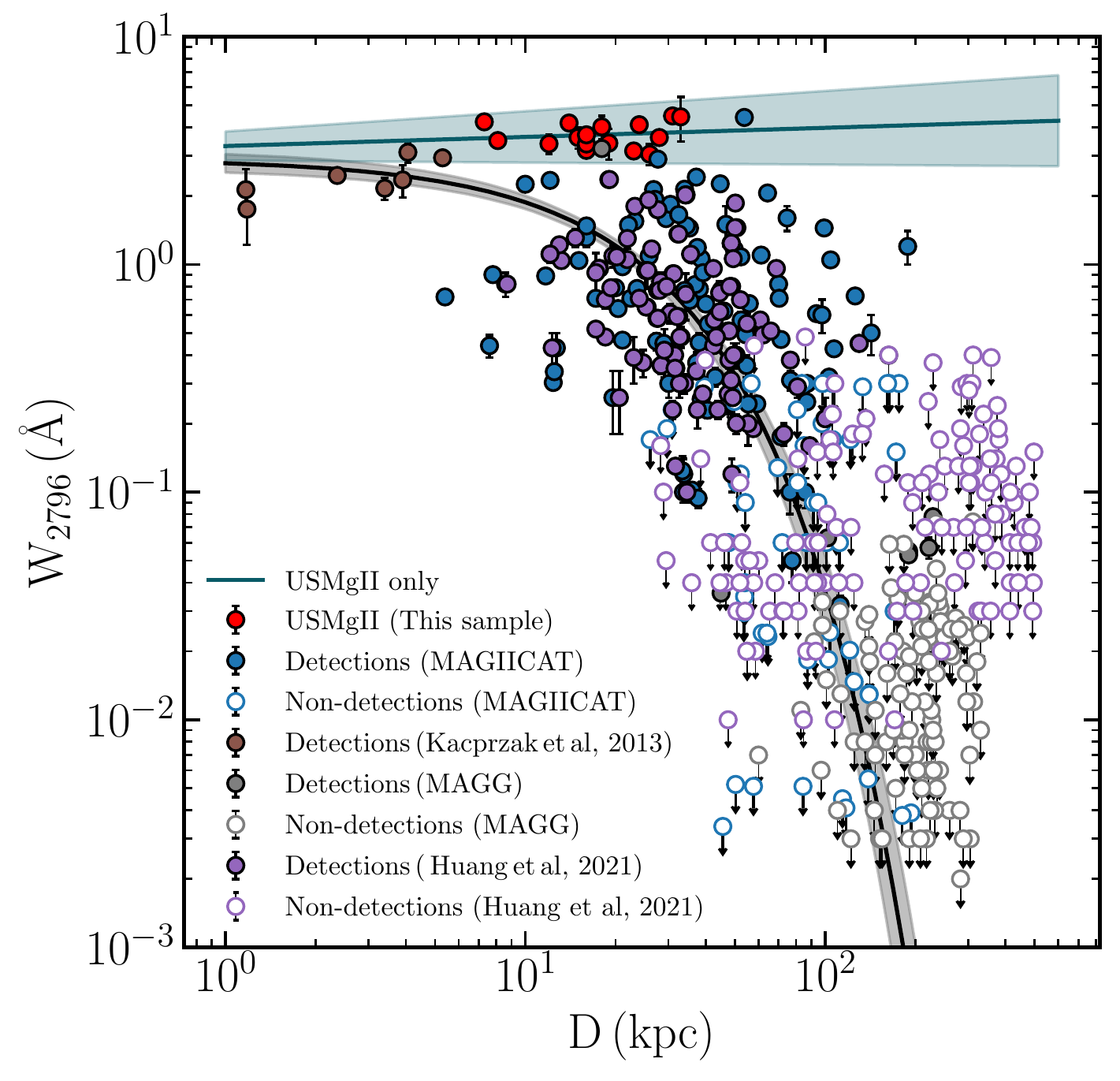}
   %     \end{subfigure}
    \caption{The scatter plot of $\mathrm{W_{2796}}$ vs D. Our \usmg\ systems are shown in red and literature data are shown in different  symbols (as indicated in the panel). Black (cyan-green) solid curve  with gray shaded region shows the best fit to the full sample (\usmg\ systems) and the associated error. It is evident that there is no anti-correlation between $\mathrm{W_{2796}}$ and D 
    for the \usmg\ absorbers (also see Figures~\ref{fig:w_vs_rho_appendix} and~\ref{fig:usmg_w_vs_rho} in the appendix). 
    }
    \label{fig:w_vs_rho}
\end{figure}

\begin{table}
\caption{Best fit to the relationship between ${\mathrm W_{2796}}$ vs D}
\begin{tabular}{l l}
\hline
\multicolumn{1}{c}{$\log(W_{2796})$} & \multicolumn{1}{c}{Reference}\\
\hline
\hline
$(-0.015 \pm 0.002) D(kpc) + (0.27\pm0.11)$   & \cite{Nielsen_2013b} \\
%  $\sigma = $ & \\
$(-0.005 \pm 0.002) D(kpc) - (0.09\pm0.12)$     & \cite{Rubin2018}\\
$(-1.17\pm0.10)\log(D(h^{-1} kpc))+(1.28\pm0.13)$& \citet{Chen_2010}\\
$(-0.72\pm0.25)\log(D(kpc))+(0.83\pm0.38)$& \citet{Huang2021}\\
$(-0.019 \pm 0.002) D(kpc) + (0.464\pm0.039)$   & This work \\
$(-0.016 \pm 0.003) D(kpc) + (0.702\pm0.067)$   & This work ($0.4\le z \le 0.6$)\\
\hline
\end{tabular}
\label{tab:w_vs_rho}
\end{table}

It has now been well-established that $W_{2796}$  anti-correlates with  $D$, albeit with  a large scatter \citep[][]{ Bergeron1991, Steidel1995, Chen_2010, Nielsen_2013b}.
In Figure~\ref{fig:w_vs_rho}, we plot $\mathrm{W_{2796}}$ vs. D for our sample as well as data from the literature. 
It is a common practice to characterize this anti-correlation, either by a log-linear or a power law dependence.
Best fit relationship obtained for different samples in the literature are summarized in Table~\ref{tab:w_vs_rho}. We obtained the best fit to the data plotted in Figure~\ref{fig:w_vs_rho} using the standard procedure discussed in Appendix~\ref{wvsrhofit}. The best fit relationship for the full sample is also shown in Figure~\ref{fig:w_vs_rho} and provided in Table~\ref{tab:w_vs_rho}. This table also provides results for absorbers at $0.4\le z\le 0.6$
(see Figure~\ref{fig:w_vs_rho_appendix}).

It is apparent that \usmg\ systems occupy a distinct region in the plot and do not follow the general anti-correlation between $\mathrm{W_{2796}}$ and D (also see Figure~\ref{fig:usmg_w_vs_rho}).
For example, for the power law fit given by \cite{Chen_2010} $W_{2796} \geqslant 3$\ang is realised for $D \leqslant 4.85$ kpc. For the log-linear fit given by \citet{Nielsen_2013b}, $W_{2796}$ saturates to a value of $1.86$\ang  for $D = 0$ kpc. Fit saturating to lower $W_{2796}$ is also the case for the fit discussed in Figure 12 of \citet{Dutta2020}. Even for our best fit for $0.4\le z \le 0.6$ absorbers (that also includes \usmg systems in the fit) \usmg\ systems are expected to have D$\le$14~kpc. This is lower than the median values discussed above. Clearly the distribution of D for the \usmg\ is inconsistent with the predictions from the general population of Mg~{\sc ii} absorbers.

The large scatter seen around the best fit is most likely to be driven by various galaxy properties such as \zabs, luminosity \citep{Nielsen_2013b, chen2010}, stellar mass \citep{churchill2013}, halo radius and the orientation \citep{bouche2012}.  Indeed, \citet{Huang2021} have explicitly shown that the scatter decreases when one takes the galaxy luminosity, stellar mass and halo radius into account for the star-forming isolated galaxies  \citep[see also,][]{Chen_2010}. Recent wide-field IFU surveys using MUSE like the \textsc{MUDF} \citep{Fossati2019}, \textsc{MUSE-ALMA} Halos \citep{Hamanowicz2020}, as well as grisms surveys using HST\citep{Dutta2020, Lundgren2021} are adding further scatter to the $\rm{W_{2796} - D}$ plot, by identifying multiple galaxies associated with \MgII\ systems out to large impact parameters. Therefore, the environment and ambiguity in absorber-galaxy association are other factors that enhance the scatter.
In what follows, we explore how the above mentioned properties of \usmg\ host galaxies compare with that of the host galaxies in the general population of Mg~{\sc ii} absorbers.

\begin{figure}
        \centering
        \includegraphics[width=0.45\textwidth]{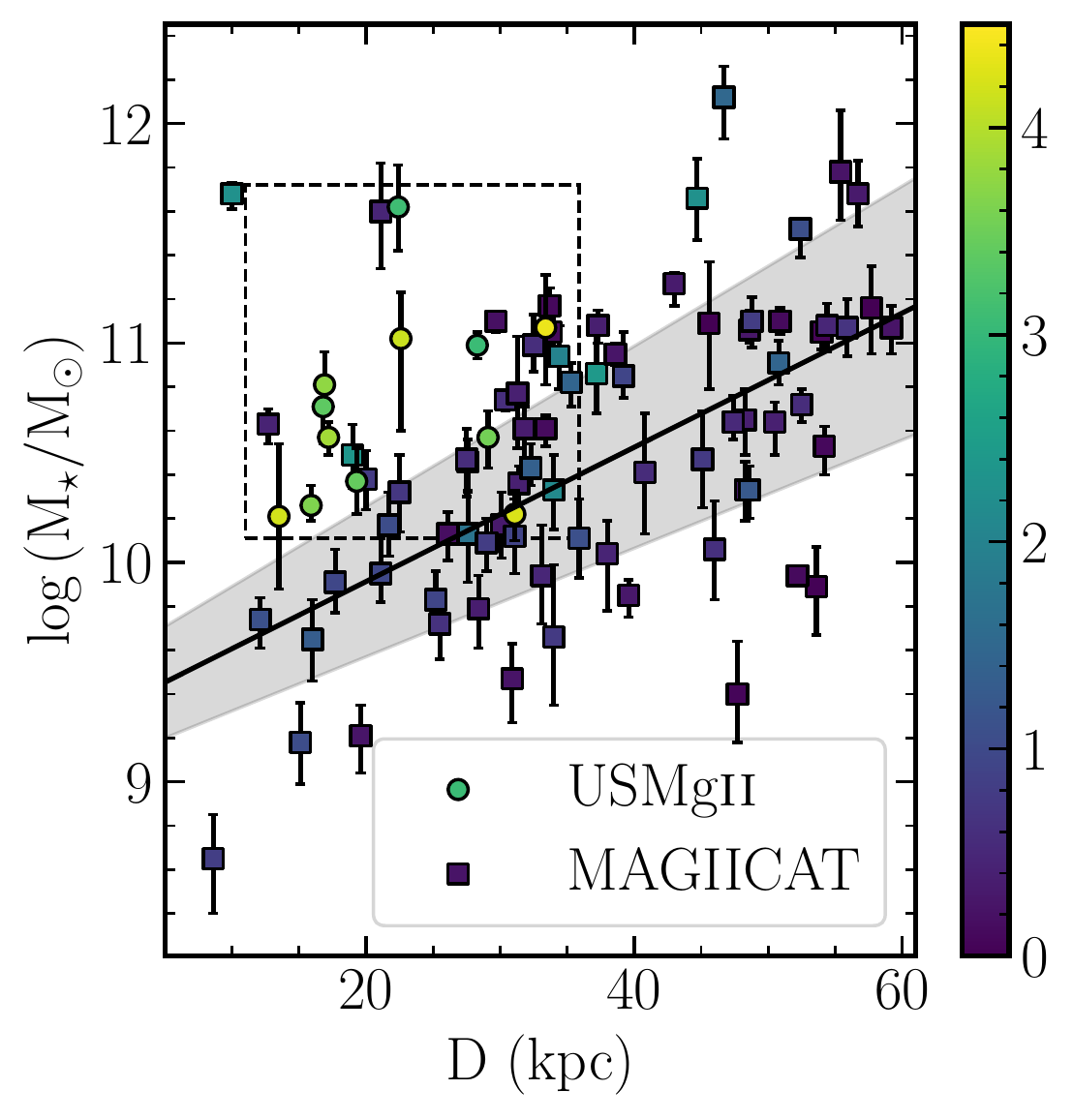}
    \caption{%{\it Right panel:} 
    The stellar mass of the \MgII\ host galaxies against the impact parameter. Points are color-coded based on the \MgII\ rest equivalent width, $W_{2796}$, as shown in the sidebar. The squares and circles represent the host galaxies taken from the \Magiicat sample and our \usmg\ sample, respectively. Region occupied by the \usmg\ systems is demarcated by the box in dashed line.}
    \label{fig:MC_gal_compared}
\end{figure}

{
%Using SED fitting we have obtained the stellar masses $\mathrm{(M_\star)}$ of both our \usmg host galaxies well separated spatially from the quasar (see column 4 of Table~\ref{tab:photometric_props}) and of the \MgII\ host galaxies from \Magiicat absorbers with D$<$60 kpc. 
%

In Figure.~\ref{fig:MC_gal_compared}, we plot $M_\star$ vs D color coded using $W_{2796}$. When we consider the \Magiicat\ galaxies, for a given $W_{2796}$ massive galaxies tend to be at a larger impact parameter. We also find a correlation between $M_\star$ and D that can be approximated by 
$\log\, (M / M_\odot) = (0.036\pm0.006) D(\rm{kpc}) + (9.301\pm0.225)$.
The solid black line and the gray shaded region around it in the figure correspond to this fit and the associated $\rm{1\sigma}$ uncertainty, respectively. 
The dashed rectangular box in this figure encompasses the points from the \usmg\ sample. It is evident that the \usmg galaxies are more massive than the \Magiicat\ host galaxies for the same impact parameter \citep[in line with the findings of][]{Rubin2018, Dutta2020}.
Note only 2 out of 12 \usmg host galaxies are within 1$\sigma$ region of the fit. We also see similar trend when we use $M_B$ instead of M$_\star$, i.e., there is a correlation between $M_B$ and D, and \usmg\ galaxies tend to be brighter compared to the relationship defined by \Magiicat\ points (see Appendix~\ref{sec:MB_dist}).
The correlation between D and $M_B$ and/or $M_*$ are typically used in the literature \citep{Chen_2010, Huang2021} 
to reduce the scatter in the D vs $W_{2796}$ plots.
For the median impact parameter and M$_*$ of our host galaxies, equation 16 of \citet{Huang2021} predicts $W_{2796}$ = 1.5\AA. Thus it appears that \usmg\ absorbers do not follow the general trend and the scatter in D vs $\rm{W_{2796}}$ plot can not explained by previously reported correction to the host galaxy stellar mass.
}

\subsection{[O~{\sc ii}] Luminosity and Star Formation Rate:}
\label{sec:loii}

\begin{figure*}
    \centering
    \begin{subfigure}{0.495\textwidth}
        \includegraphics[width=\textwidth]{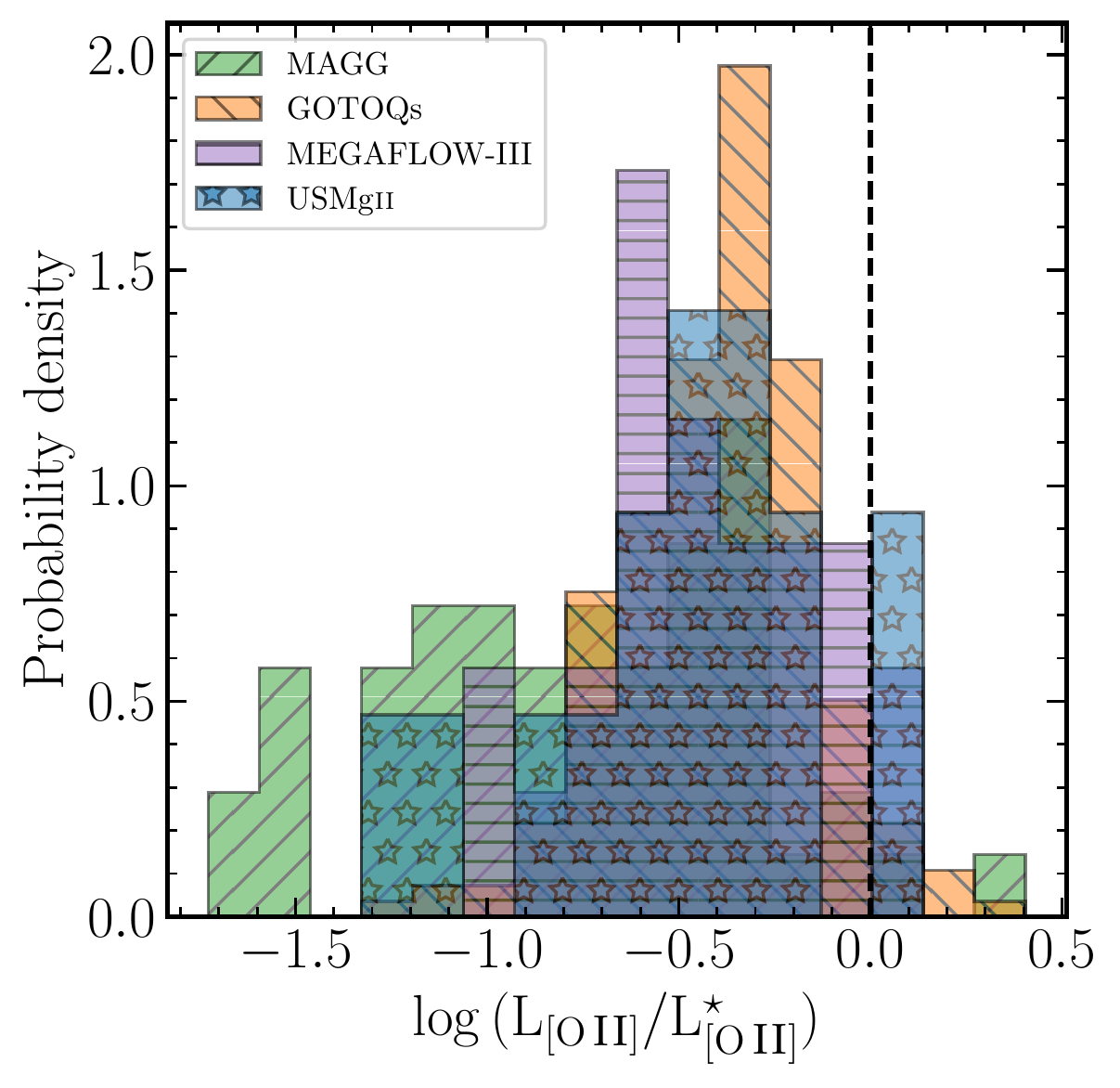}
    \end{subfigure}
    \begin{subfigure}{0.495\textwidth}
        \includegraphics[width=\textwidth]{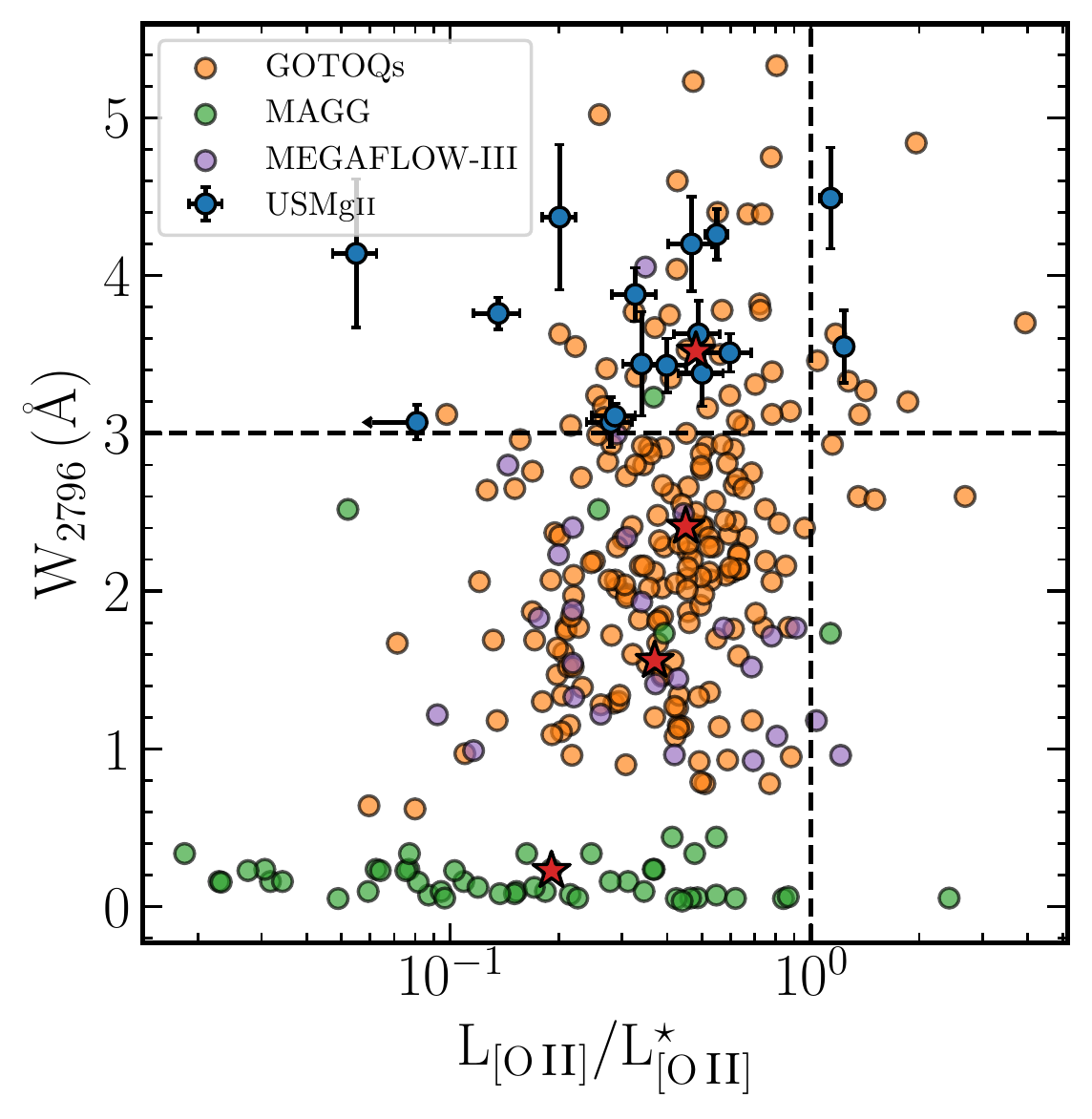}
    \end{subfigure}
    
    \caption{{\it Left panel:}
    Distribution of \OII\  luminosity scaled by the characteristic \OII\ luminosity of the field galaxies \citep{Comparat2016} at the redshift of the galaxies.
    The blue histogram with stars, orange histogram tilted vertical lines, the magenta histogram with the horizontal lines and the green histogram with tilted lines on the left corresponds to the \OII\  luminosity distribution of the \usmg host galaxies from our \usmg sample, of the GOTOQs \citep{Joshi2017}, galaxies from the MEGAFLOW-III survey \citep{Schroetter2019}, MAGG survey \citep{Dutta2020} respectively. {\it Right panel:} $\rm{W_{2796}}$ against the scaled \OII\ luminosity. Color assignment is same as that of the left panel. The red stars correspond to the median values of the $\rm{W_{2796}}$ and $\rm{L_{O\textsc{ii}} / L_{O\textsc{ii}}^\star}$ in 4 sub-sample based on $\rm{W_{2796}}$ (see text for details). The horizontal dashed black line corresponds to $\rm{W_{2796} = 3\AA}$, i.e., limiting value for the \usmg systems. The vertical dashed line in both the panels corresponds to the characteristic \OII\ luminosity, $\rm{L_{[O\textsc{ii}]}^\star}$.}
    \label{fig:o2luminosity_dist}
\end{figure*}

The measured \OII\ line flux for all the host galaxies of our \usmg\ systems are given in column 6 of Table~\ref{tab:spectroscopic_properties}. These values are obtained after applying the corrections as explained above (see Section~\ref{sec:observations}) but without applying any dust correction.
Left panel in Figure~\ref{fig:o2luminosity_dist}, we plot the distribution of \OII\  luminosity ($\mathrm{L_{[O\textsc{ii}]}}$) scaled by the characteristic \OII\ luminosity ($\,L_{[O\textsc{ii}]}^\star$) of the field galaxies \citep{Comparat2016} at the host galaxy redshift. The blue histogram with star, orange histogram with tilted vertical lines, the red histogram with the horizontal lines and the green histogram with tilted lines on the left corresponds to the distribution of the scaled $\mathrm{L_{[O~\textsc{ii}]}}$
of the \usmg host galaxies from our \usmg sample, 
GOTOQs \citep{Joshi2017}, MEGAFLOW-III survey \citep{Schroetter2019}, 
MAGG survey \citep{Dutta2020} respectively. The \OII\ luminosity of our galaxies ranges from 0.06 to 1.43 $\,L_{[O\textsc{ii}]}^\star$, with a median value of 0.39 $\,L_{[O\textsc{ii}]}^\star$ with $\,L_{[O\textsc{ii}]}^\star(z=0.5) = 3.2\times10^{41}~erg~s^{-1}$.
Note that, in the case of \zabs = 0.5896 towards J0218-0832, the host galaxy does not show detectable \OII\ emission with $\,L_{[O\textsc{ii}]} \le 1.8\times10^{41}~erg~s^{-1} (3\sigma)$. In the case of galaxies identified by \citet{Gauthier2013} for the \zabs = 0.5623 system towards J2207-0901, the observed \OII\ luminosity of the galaxy with the lowest D is 0.56 $\,L_{[O\textsc{ii}]}^\star$. 

Using the \OII\ luminosity function  \citep{Comparat2016} we calculate the median expected $L_{[O\textsc{ii}]}$ and the number of galaxies having  $L_{[O\textsc{ii}]} \ge\mathrm{L^\star_{[O\textsc{ii}]}}$.
When we consider $\mathrm{L_{min} = 0.06\% ~(or~0.1\%) L^\star_{[O\textsc{ii}]}}$ based on the lowest luminosity observed in our sample,
we expects about 0.6\% (or 1.5\%)  of  galaxies in our sample will be super-$\rm{L_{[O\textsc{ii}]}^\star}$. Also expected median \OII\ luminosity is less than 0.2 $\mathrm{L^*_{[O\textsc{ii}]}}$. Therefore, out of 18 \usmg host galaxies in our sample,
we expect about 0.2 of them to be super-$\,L_{[O\textsc{ii}]}^\star$ galaxies.
{ In our sample there are two host galaxies with $\,L_{[O\textsc{ii}]}$ in excess of $\,L_{[O\textsc{ii}]}^\star$ and
the observed median value (0.39$\,L_{[O\textsc{ii}]}^\star$) is higher than that predicted ($<0.2L_{[O\textsc{ii}]}^\star$) from the luminosity function. Therefore, we can conclude that $\,L_{[O\textsc{ii}]}$ of the \usmg host galaxies are significantly higher than what we expect from random sample of galaxies and 
\usmg\ selection  preferentially picks galaxies with higher $\,L_{[O\textsc{ii}]}$.}

%Next, we use the \OII\ line luminosity to estimate the SFR. 
The SFR is assumed to be proportional to the \OII\ luminosity of the galaxy and is given by \cite{Kennicat1998},
 $  SFR_{[O\textsc{ii}]} = (1.4\pm0.4)\times 10^{-41}L_{[O\textsc{ii}]}\,\rm{ergs\, s^{-1}}$.
Measured $SFR_{[O\textsc{ii}]}$
are also given in Table~\ref{tab:photometric_props}.
{ As can be seen from this table,
SFR based on SED fitting ranges from 2.6 $\rm{M_\odot yr^{-1}}$ to 33.5 $\rm{M_\odot yr^{-1}}$, with a median value of 7.7 $\rm{M_\odot yr^{-1}}$}, whereas the
$SFR_{[O\textsc{ii}]}$
ranges from 0.26 $\rm{M_\odot yr^{-1}}$ to 6.50 $\rm{M_\odot yr^{-1}}$, with a median value of 1.61 $\rm{M_\odot yr^{-1}}$.  It is well documented that the SFR obtained based on \OII\ emission line have systematic uncertainties that depend on variation in reddening, chemical abundance (stellar mass through the mass metallicity relations) and ionization \citep[][]{Moustakas2006, Gilbank2010}. A considerable contribution to the underestimation of $SFR_{[O\textsc{ii}]}$ 
originates from the dust corrections that we have not applied for our estimation. Also it has been found that in galaxies with 12+[O/H]$>$8.7 (or B-band luminosity in excess $10^{11} \mathrm{L_{B\odot}}$) one tends to underestimate the $SFR_{[O\textsc{ii}]}$ \citep{Moustakas2006}. 
Therefore, we have used SFR from the SED fitting in the analysis presented below.

\subsection{\OII\ luminosity vs $\rm{W_{2796}}$}
If the large $\rm{W_{2796}}$ of the \usmg\ absorbers is related to feedback processes in the host galaxy, then we expect a correlation between $\rm{W_{2796}}$ and $\,L_{[O\textsc{ii}]}$.
Such a correlation 
is also found between the $\,L_{[O\textsc{ii}]}$  and $\rm{W_{2796}}$ in the stacked SDSS spectrum \citep{noterdaeme2008b,Menard2011}. It was suggested that if this correlation is driven by any physical mechanism then one can use $\rm{W_{2796}}$ as a proxy to pick high-luminosity star-forming galaxies. However, as pointed out by \citet{Joshi2018}, the main driver for this correlation could be the impact parameter vs. $\rm{W_{2796}}$ correlation \citep[also see][]{Lopez2012}. In that case, there is a high probability for emission from the galaxies hosting the high $\rm{W_{2796}}$ systems to fall inside the SDSS fiber compared to the host galaxies of the low $\rm{W_{2796}}$ systems. 

In their SIMPLE sample, \citet{Bouche2007} have found a 2$\sigma$ correlation between the SFR derived from the H$\alpha$ luminosity and $\rm{W_{2796}}$. They %suggested that this could be 
explained this by assuming the absorbing gas being part of the wind from the host galaxies.
As we now have larger number of  $\,L_{[O\textsc{ii}]}$ measurements for host galaxies of \MgII\ absorbers, we revisit this correlation.
In the right panel of Figure~\ref{fig:o2luminosity_dist}, we plot the scaled \OII\ luminosity, 
i.e., $\rm{L_{[O\textsc{ii}]} / L_{[O\textsc{ii}]}^\star }$, vs. $\rm{W_{2796}}$. 
We do not find any trend between $\rm{W_{2796}}$ and $\rm{L_{[O\textsc{ii}]} / L_{[O\textsc{ii}]}^\star}$ in our \usmg\ sample alone.
In the same plot, we also show the points from the GOTOQ, MAGG and MEGAFLOW-III samples.
It is evident from the figure that most of the galaxies with $\rm{L_{[O\textsc{ii}]} / L_{[O\textsc{ii}]}^\star} \ge 1$ are associated with high $\rm{W_{2796}}$ systems (i.e. 58\% and 79\% of such galaxies are associated with $\rm{W_{2796}}\ge$ 3 and 2.5\AA\ absorbers respectively). Similarly, galaxies with $\rm{L_{[O\textsc{ii}]} / L_{[O\textsc{ii}]}^\star} \le 0.1$ are predominantly associated with low equivalent width (i.e. $\rm{W_{2796}}\le 1$\AA) absorbers from \citet{Dutta2020}.

Thus, when we perform the rank-correlation analysis for the combined sample, we do find a correlation (Spearman rank correlation of $r_S$ = 0.34 and a $p$-value of $<10^{-4}$) between the quantities plotted in Figure~\ref{fig:o2luminosity_dist}.
When we restrict ourselves to stronger \MgII\ absorption systems, the strength of this correlation decreases significantly. For example, when we consider systems with $\rm{W_{2796} \geqslant 0.5 \AA}$, the rank correlation coefficient, $\rm{r_S}$ decreases to 0.21 ($p$ value $ = 6\times10^{-4}$), and for systems with $\rm{W_{2796} \geqslant 2 \AA}$, $\rm{r_S}$ further decreases to 0.13 ($p$ value $ = 0.1$). When we exclude the GOTOQs, which are biased towards high $\,L_{[O\textsc{ii}]}$ galaxies, $r_S$ drops from 0.34 to 0.23 ($p$-value $ = 0.03$) for the full \MgII\ sample. Additionally, if we restrict ourselves to host galaxies having impact parameter less than 50 kpc (excluding GOTOQs), we find $r_S$ = 0.20 ($p$-value $ = 0.23$), while considering a maximum impact parameter of 100 kpc, $r_S$ = 0.19 ($p$-value $ = 0.18$). 

For better visualization, we divide the sample into 4 sub-samples of \MgII\ absorbers 
based on the rest equivalent width of the \MgII\ absorption ($\rm{W_{2796} \leqslant 1\AA}$, $\rm{1\AA < W_{2796} \leqslant 2\AA}$, $\rm{2\AA < W_{2796} \leqslant 3\AA}$, $\rm{W_{2796} > 3\AA}$). For each of these sub-samples, we calculate the median values of $\rm{W_{2796}}$ and $\rm{L_{[O\textsc{ii}]} / L_{[O\textsc{ii}^\star] }}$. The four blue stars in the right panel of Figure \ref{fig:o2luminosity_dist} corresponds to the median values for each sub-sample. 
%These points confirm our correlation results. 
{ The analysis presented here suggests a correlation between ${\rm W_{2796}}$ and $\rm{L_{[O\textsc{ii}]}}$ for the full sample. While \usmg\ selection statistically picks high luminosity host galaxies, the scatter in $\rm{L_{[O\textsc{ii}]}}$ is large at a given ${\rm W_{2796}}$. }

\begin{figure*}
    \centering
    \begin{subfigure}{0.495\textwidth}
        \includegraphics[width=\textwidth, keepaspectratio]{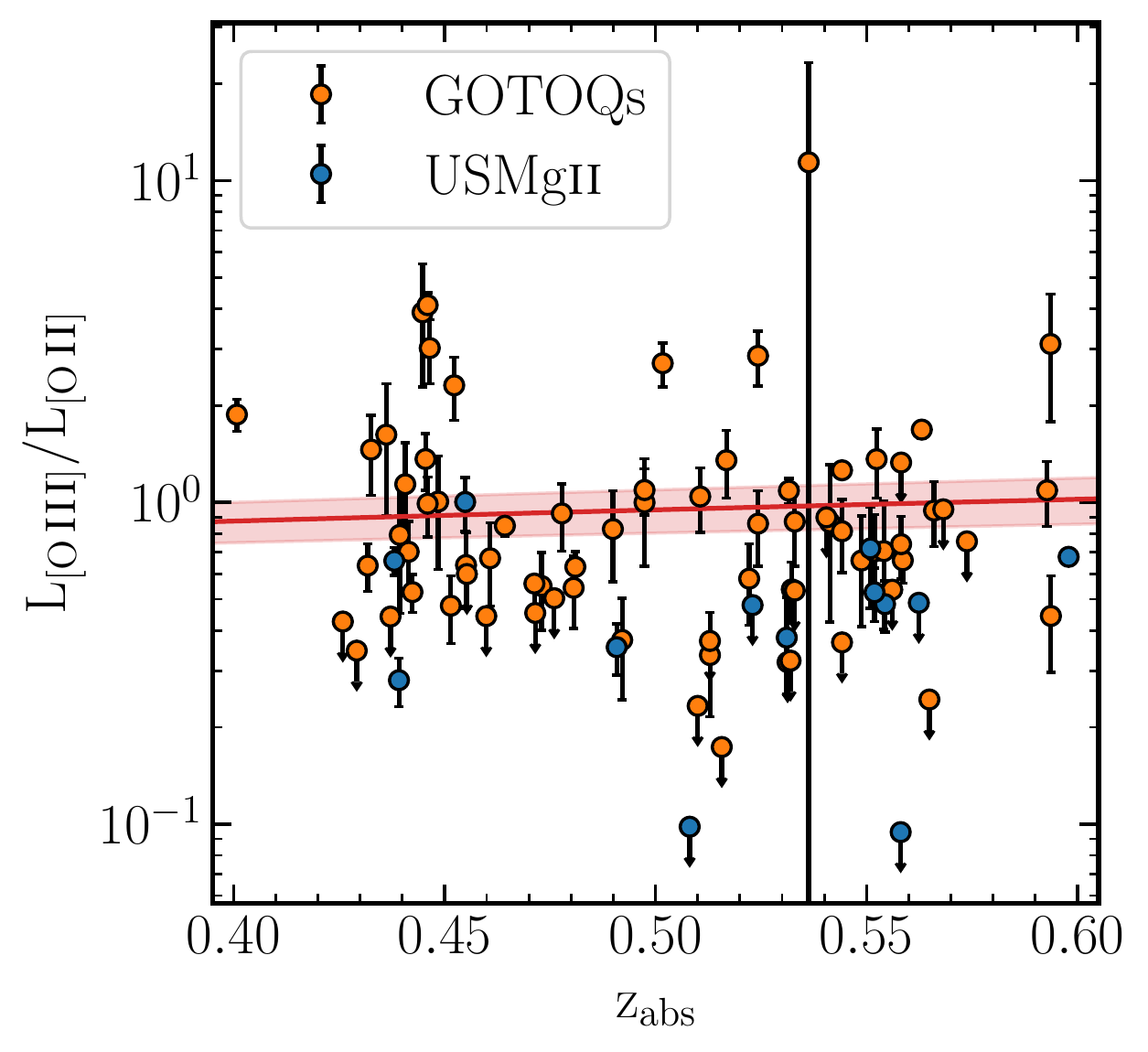}
    \end{subfigure}
    \begin{subfigure}{0.495\textwidth}
        \includegraphics[width=\textwidth, keepaspectratio]{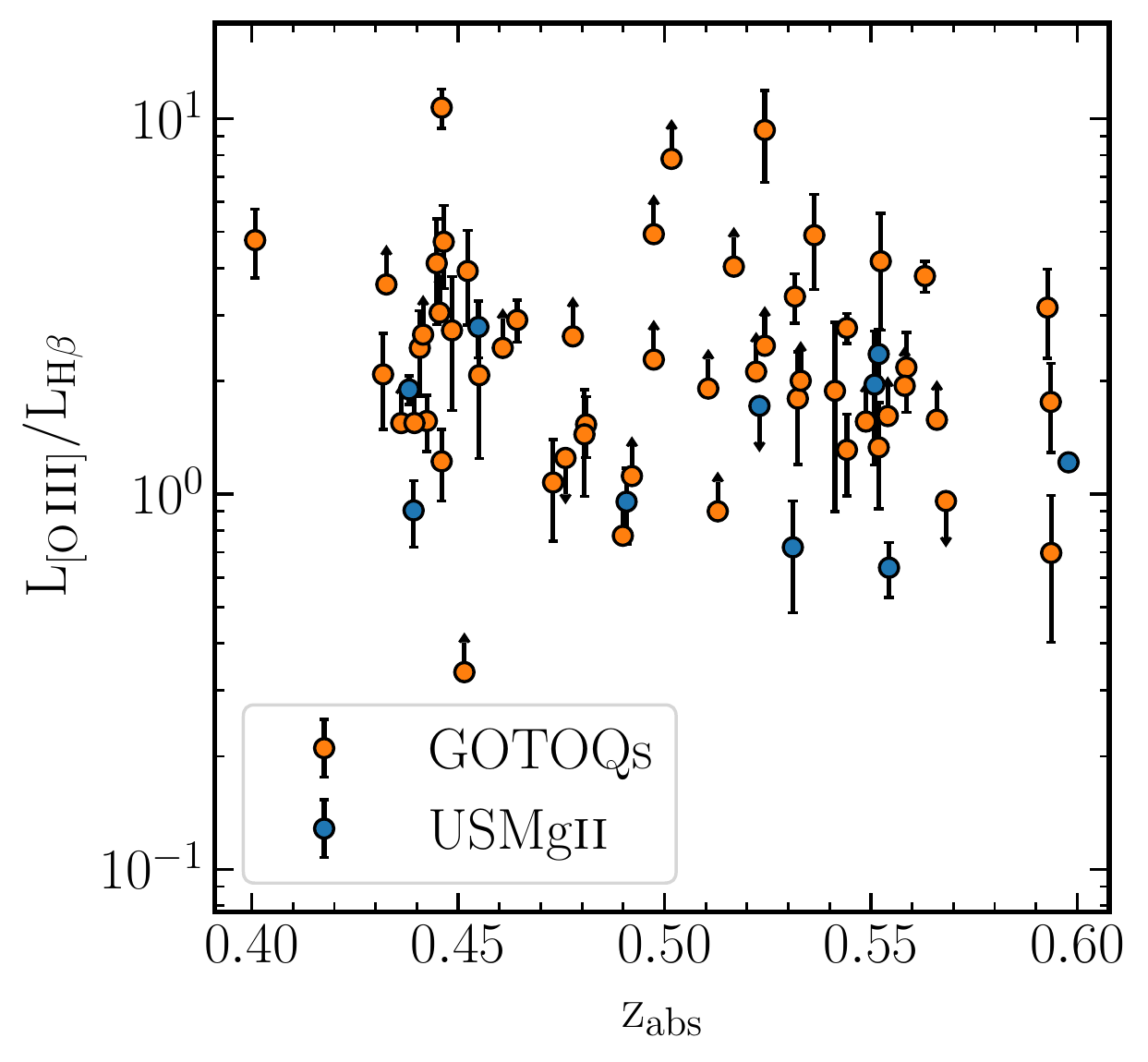}
    \end{subfigure}
    
    \caption{{\it Left panel:}  Ratio of \OII\ and \OIII\ line luminosities associated with \MgII\ host galaxies
    as a function  the absorption redshift. Blue points correspond to our \usmg sample while the orange points are taken from GOTOQs (with $\rm{0.4\, \leqslant z_{abs} \leqslant\, 0.6}$) \citep{Joshi2017}. 
    The red line gives the observed redshift evolution of $L_{\OII}$/$L_{\OIII}$, for the star-forming galaxies \citep{Khostovan2016}. {\it Right panel:} Ratio of \OIII\ and $\rm{H\beta}$  line luminosities as a function of \zabs.
    }
    \label{fig:lineratios}
\end{figure*}

\subsection{Nebular Emission Line Ratios: metallicity and ionization parameter}

The \OIII/\OII\ as well as the \OIII/{H$\beta$} line ratios are very sensitive to the hardness of the ionizing radiation field (i.e. temperature of the stars), and can distinguish between the ionization mechanisms of the nebular gas in a galaxy \citep[][]{Baldwin1981,Kewley2001}. Over the redshift range $\rm{z = 0 - 5}$, a rise in both the ratios, \OIII/\OII\ and \OIII/$\rm{H\beta}$, of star-forming galaxies has been observed \citep{Nakajima2014,steidel2014,Kewley2015,Khostovan2016}. This implies that compared to the local galaxies, galaxies at higher redshifts have higher ionization parameter, lower metallicity, harder stellar ionizing radiation field, and higher electron densities.

Left panel of Figure \ref{fig:lineratios} shows the line luminosity ratio  $L_{[O\textsc{iii}]}/L_{[O\textsc{ii}]}$ vs. redshift. Host galaxies of our \usmg\ absorption systems and 
GOTOQs (at $\rm{0.4\, \leqslant z_{abs} \leqslant\, 0.6}$) are plotted as blue and orange points, respectively.
There are 13 \usmg\ systems for which these measurements are possible. In cases where the \OIII\ emission is not clearly detected, we give $3\sigma$ upper limits.
The red line corresponds to the evolution of the line ratio  $L_{[O\textsc{iii}]}/L_{[O\textsc{ii}]}$ for the star-forming galaxies \citep{Khostovan2016}.
It is evident from this figure that measurements from GOTOQs follow the general trend of the field galaxies.  However, the measured ratio in the host galaxies of \usmg\ systems are typically lower than the best fit to the star-forming galaxies. The KS-test (with D = 0.542 with p-value = 0.013) also confirms the distribution of $L_{[O\textsc{iii}]}/L_{[O\textsc{ii}]}$ in these two populations are different. There are 7 GOTOQs that have $W_{2796}>3$\AA\ and the measured $L_{[O\textsc{ii}]}/L_{[O\textsc{iii}]}$ are distributed equally around the fit for star-forming galaxies. 
We see a negative trend between $L_{[O\textsc{iii}]}/L_{[O\textsc{ii}]}$  and $W_{2796}$ in the GOTOQ sample (i.e. largest ratios are seen among the host galaxies of absorbers with low $W_{2796}$). However, this anti-correlation  is not statistically significant (see Figure~\ref{fig:o3bo2vswmg2} in appendix~\ref{sec:o3bo2vswmg2}). Thus the difference between GOTOQs and \usmg\ can not be simply attributed to the presence of such a trend.

In the right panel of Figure~\ref{fig:lineratios}, we show the redshift evolution of the ratio $L_{[O\textsc{iii}]}/ \rm{H\beta}$ of the \usmg\ host galaxies compared to that of the GOTOQs over the same redshift range. 
%Blue points corresponds to the \usmg\ host galaxies from our sample while the orange points are taken from GOTOQs sample of \citet{Joshi2017}.
Simultaneous measurements of $L_{[O\textsc{iii}]}/ \rm{H\beta}$ are possible for only 9 galaxies in our sample. 
Like for $L_{[O\textsc{iii}]}/L_{[O\textsc{ii}]}$, the $L_{[O\textsc{iii}]} /\rm{H\beta}$ measured in the case of \usmg\ tend to be lower compared to the measurements from the GOTOQs { and general population of galaxies}.
The KS-test gives D = 0.487 and a p-value of 0.050.
This could indicate that the ionizing radiation in the star-forming field galaxies is harder than that in the \usmg host galaxies. To explore this further, in { what follows,} we use the line ratios to measure the metallicity and ionization parameter of the gas.

%\subsection{Nebular metallicity and ionization parameter}

\begin{figure}
    \centering
    \includegraphics[width=0.475\textwidth]{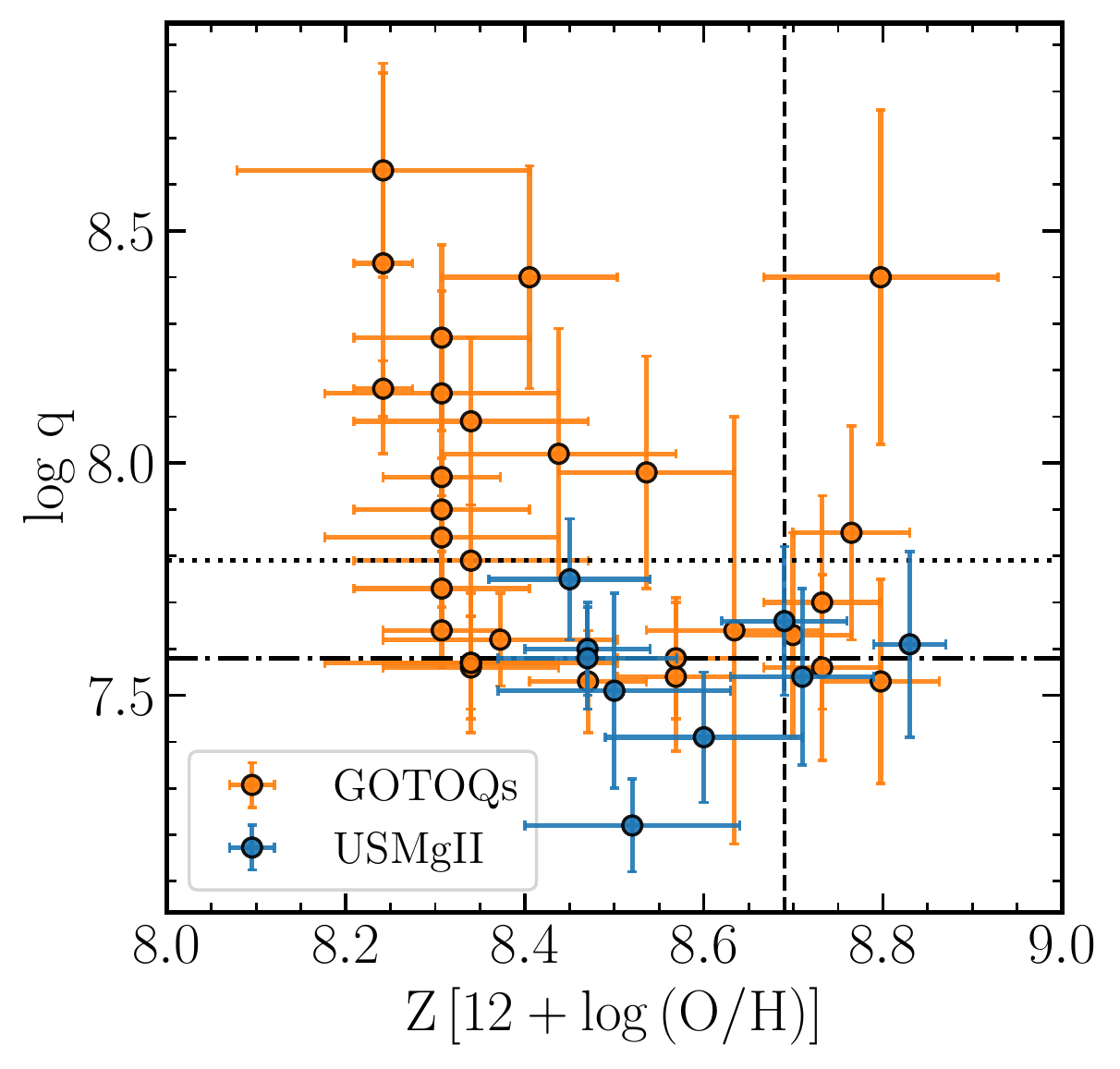}
    \caption{Ionization parameter (q) vs. the gas phase metallicity (Z) for the GOTOQs (orange) and the \usmg (blue) host galaxies in the redshift range $\rm{0.4\, \leqslant\, z\, \leqslant\, 0.6}$. The black dotted and dash-dotted horizontal lines corresponds median ionization parameter for the GOTOQs and the \usmg host galaxies respectively. The host galaxies of \usmg preferentially occupy  low q region. The dashed vertical line corresponds to the solar oxygen abundance.
    }
    \label{fig:q_vs_z}
\end{figure}

Gas phase metallicity is usually calculated from the metallicity-sensitive line luminosity ratios of strong emission lines present in the galaxy spectrum \citep[see e.g., ][for various such calibrations against different line ratios]{Nagao2006}.
Since we have the coverage of different nebular emission lines like \OII\ $\rm{\lambda\lambda\, 3727, 3729}$, $\rm{H\beta\, \lambda\, 4862}$, and \OIII\ $\rm{\lambda\lambda\, 4960, 5008}$, we make use of the $\rm{R_{23}}$ metallicity estimator, originally proposed by \citet{Pagel1979} and defined as,
$
%\begin{align*}
%\centering
    R_{23} = (L_{[O\textsc{ii}]\lambda\lambda 3727, 3729} + L_{[O\textsc{iii}]\lambda\lambda 4960, 5008})/{L_{H\beta}}. $
%\end{align*}
%
Based on various photoionization models, this ratio has been calibrated against oxygen abundances \citep[see e.g.,][]{McGaugh1991, Kewley2002}. However, 
the estimator $\rm{R_{23}}$ is known to have a strong dependence 
on the ionisation parameter (q), i.e., the inferred metallicity will be different for different assumed values of q. 
Therefore, we measure the gas phase metallicity ($\rm{Z}$) and the ionization parameter ($\rm{q}$) simultaneously based on the emission lines (\OII, \OIII\ and H$\rm{\beta}$) present in the galaxy spectra and the photoionization model provided by \citet{Levesque2010} using the python fork \citep{Mingozzi2020} of the IDL code \textsc{izi} \citep[Inferring metallicity and ionization parameters,][]{Blanc2015}. Apart from two cases, we do not detect the \OIII\ $\rm{\lambda\, 4960}$ emission at $\rm{\ge3\sigma}$ significance. We therefore take the associated flux of this line to be one-third of the flux associated with \OIII\ $\rm{\lambda\, 5008}$ line as suggested by theoretical studies \citep{Storey2000}.

Another difficulty in the use of $\rm{R_{23}}$ is that the ratio is doubly degenerate in terms of the oxygen abundance. This is because, at low abundance the intensity of the forbidden lines scales roughly with the chemical abundance, while at high abundance the nebular cooling is dominated by infrared fine-structure lines, and the electron temperature drops too low to collisionally excite the optical forbidden lines. Nebular emission lines like [N\textsc{ii}] $\lambda\, 6564$ are usually used to break this degeneracy \citep{Kewley2002, Pettini2004}. However, in our case, we do not have spectral coverage of the [N\textsc{ii}] $\lambda\, 6564$ line. As a result, to avoid any confusion, we will be considering only the upper branch metallicities. We detect all the three nebular lines only in the case of 10 host galaxies. The obtained gas phase metallicities and the ionization parameters are given in columns 8 and 9 of Table \ref{tab:spectroscopic_properties}. For these cases, we find the gas-phase metallicities, 12 + log(O/H), to be in the range from 8.45 to 8.83 (0.58-1.41 times the solar abundance for log(O/H)$_\odot$ = 8.69).
Given the mass-metallicity relation \citep{Ma2016}, this corresponds to a stellar mass in the range 9.23 $\rm{\leqslant\, \log\, M_\star\,[M_\odot] \leqslant}$ 10.42. 

We performed the same exercise for the GOTOQs in the redshift range of our interest. In this redshift range, for only 29 GOTOQs, all the three emission lines are detected with at least $\rm{3\sigma}$ significance. In Figure \ref{fig:q_vs_z}, we show the scatter plot of the ionization parameter versus the gas phase metallicity for both these 29 GOTOQs as well as the \usmg host galaxies from our sample. From this plot, it is quite apparent that the \usmg galaxies tend to have a lower ionization parameter compared to the GOTOQs
(KS-test gives D = 0.52  and a p-value of 0.03).
This indicates that, compared to the GOTOQs, the \usmg host galaxies are dominated by softer ionizing field.

\section{Discussion}
\label{sec:discuss}
\subsection{Gas kinematics}
\begin{figure}
    \centering
    \includegraphics[width=0.49\textwidth]{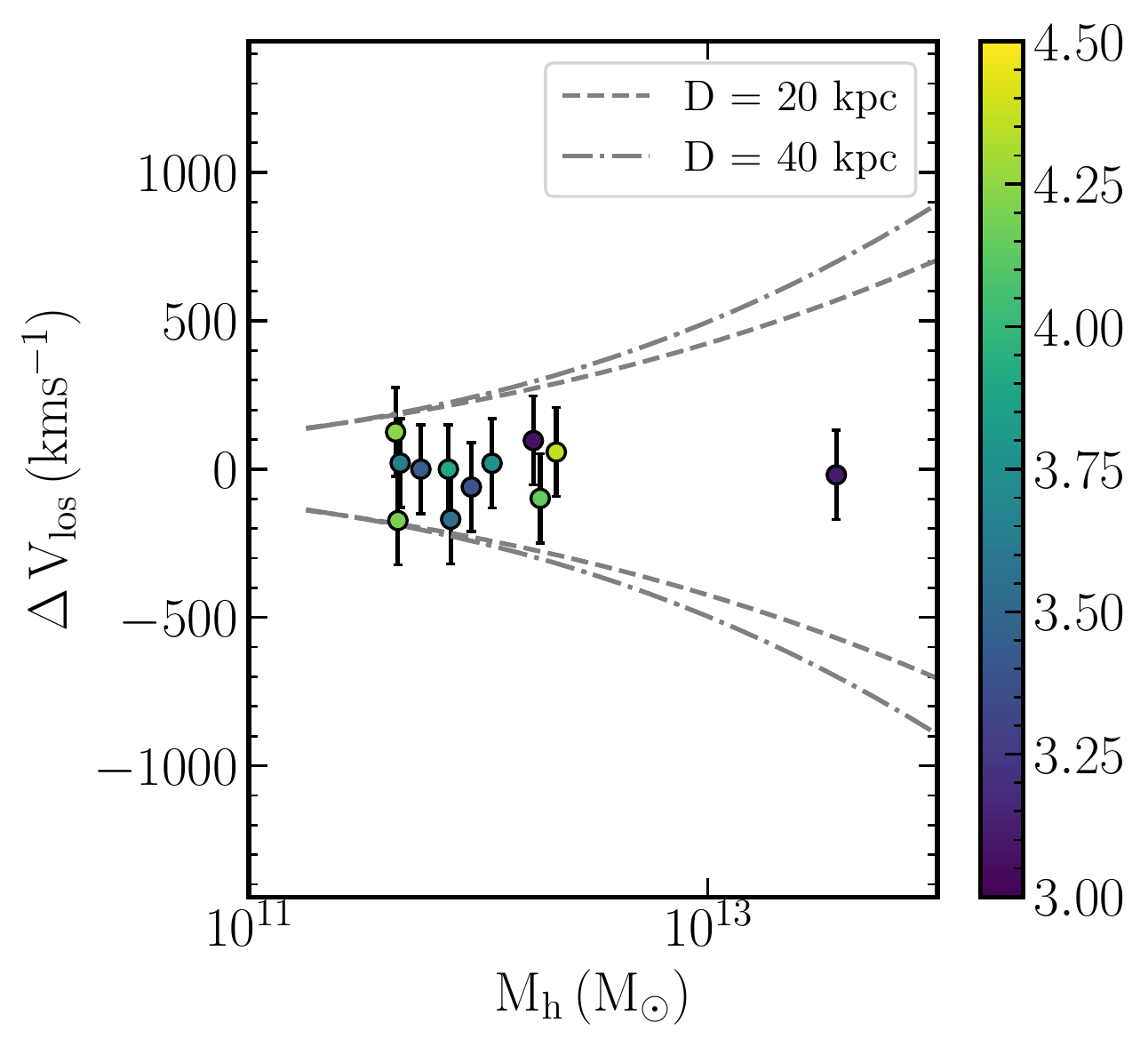}
    \caption{
    The velocity difference between the \MgII\ absorbing gas along the quasar sightline and the \usmg host galaxies is plotted against their dark matter halo mass color-coded following $\rm{W_{2796}}$ as indicated on the right side of the figure. The gray dashed and dot-dashed lines correspond to the escape velocity for the dark matter halo potential at the impact parameters of 20 kpc and 40 kpc respectively. For each measured velocity difference, we associate an error bar of $\pm$150 \kms. 
    }
    \label{fig:veldiff_potential}
\end{figure}
%\end{comment}

From the virial radius of the \usmg host galaxies in our sample given in column 10 of Table \ref{tab:photometric_props}, it is evident that the impact parameter is typically one-tenth of the virial radius. This (together with lack of additional galaxies within 100 kpc in most cases, see Table~\ref{tab:observation_log}) provides a strong reason to believe that the kinematics of the absorbing gas is mostly governed by the physical processes operating in the halo of the identified host galaxy itself.

In Figure~\ref{fig:veldiff_potential}, we plot the rest frame velocity difference between the absorbing gas and the host galaxy along our line of sight ($\Delta v_{los}$) against the halo mass color-coded as the $\rm{W_{2796}}$. 
Here, we assume the dark matter halo follows the NFW profile \citep{Navarro1997} and halo concentration of $\rm{c_h}$ = 10. 
The dashed and the dot-dashed gray lines corresponds to the escape velocity against the halo mass at a distance of 20 kpc and 40 kpc respectively from the center of the halo.
{ This figure suggests the measured velocity offsets are consistent with the absorbing gas being bound to the galaxy.}

{ However, in order to  quantify what fraction of the absorbing gas is bound to the galaxy we also need to know the line of sight velocity spread.} Recall that a large $W_{2796}$ of \usmg\ absorption will correspond to large velocity spread, $\Delta v_{los}$, along our line of sight \citep[see][]{Ellison2006,Zou2018}. It is well known that $W_{2796}$ is strongly correlated with $\Delta v_{los}$. This correlation will imply $\Delta v_{los}>300$ \kms for our sample. 
It is also suggested that velocity widths in excess of 300 \kms may originate from multiple components \citep[see for example,][]{Ledoux1998,Zou2018}. As mentioned before, in three systems in our sample we see multiple component structure even at SDSS resolution. Also in three cases we see more than one galaxy at the same redshift as the \usmg\ absorber.  The errorbars in Figure~\ref{fig:veldiff_potential} indicate a spread of $\pm$150 \kms\ around the measured velocity offset. If velocity spreads are of this order then in most cases the absorbing gas will be bound to the galaxy. However, { high resolution spectroscopic observations are needed to directly measure the fraction of absorbing gas bound to the galaxies.}
%the discussions presented above clearly indicate the importance of measuring the velocity fields accurately using high resolution spectra to draw firm conclusions on this issue.

%
\subsection{Are \usmg\ host galaxies starbursts?}
\label{sec:Mstar}

\begin{figure}
    \centering
    \includegraphics[width=.495\textwidth]{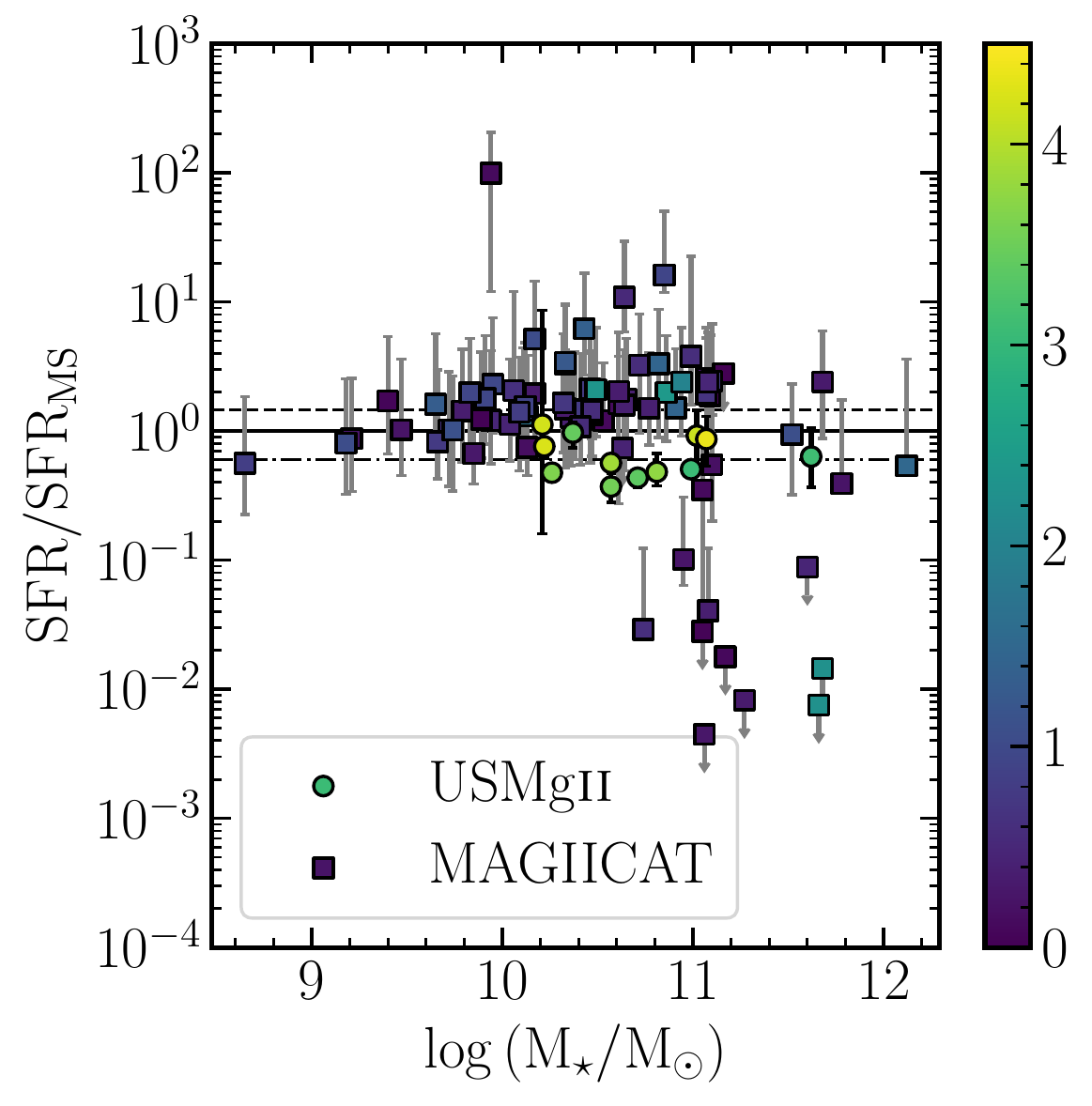}
    \caption{Scatter plot of SFR scaled by the SFR of a main sequence galaxy of the same stellar mass at \zabs\ (SFR$_{\rm MS}$) vs stellar mass for \usmg (circle) and \Magiicat\ host galaxies (squares).
    Color-coding is based on the rest equivalent width, $\rm{W_{2796}}$. The black solid line corresponds to SFR of the main sequence. The dashed black line corresponds to the median SFR of the \Magiicat\ galaxies while the the dot-dashed line corresponds to that of the \usmg host galaxies.
    }
    \label{fig:sfr_ms}
\end{figure}
To investigate whether the \usmg host galaxies identified in our sample are regular star-forming galaxies, or star-burst or post star-burst galaxies, we put these galaxies in the SFR scaled by the main sequence SFR at the redshift of the galaxy \citep{Speagle_2014} 
versus stellar mass plot as shown in Figure \ref{fig:sfr_ms}.
In the same plot, we have also shown the isolated \MgII\ absorbers within the impact parameter of 60 kpc from the \Magiicat\ sample. 
We identify that the \usmg host galaxies have systematically less SFR (i.e. median SFR = 0.6 $\mathrm{SFR_{MS}}$) compared to the star-forming main sequence \citep{Speagle_2014} as well as the \Magiicat\ host galaxies (median SFR = 1.5 $\mathrm{SFR_{MS}}$). 
Thus \usmg\ systems considered here are not star bursting galaxies and typically form stars at a rate slightly lower than the star forming main sequence.

Next, we test whether the SFR in the \usmg host galaxies is sufficient to launch galactic winds for the inferred gravitation potential. Given the stellar mass of $\rm{M_\star}$, the { theoretical threshold SFR for launching winds is given by \citet{Murray2011}}, 
\begin{align*}
    \rm{SFR_{threshold}\, =\, 1.7\left(\frac{M_\star}{10^{10}M_\odot}\right)^{0.56}\, M_\odot\, yr^{-1}}.
\end{align*}
For the 12 \usmg host galaxies for which the $M_\star$ and SFRs are inferred through the SED fitting, {{ we find that the SFRs are higher than the corresponding threshold values. However, \citet{Rubin2014} does found several galaxies with SFR above this threshold that do not show any wind signatures.}}
%
%Hence it is very likely that the gas in absorption may have come from galactic winds itself. 
The definite way of identifying the presence of galactic winds in a galaxy is to confirm whether there is a blue-shifted component present in the stellar absorption profiles. 
However, detecting galactic winds through such analysis is not possible with our galaxy spectra that does not cover Mg~{\sc ii} wavelength range. { Detecting wind signatures with future observations is important to quantify the role played by winds from ongoing as well as past starburst activities in producing \usmg\ absorption. }
%}}

\subsection{Fraction of galaxies producing USMgII absorption}

    From Figure \ref{fig:w_vs_rho}, we see that \usmg host galaxies can have impact parameters up to 40 kpc. However, only a fraction of galaxies within D$\sim$40 kpc will produce an \usmg absorption. To find this fraction ($f_{USMg\textsc{ii}}$),
    we consider the isolated galaxies in galaxy-centric \MgII\ absorption surveys conducted by \citet{Dutta2020}, \citet{Huang2021} and also the \Magiicat.  We find only one out of 34 galaxies (i.e. $f_{USMg\textsc{ii}}\sim 0.03$) at $D\leqslant 20$ kpc and none of the 106 galaxies at $20 \leqslant D (kpc)  \leqslant 40$ produce \usmg\ absorption. For the full range, D$\leqslant 40$, we get an average $f_{USMg\textsc{ii}} \sim 0.007$.  Usually, $f_{USMg\textsc{ii}}$ values are interpreted in terms of covering factor of the gas.
    { In this picture, the projected area around galaxy centres covered by the \usmg\ absorber is $\sim$1\% of the total area. The large $W_{2796}$ seen in the \usmg\ absorbers suggest large velocity spread among multiple clouds along our line of sight. 
    %The above two pictures are difficult to reconcile. 
    Another possible interpretation of low $f_{USMg\textsc{ii}}$ is to say \usmg\ originate from a different population of galaxies.
    %
    %However, the small values of $f_{USMg\textsc{ii}}$ found here together with 
    The impact parameter distribution of \usmg host galaxies not following the general population (see section~\ref{sec:wvsd}) may suggest that \usmg\ selection picks a different population (high $M_\star$ with relatively low SFR/SFR$_{\rm MS}$ and low ionization parameter with large reservoir of CGM gas) of host galaxies.
    }

\subsection{Galaxy environments of \usmg absorbers}
From the recent studies \citep{Fossati2019,Dutta2020, Hamanowicz2020} of \MgII\ absorption systems with integral field units (IFU) like MUSE, there is growing evidence that a significant fraction of \MgII\ absorbing gas might not be associated with isolated galaxies. As mentioned before, there is one \usmg system present in the sample of \citet{Dutta2020} and three in the sample of \citet{Schroetter2019}. Interestingly,
for all these 4 systems the host galaxies are found to be isolated (i.e., no other galaxy is found within 100 kpc from the quasar sightline and a maximum velocity separation of 500 $km\,s^{-1}$). 
{ As discussed in Section~\ref{sec:completeness}, there are 7 \usmg\ absorbers with only one potential host galaxy (with $m_r<23.6$ mag) within an impact parameter of 100 kpc. Therefore, at least 33\% of \usmg\ absorbers in our sample are hosted by isolated galaxies. This fraction is 57\% if we consider the limiting magnitude of $m_r=22.5$ mag. 
}
We identify three cases (i.e. $\sim$14\%) that are likely to be associated with more than one galaxy. These systems are towards J0156+0343, J2207-0901 and, J2338-0056. For the first case (J0156+0343),  we detect extended \OII\ emission consistent with the host galaxy(ies) being part of merging systems (see Figure~\ref{fig:J0156Merger}). The second case (J2207-0901) has already been studied by \citet{Gauthier2013} and the \usmg absorption is likely to be originating from intra-group gas. For the third system (J2338-0056),  we identify a \usmg host galaxy at an impact parameter of 79 kpc, however there are two more galaxies (with consistent photometric redshift) present for which we could not get the spectrum and these are likely to be part of a galaxy group.

\section{Conclusions}
\label{sec:conclusion}
In this paper, we have explored the nature of the galaxies hosting \usmg absorption at redshift, z $\sim$ 0.5.{ Main summary of this work is provided below}.

1. Using the SDSS \MgII/\FeII\ absorber catalog \citep{Zhu2013}, we have created a sample of 109 \usmg absorbers accessible to SALT  in the redshift range $0.4 \leqslant z_{abs} \leqslant 0.6$. { From this sample, we confirm only 27 absorbers to be \usmg systems}. Out of these 27 systems, 21 of them were observed with SALT and for 18 of them, the \usmg host galaxies are identified.
Among the rest of the 5 unobserved systems, two of them are GOTOQs, i.e, emission is detected in the background quasar spectra. Therefore, among the 27 \usmg systems, 
we have identified the host galaxies for 20 of them.

2. Given the power-law fit \citep{Chen_2010} between the impact parameter (D) and the rest equivalent width ($\rm{W_{2796}}$), one expects the \usmg absorbers to reside within D$\sim$5 kpc. However, for our sample,
the impact parameters ranges from 7.3 to 79 kpc with a median of 18 kpc. Using the log-linear fit to the combined sample \citep{Nielsen_2013}, we argue that the \usmg absorbers constitute statistically distinct population in the $W_{2796} - D$ plane.

3. Among the the 23 \usmg systems (20 detections and 3 non-detections), we have detected \OII\ emission in 18 cases. We conclude that, including the two unobserved GOTOQs, at least about 70\% of the host galaxies are star-forming (i.e. SFR$\ge$1 M$_\odot$ yr$^{-1}$). The measured \OII\ luminosities  are in the range of $0.06 L_{[O\textsc{ii}]}^\star$ to $1.43 L_{[O\textsc{ii}]}^\star$ with a median value of $0.39 L_{[O\textsc{ii}]}^{\star}$.

4. When we combined our sample with those from the literature,
we find a positive correlation between $W_{2796}$ and $L_{\OII}$ ($r_S = 0.34$ and p-value of $<10^{-4}$) for \usmg host galaxies as well as from the literature (see Figure~\ref{fig:o2luminosity_dist}). However, this correlation is seen to be driven by weak \MgII\ absorbers (i.e. $W_{2796}<1$\AA) which tend to be associated with low luminosity galaxies with higher D.

5. Compared to the field galaxies and the GOTOQs, we find the measured emission lines ratios \OIII\ / \OII\ and \OIII\ / $\rm{H\beta}$ to be smaller for the \usmg host galaxies. Using photoionization models we find this trend to be driven by the ionizing radiation in these \usmg host galaxies being softer compared to the field galaxies. 

6. We measure galaxy parameters like,{{  stellar mass ($10.21\leqslant log~[M_\star/M_\odot]\leqslant 11.62$), B-band magnitude ($-21.68\leqslant M_B \leqslant -19.05$), star formation rate ($2.59\leqslant SFR [M_\odot yr^{-1}]\leqslant 33.51$) and  halo mass ($11.64\leqslant log~[M_h/M_\odot]\leqslant 13.56$) using SED fitting. The inferred virial radius of halos ($145\leqslant R_h[kpc] \leqslant 624$) are at least 5 times larger than the impact parameter measured.}} The measured $M_\star$ for \usmg are higher than those found for general \MgII\ absorbers at a given impact parameter. However the measured SFR are slightly lower than what has been predicted by star forming main sequence.

7. The general population of \MgII\ absorbers show a correlation between D and $M_\star$. We find the \usmg\ systems to not follow this trend. For the same impact parameter, the \usmg host galaxies are more massive and luminous compared to the host galaxies of relatively weak \MgII\ absorbers.

8. We find the mean velocity difference between the \MgII\ absorption redshift and galaxy redshift from \OII\ emission to $\mu = 3.48\pm5.82$ \kms with a $\sigma=71.02\pm4.83$ \kms. This velocity difference is well within the expected escape velocities for the inferred range of $M_h$. However, we argue that $W_{2796}$ measured along the line of sight should correspond to a velocity width in excess of 300 \kms. Therefore, what fraction of the observed gas in \usmg systems is bound can be answered only using high resolution spectra. Additionally,  large velocity widths could indicate multiple components (i.e. merging systems) contributing to the absorption. This can also be probed using high resolution spectroscopy.

9. In the literature sample of \usmg\ systems we find all the 4 known host galaxies are isolated (no companion galaxy within 100 kpc and  $\pm$500 \kms). In our sample we find that {{ at least $33\%$ of host galaxies of \usmg\ are isolated up to a projected distance of 100 kpc and the r band limiting magnitude of 23.6.}} Measuring the galaxy orientations using high spatial resolution imaging of the isolated galaxies together with a high-SNR spectrum probing the 
down the barrel absorption will be very useful to quantify the role of strong outflows in the case of \usmg systems.

\section*{Acknowledgments}
All the new observations reported in this paper were obtained with the Southern African Large Telescope (SALT). RD gratefully acknowledges support from the European Research Council (ERC) under the European Union's Horizon 2020 research and innovation programme (grant agreement No 757535). This project makes use of the following softwares : NumPy \citep{numpy2020}, SciPy \citep{scipy2020}, Matplotlib \citep{matplotlib2007}, and AstroPy \citep{astropy:2013, astropy:2018}. LKG  thanks Aromal P for helpful discussions on data reductions and data analysis.We thank the anonymous referee for the comments and suggestions that significantly improved the presentations of this paper.

This paper makes use of SDSS observational data. Funding for the Sloan Digital Sky Survey IV has been provided by the Alfred P. Sloan Foundation, the U.S. Department of Energy Office of Science, and the Participating Institutions. SDSS-IV acknowledges support and resources from the Center for High Performance Computing  at the University of Utah. The SDSS website is www.sdss.org. SDSS-IV is managed by the Astrophysical Research Consortium for the Participating Institutions of the SDSS Collaboration including  the Brazilian Participation Group, the Carnegie Institution for Science, Carnegie Mellon University, Center for Astrophysics | Harvard \& Smithsonian, the Chilean Participation 
Group, the French Participation Group, Instituto de Astrof\'isica de 
Canarias, The Johns Hopkins University, Kavli Institute for the 
Physics and Mathematics of the Universe (IPMU) / University of Tokyo, the Korean Participation Group, Lawrence Berkeley National Laboratory, Leibniz Institut f\"ur Astrophysik Potsdam (AIP),  Max-Planck-Institut 
f\"ur Astronomie (MPIA Heidelberg), Max-Planck-Institut f\"ur Astrophysik (MPA Garching), Max-Planck-Institut f\"ur Extraterrestrische Physik (MPE), National Astronomical Observatories of China, New Mexico State University, 
New York University, University of Notre Dame, Observat\'ario Nacional / MCTI, The Ohio State University, Pennsylvania State University, Shanghai 
Astronomical Observatory, United Kingdom Participation Group, 
Universidad Nacional Aut\'onoma de M\'exico, University of Arizona, University of Colorado Boulder, University of Oxford, University of Portsmouth, University of Utah, University of Virginia, University of Washington, University of 
Wisconsin, Vanderbilt University, and Yale University.
\section*{Data Availability}
Data used in this work are obtained using SALT. Raw data will become available for public use 1.5 years after the observing date at https://ssda.saao.ac.za/.

%\subsection{Other Potential Absorbing Galaxies}

%%%%%%%%%%%%%%%%%%%%%%%%%%%%%%%%%%%%%%%%%%%%%%%%%

%\newpage
%\mbox{~}
%\clearpage
%\newpage
\bibliographystyle{mnras}
\bibliography{mybib} % if your bibtex file is called example.bib
\mbox{~}
\clearpage
\newpage

\appendix
\section{Gaussian fits to absorption lines}

\begin{figure*}
    \centering
    \includegraphics[height=9.2in, width=\textwidth]{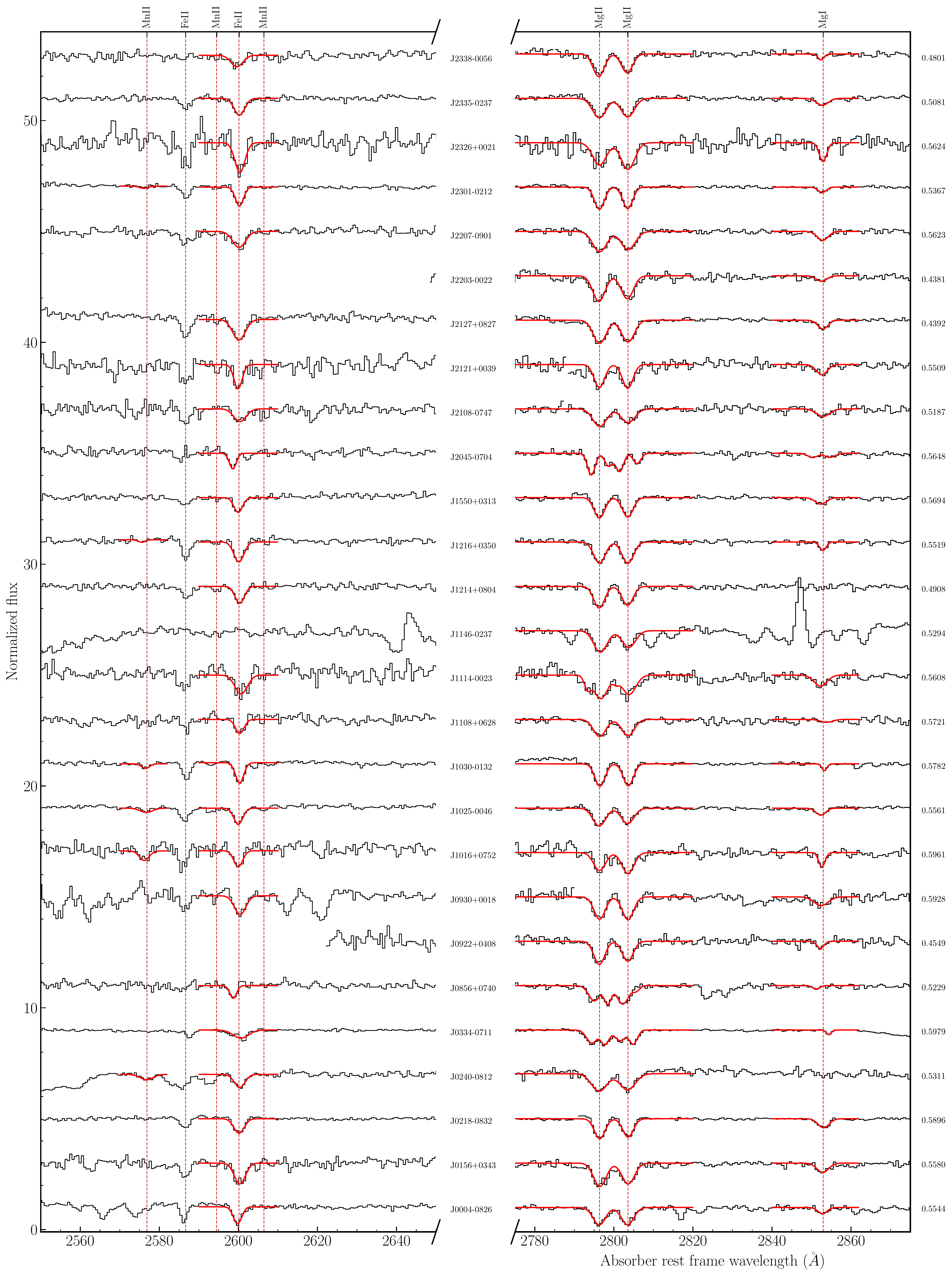}
    \caption{Gaussian fits to absorption profiles of \MnII, \FeII, \MgII, and, \MgI\ for the all \usmg\ systems in our sample in the rest frame of the absorbing gas. Normalized quasar spectrum is shown in black. Solid red lines overlaid on the normalized spectra are obtained Gaussian fits. The red dashed vertical lines drawn according to the \usmg absorption redshift corresponds to the different transitions marked on the top panel. }
    \label{fig:abs_profiles}
\end{figure*}

\begin{figure}
    \centering
    \includegraphics[width=0.5\textwidth]{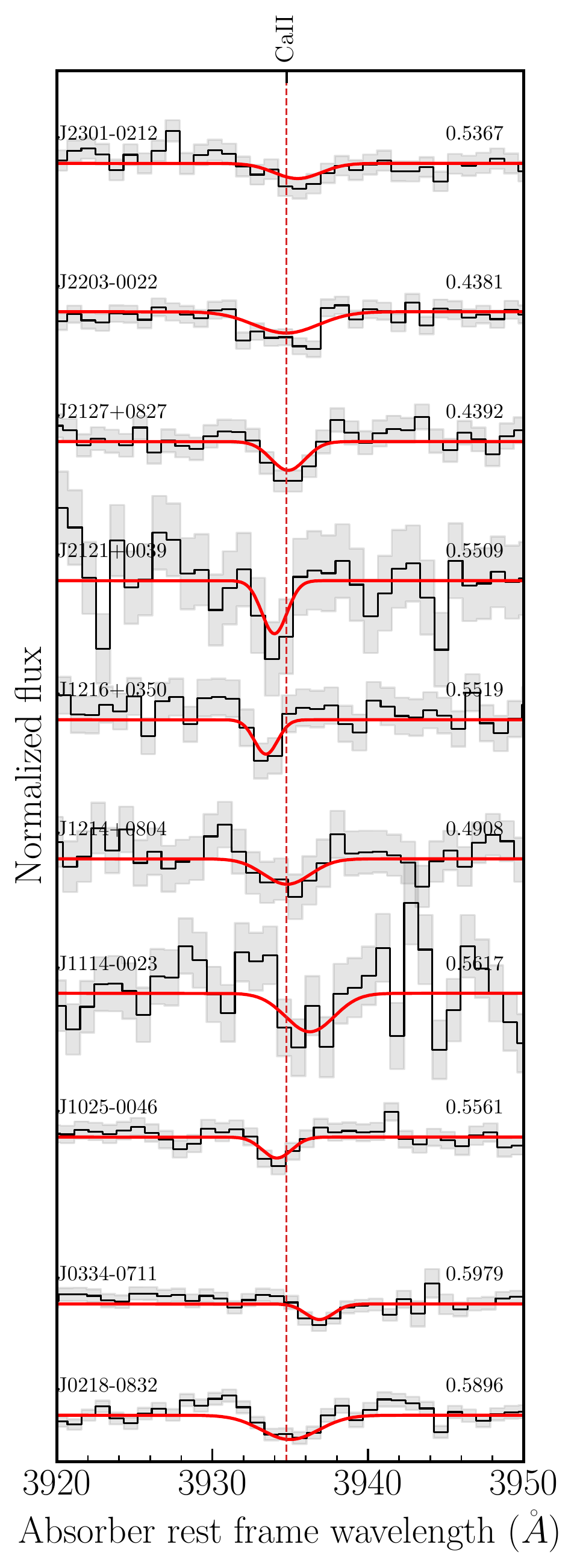}
    \caption{Gaussian fit to the \CaII\ absorption associated with the \usmg absorptions in the absorber rest frame. The vertical red dashed line corresponds to the expected location of \CaII\ absorption based on redshift of the \usmg absorption.}
    \label{fig:caii_absorption}
\end{figure}

{ We computed the  redshift and rest equivalent widths of absorption lines using Gaussian fits. In Figure~\ref{fig:abs_profiles} we summarise the fits to \MgII, \MgI, \FeII\ and \MnII. The corresponding plots for cases with clear  \CaII\ detections are shown in Figure~\ref{fig:caii_absorption}.}

\section{BAL quasars as False USMg\textsc{II} Detection in the Absorber Catalog}
\label{sec:append_a}
With our selection criteria for \usmg absorption systems, i.e. accessible to SALT and having absorption redshift in the interval 0.4$\le$\zabs$\le$0.6, the JHU-SDSS Metal Absorber Catalog of \citet{Zhu2013} yields a total of 109 \usmg\ systems. As discussed in Section\ref{sec:sample}, a careful visual inspection of individual spectrum resulted in only a total of 27 \usmg systems being secure. { Table~\ref{tab:append_1} summarises the details of all the 109 \usmg\ systems. 
%(see section \ref{sec:sample}, for more details). 
In the last column of this table, we describe our reasoning for the inclusion or the exclusion of all of these 109 systems in our \usmg sample.}

\begin{table*}
\caption{Details of all the 109 \usmg\ absorber from \citet{Zhu2013}}
\begin{tabular}{lccccccc}
\hline

No. & Quasar & Plate & MJD   & Fiber & $z_{qso}$ & $z_{abs}$ & Comment\\
\hline
1	&  J000413.73-082625.4  &  7148  &  56591  &  326  &  2.2470  &  0.5544  &  Secured \usmg system. Clear detection  of \MgII\ and   \FeII\ doublets.\\
2	&  J000510.28-042216.9  &  7035  &  56568  &  606  &  2.1950  &  0.5014  &  False detection  of  \usmg due to blending.\\
3	&  J000701.42-011459.1  &  4217  &  55478  &  178  &  2.3220  &  0.5742  &  \MgII\ identification is not secured.\\
4	&  J000828.36-004326.7  &  4217  &  55478  &  224  &  2.0530  &  0.4876  &  False detection  of  \usmg due to blending.\\
5	&  J001239.55+091650.1  &  4536  &  55857  &  654  &  2.3880  &  0.5300  &  False detection  of  \usmg.\\
6	&  J001409.26-070806.0  &  7150  &  56597  &  510  &  2.1950  &  0.4547  &  False detection  of  \usmg due to blending.\\
7	&  J003459.49+034456.9  &  4303  &  55508  &  414  &  2.2350  &  0.5917  &  BAL quasar.\\
8	&  J003941.88+054253.1  &  4419  &  55867  &  660  &  2.1050  &  0.5955  &  BAL quasar.\\
9	&  J004502.79+003604.1  &  3589  &  55186  &  806  &  2.0940  &  0.5221  &  BAL quasar. \\
10	&  J005233.67+014040.8  &  4306  &  55584  &  74   &  2.3010  &  0.5660  &  False detection  of  \usmg due to blending.\\
11	&  J010515.45+004742.3  &  4225  &  55455  &  955  &  2.4060  &  0.5659  &  False detection  of  \usmg due to blending.\\
12	&  J010749.31+033913.5  &  4309  &  55528  &  104  &  0.5070  &  0.5000  &  BAL quasar.\\
13	&  J012606.00-085651.9  &  2878  &  54465  &  434  &  1.8314  &  0.5197  & BAL quasar.\\
14	&  J015035.94+083512.4  &  4532  &  55559  &  930  &  1.9950  &  0.4903  &  BAL quasar.\\
15	&  J015345.35-100839.6  &  665   &  52168  &  209  &  1.9148  &  0.5655  &  BAL quasar.\\
16	&  J015635.18+034308.1  &  4270  &  55511  &  144  &  1.3710  &  0.5580  &  Secured \usmg\ system. Clear detection  of \MgII\ and   \FeII\ doublets.\\
17	&  J015931.02+005449.6  &  4234  &  55478  &  588  &  2.4500  &  0.5740  &  False detection  of  \usmg.\\
18	&  J020102.94-055819.2  &  4398  &  55946  &  525  &  1.9440  &  0.4630  &  BAL quasar.\\
19	&  J020630.60-061124.4  &  7238  &  56660  &  227  &  2.0190  &  0.5351  &  BAL quasar. \\
20	&  J020959.47-055125.7  &  4394  &  55924  &  566  &  2.5170  &  0.5849  &  False detection  of  \usmg. \\
21	&  J021035.82-010524.2  &  1074  &  52937  &  5    &  1.9242  &  0.5838  &  BAL quasar.\\
22	&  J021717.53-005148.9  &  4236  &  55479  &  140  &  2.2360  &  0.5903  &  BAL quasar.\\
23	&  J021820.10-083259.4  &  4395  &  55828  &  780  &  1.2190  &  0.5896  &  Secured \usmg system. Clear detection  of \MgII\ and   \FeII\ doublets.\\
24	&  J022234.71-040417.6  &  6369  &  56217  &  739  &  2.0430  &  0.5029  &  False detection  of  \usmg due to blending.\\
25	&  J023740.12-001231.0  &  4240  &  55455  &  476  &  1.9800  &  0.4503  &  BAL quasar.\\
26	&  J024008.21-081223.4  &  7057  &  56593  &  480  &  2.2310  &  0.5311  &  Secured \usmg system. Clear detection  of \MgII\ and   \FeII\ doublets.\\
27	&  J024426.85-005440.1  &  3650  &  55244  &  8    &  2.3130  &  0.4907  &  False detection  of  \usmg.\\
28	&  J024459.75+005411.1  &  7337  &  56686  &  756  &  2.3790  &  0.5958  &  False detection  of  \usmg.\\
29	&  J024747.59-004810.1  &  6782  &  56576  &  54   &  1.9300  &  0.5780  &  BAL quasar.\\
30	&  J025216.20-055801.9  &  7054  &  56575  &  324  &  2.3250  &  0.5082  &  False detection  of  \usmg due to blending.\\
31	&  J030443.10-082537.5  &  458   &  51929  &  98   &  1.9057  &  0.5512  & BAL quasar.\\
32	&  J033438.28-071149.0  &  461   &  51910  &  74   &  0.6358  &  0.5979  &  Secured \usmg system. Clear detection  of \MgII\ doublets. \\
33	&  J080853.34+084509.2  &  2570  &  54081  &  444  &  1.9814  &  0.5013  & False identification of \MgII\ doublets.\\
34	&  J081242.14+082832.5  &  2570  &  54081  &  616  &  2.3181  &  0.5103  & BAL quasar. \\
35	&  J082126.38+092529.1  &  2571  &  54055  &  518  &  1.8351  &  0.5358  & BAL quasar. \\
36	&  J083033.10+075438.9  &  4866  &  55895  &  754  &  1.9600  &  0.5579  &  BAL quasar.\\
37	&  J084031.86+053419.2  &  4903  &  55927  &  226  &  1.8930  &  0.4366  &  False detection  of  \usmg.\\
38	&  J085627.09+074031.7  &  1299  &  52972  &  173  &  1.8932  &  0.5224  &  Secured identification of \usmg. Clear detection of \MgII\ and \FeII\ doublets.\\
39	&  J090653.58+095930.3  &  1739  &  53050  &  286  &  1.6885  &  0.4504  &  BAL quasar. \\
40	&  J091532.66+073734.3  &  4869  &  55896  &  944  &  2.3510  &  0.5176  &  False detection  of  \usmg.\\
41	&  J092100.99+045311.7  &  991   &  52707  &  217  &  1.8163  &  0.4924  &  BAL quasar.\\
42	&  J092222.53+040858.6  &  568   &  52254  &  372  &  0.7146  &  0.4549  &  Secured \usmg system. Clear detection  of \MgII\ doublets.\\
43	&  J093020.60+001828.0  &  3823  &  55534  &  996  &  2.4300  &  0.5928  &  Secured \usmg system. Clear detection  of \MgII\ and   \FeII\ doublets.\\
44	&  J093512.75-013724.6  &  3767  &  55214  &  776  &  2.1490  &  0.4770  &  False detection  of  \usmg. No  trace  of \FeII\ doublets.\\
45	&  J100726.26+051300.7  &  573   &  52325  &  445  & 1.9052   &  0.4488  &  False identification of \usmg.\\
46	&  J101610.82+075209.1  &  5334  &  55928  &  90   &  2.1870  &  0.5961  &  Secured \usmg system. Clear detection  of \MgII\ and   \FeII\ doublets.\\
47	&  J102510.09-004644.9  &  3832  &  55289  &  409  &  2.2120  &  0.5561  &  Secured \usmg system. Clear detection  of \MgII\ and   \FeII\ doublets.\\
48	&  J103015.07-025032.8  &  3785  &  55241  &  724  &  2.1480  &  0.5783  &  Secured \usmg system. Clear detection  of \MgII\ and   \FeII\ doublets.\\
49	&  J103059.75-013237.7  &  3785  &  55273  &  338  &  2.1680  &  0.5714  &  BAL quasar\\
50	&  J103326.53+015841.7  &  4739  &  55644  &  18   &  2.1260  &  0.4720  &  \MgII\ identification is not secured.\\
51	&  J104652.17+063404.0  &  4853  &  55928  &  318  &  2.3650  &  0.5572  &  False detection  of  \usmg.\\
52	&  J110338.72+084101.1  &  5359  &  55953  &  928  &  2.3900  &  0.5859  &  BAL quasar.\\
53	&  J110817.93+062832.9  &  1003  &  52641  &  19   &  1.2049  &  0.5721  &  Secured \usmg system. Clear detection  of \MgII\ and   \FeII\ doublets.\\
54	&  J111249.66+005310.1  &  278   &  51900  &  619  &  1.6822  &  0.4274  &  BAL quasar.\\
55	&  J111359.99-002342.7  &  279   &  51984  &  262  &  0.9535  &  0.5617  &  Secured \usmg system. Clear detection  of \MgII\ and   \FeII\ doublets.\\
56	&  J111816.33+083436.1  &  2413  &  54169  &  16   &  0.5830  &  0.5821  & \usmg absorption from the host galaxy of the quasar.\\ 
57	&  J112021.40+084423.0  &  5369  &  56272  &  766  &  2.1140  &  0.5505  &  BAL quasar.\\
58	&  J113826.30-015428.9  &  3775  &  55207  &  699  &  1.9230  &  0.5770  &  BAL quasar.\\
59	&  J114546.21+032251.9  &  514   &  51994  &  458  &  2.0091  &  0.5441  &  False identification of \usmg.\\
60	&  J114614.24-023716.1  &  3790  &  55208  &  420  &  2.1500  &  0.5295  &  Secured detection of \usmg absorption.\\
\hline
\end{tabular}
   \label{tab:append_1}
\end{table*}

\begin{table*}
\ContinuedFloat
\caption{Continued.}
\begin{tabular}{lccccccc}
\hline
61	&  J115820.00+065415.3  &  1622  &  53385  &  64   &  1.7105  &  0.4651  &  BAL quasar.\\
62	&  J120359.81+095419.0  &  5391  &  56000  &  552  &  2.0240  &  0.4711  &  BAL quasar.\\
63	&  J121453.29+080457.7  &  1624  &  53386  &  505  &  1.4325  &  0.4908  &  Secured \usmg system. Clear detection  of \MgII\ and   \FeII\ doublets.\\
64	&  J121628.03+035031.8  &  4751  &  55646  &  614  &  0.9960  &  0.5519  &  Secured \usmg system. Clear detection  of \MgII\ and   \FeII\ doublets.\\
65	&  J122043.21-013215.3  &  333   &  52313  &  540  &  0.4482  &  0.4453  &  \usmg absorption from the host galaxy of the quasar.\\
66	&  J122354.57+095935.3  &  5401  &  55949  &  518  &  2.3530  &  0.5692  &  BAL quasar.\\
67	&  J123256.70-000604.3  &  3848  &  55647  &  388  &  2.3750  &  0.5211  &  False detection  of  \usmg.\\
68	&  J124602.10+075425.6  &  5407  &  55926  &  39   &  2.1390  &  0.5432  &  BAL quasar.\\
69	&  J124628.80+061039.2  &  847   &  52426  &  366  &  1.9588  &  0.5925  &  BAL quasar.\\
70	&  J124644.46-025122.3  &  336   &  51999  &  172  &  1.7713  &  0.4961  &  BAL quasar.\\
71	&  J133748.81-020102.7  &  4047  &  55352  &  710  &  0.5780  &  0.5742  &  BAL quasar.\\
72	&  J140301.31+032531.9  &  4783  &  55652  &  446  &  2.2560  &  0.5212  &  BAL quasar.\\
73	&  J140601.18-012443.1  &  4038  &  55363  &  796  &  1.8350  &  0.5186  &  BAL quasar.\\
74	&  J141206.73+071325.6  &  1824  &  53491  &  625  &  2.0192  &  0.5996  &  False identification of \usmg.\\
75	&  J142717.70+084255.2  &  5462  &  55978  &  222  &  2.1970  &  0.5313  &  False detection  of  \usmg due to blending.\\
76	&  J143830.68-014807.4  &  4026  &  55325  &  696  &  2.2200  &  0.5914  &  BAL quasar.\\
77	&  J144959.96+003225.3  &  309   &  51994  &  357  &  1.7065  &  0.4648  &  BAL quasar.\\
78	&  J144959.96+003225.3  &  310   &  51990  &  305  &  1.9253  &  0.5804  &  BAL quasar.\\
79	&  J145704.53+072033.3  &  1815  &  53884  &  299  &  2.0192  &  0.5966  &  False identification of \usmg.\\
80	&  J145913.71+000215.9  &  4019  &  55363  &  36   &  1.9300  &  0.5807  &  BAL quasar.\\
81	&  J150919.01-013353.4  &  922   &  52426  &  586  &  2.0656  &  0.5128  &  BAL quasar.\\
82	&  J150919.01-013353.4  &  922   &  52426  &  586  &  2.0656  &  0.5244  &  BAL quasar.\\
83	&  J150938.28+050524.4  &  4776  &  55652  &  782  &  2.3270  &  0.5632  &  False detection  of  \usmg.\\
84	&  J152218.01+051007.0  &  4803  &  55734  &  558  &  2.4540  &  0.5863  &  False detection  of  \usmg. \\
85	&  J153200.65+090540.0  &  5209  &  56002  &  812  &  2.1560  &  0.5633  &  BAL quasar.\\
86	&  J155003.71+031325.0  &  594   &  52045  &  611  &  1.7879  &  0.5694  &  Secured \usmg system. Clear detection  of \MgII and   \FeII doublets. \\
87	&  J162902.98+091322.5  &  2532  &  54589  &  494  &  1.9913  &  0.4470  & False identification of \usmg.\\
88	&  J204333.20-001104.2  &  981   &  52435  &  44   &  0.5469  &  0.5293  & BAL quasar.\\
89	&  J204501.33-070452.6  &  1917  &  53557  &  205  &  0.6702  &  0.5635  &  Secured \usmg system. Clear detection  of \MgII\ and   \FeII\ doublets.\\
90	&  J210851.54-074726.5  &  638   &  52081  &  126  &  1.4890  &  0.5187  &  Secured \usmg system. Clear detection  of \MgII\ and   \FeII\ doublets.\\
91	&  J212119.15+012039.6  &  5142  &  55825  &  776  &  2.1980  &  0.4906  &  False detection  of  \usmg due to blending.\\
92	&  J212143.98+003954.2  &  5142  &  55825  &  220  &  1.3480  &  0.5509  &  Secured \usmg system. Clear detection  of \MgII\ and   \FeII\ doublets.\\
93	&  J212727.19+082724.6  &  4089  &  55470  &  402  &  0.7450  &  0.4392  &  Secured \usmg system. Clear detection  of \MgII\ and   \FeII\ doublets.\\
94	&  J213157.89+031459.0  &  4080  &  55471  &  29   &  2.2800  &  0.5299  &  False detection  of  \usmg.\\
95	&  J213555.40-063553.6  &  641   &  52199  &  567  &  1.1724  &  0.4443  &  False identification of \MgII\ doublets.\\
96	&  J220131.89-001543.4  &  4198  &  55480  &  118  &  1.9300  &  0.4462  &  BAL quasar.\\
97	&  J220330.04-002211.4  &  372   &  52173  &  26   &  1.7640  &  0.4381  &  Secured \usmg system. Clear identification of \FeII\ and \MgII\ doublets.\\
98	&  J220619.09-031423.1  &  4377  &  55828  &  296  &  2.2080  &  0.5970  &  BAL quasar. \\
99	&  J220702.53-090127.8  &  718   &  52206  &  209  &  1.2965  &  0.5623  &  Secured \usmg system. Clear detection  of \MgII\ and   \FeII\ doublets.\\
100	&  J222848.61+000633.7  &  4202  &  55445  &  940  &  2.3280  &  0.5352  &  False detection  of  \usmg.\\
101	&  J222848.61+000633.7  &  4214  &  55451  &  10   &  2.2370  &  0.5078  &  BAL quasar.\\
102	&  J223308.78-085445.4  &  721   &  52228  &  71   &  1.8525  &  0.5238  &  BAL quasar.\\
103	&  J230101.29-021200.0  &  4362  &  55828  &  814  &  0.6190  &  0.5367  &  Secured \usmg system. Clear detection  of \MgII\ and   \FeII\ doublets.\\
104	&  J230124.06+091704.6  &  5056  &  55921  &  110  &  2.1250  &  0.5555  &  BAL quasar.\\
105	&  J232627.39+034627.2  &  4283  &  55864  &  538  &  2.0050  &  0.4735  &  BAL quasar.\\
106	&  J232653.14+002142.9  &  1095  &  52521  &  549  &  2.1959  &  0.5624  &  Secured identification of \usmg.\\
107	&  J233548.62-023734.3  &  4357  &  55829  &  434  &  1.2250  &  0.5081  &  Secured \usmg system. Clear detection  of \MgII\ and   \FeII\ doublets.\\
108	&  J233818.25-005610.5  &  385   &  51877  &  295  &  0.8946  &  0.4801  &  Secured \usmg system. Clear detection  of \MgII\ doublets. \\
109	&  J235859.47-002426.2  &  387   &  51791  &  181  & 1.7592   &  0.4930  &  BAL quasar.\\
\hline
\end{tabular}
\end{table*}

\section{Survey Completeness}
{ In Table~\ref{tab:completeness}, we list the galaxies in DECaLS galaxy catalog \citep{Dey2019} having consistent photo-z with our \usmg\ absorbers and within an impact parameter of 100 kpc. Last column of this table provides the status of spectroscopic redshift measurements. We use the information provided in this table to quantify the survey completeness of our sample (see section 3.1). }

\begin{table*}
\caption{Details of galaxies with consistent photo-z to the \usmg\ absorbers in our sample with D$<$100 kpc}
    \centering
    \begin{tabular}{cccccccc}
    \hline
No. & Quasar Name & \zabs & Galaxy coordinates & D(kpc) & $m_r$ & Photo-z  & Comments \\
\hline
1 & J0004-0826 & 0.5544 & J000413.99-082624.27 & 26 & 19.9 & 0.558 $\pm$0.044 & Spectroscopically confirmed \\
 &  & & J000413.18-082632.34 &  69 & 22.9 & 0.648$\pm$0.161 & Consistent photo-z \\
% &  & 77 & 23.3 & 0.887 & 0.212 & Not consistent \\
2 & J0156+0343& 0.5581 & J015635.21+034306.03 &  14 & 20.9 & 0.573$\pm$0.037 & Spectroscopically confirmed \\
 & &  & J015635.19+034311.86 &  23 & 23.3 & 0.635$\pm$0.403 & Consistent photo-z \\
 &&  & J015634.97+034305.15 &  28 & 21.6 & 0.488$\pm$0.104 & Consistent photo-z \\
 & & & J015635.30+034303.22 &  34 & 23 & 0.497$\pm$0.143 & Spectroscopically confirmed \\
 & & & J015635.54+034311.76 &  41 & 22 & 0.524$\pm$0.112 & Consistent photo-z \\
 & & & J015635.55+034304.93 &  41 & 21 & 0.455$\pm$0.106 & Consistent photo-z \\
 
3 & J0218-0832 & 0.5896 & J021820.26-083258.39 &  16 & 21.1 & 0.632$\pm$0.026 & Spectroscopically confirmed \\
4 & J0240-0812 & 0.5311 & J024008.39-081222.53 &  18 & 20.7 & 0.524$\pm$0.076 & Spectroscopically confirmed \\
% &  & 72 & 21.2 & 0.881 & 0.113 & Not consistent \\
% &  & 72 & 22 & 0.7 & 0.049 & Not consistent \\
 &&  & J024007.42-081224.78 &  74 & 22.7 & 0.556$\pm$0.1 & Consistent photo-z \\
 &&  & J024008.97-081217.78  &  79 & 23.4 & 0.77$\pm$0.306 & Consistent photo-z \\
% &  & 81 & 21.7 & 0.328 & 0.091 & Not consistent \\
 &&  & J024009.08-081225.12  &  82 & 23.4 & 0.596$\pm$0.107 & Consistent photo-z \\
5 & J0334-0711 & 0.5977 & J033438.08-071152.08&  28 & 20.9 & 0.553$\pm$0.18 & Spectroscopically confirmed \\
 &  & & J033438.54-071153.17 &  37 & 22.6 & 0.595$\pm$0.074 & Consistent photo-z \\
% &  & 49 & 15.5 &  &  & Star \\
% &  & 66 & 22.5 & 0.693 & 0.05 & Not consistent \\
6 & J0856+0740 & 0.5232 & J085627.22+074031.36 &  12 & 23 & 0.627$\pm$0.141 & Consistent photo-z but no nebular emission detected \\
& &  & J085627.31+074029.51 &  24 & 21.7 & 0.577$\pm$0.112 & Spectroscopically confirmed \\
% &  & 49 & 21.8 & 0.685 & 0.029 & Not consistent \\
% &  & 59 & 19.5 &  &  & Star \\
 & &  & J085626.52+074025.70 &  65 & 23 & 0.503$\pm$0.127 & Consistent photo-z \\
% &  & 69 & 22.7 & 0.776 & 0.094 & Not consistent \\
% &  & 94 & 19 & 0.176 & 0.091 & Not consistent \\
 &  & & J085626.08 +074035.84 &  98 & 23.5 & 0.558$\pm$0.257 & Consistent photo-z \\
7 & J0922+0408 & 0.4549 & J092222.52+040856.00 &  15 & 20.8 & 0.333$\pm$0.109 & Not consistent but spectroscopically confirmed \\
% &  & 51 & 19.8 &  &  & Star \\
% &  & 81 & 22.6 & 0.964 & 0.121 & Not consistent \\
8 & J1114-0023&0.5610  & J111400.04-002341.21 & 10 & 22 & 0.893$\pm$0.18 & Not consistent \\
% &  & 19 & 20.8 & 0.229 & 0.111 & Not consistent \\
 &&  & J111400.19-002341.33 &  20 & 22.5 & 0.505$\pm$0.101 & Consistent photo-z \\
% &  & 23 & 22.5 & 0.739 & 0.113 & Not consistent \\
 & & & J111359.65-002336.94 &  49 & 23.4 & 0.725$\pm$0.205 & Consistent photo-z \\
 %&  & 78 & 20.6 & 0.532 & 0.025 & Not consistent \\
 %&  & 79 & 21 & 0.107 & 0.063 & Not consistent \\
9 & J1214+0804 & 0.4908 & J121453.46+080457.03  &  16 & 20.5 & 0.408$\pm$0.079 & Spectroscopically confirmed \\
 &  & & J121453.16+080446.44  &  68 & 23.5 & 0.869$\pm$0.57 & Consistent photo-z \\
% &  & 84 & 23.6 & 0.791 & 0.28 & Not consistent \\
% &  & 94 & 21.2 & 0.389 & 0.092 & Not consistent \\
10 & J1216+0350 &0.5519  & ... &  ...&... &... & GOTOQ with no other candidates\\
%&0.5519 &18 & 23.5 & 0.992$\pm$0.189 & Not consistent \\
% &&  & 69 & 23.5 & 1.048$\pm$0.437 & Not consistent \\
% &&  & 80 & 21.1 & 0.359$\pm$0.158 & Not consistent \\
11 & J1550+0313&0.5694 &J155003.70+031325.70  &  4 & 21.9 & 0.436$\pm$0.361 & Consistent photo-z but no nebular emission detected \\
 &  & & J155003.47+031325.77  &  23 & 20.5 & 0.485$\pm$0.085 & Spectroscopically confirmed \\
 &  & & J155002.95+031333.28 &  90 & 21.5 & 0.314$\pm$0.269 & Consistent photo-z \\
12 & J2045-0704&0.5649 & J204501.46-070452.13 &  13 & 21.8 & 0.56$\pm$0.086 & Consistent photo-z but no nebular emission detected \\
% &  & 24 & 22.2 & 0.361 & 0.177 & Not consistent \\
% &  & 45 & 20.3 & 0.321 & 0.127 & Not consistent \\
 & & & &  52 & 22.6 & 0.58$\pm$0.271 & Consistent photo-z \\
% &  & 73 & 22.7 & 0.748 & 0.112 &  \\
% &  & 85 & 18.1 &  &  & Star \\
13 & J2108-0704 & 0.5187 & J210851.43-074727.67 &  12 & 21.6 & 0.398$\pm$0.144 & Spectroscopically confirmed \\
% &  & 81 & 22.9 & 0.863 & 0.23 & Not consistent \\
% &  & 88 & 22.8 & 1.159 & 0.211 & Not consistent \\
14 & J2121+0039 & 0.5509& J212144.18+003954.25 &  19 & 20.9 & 0.566$\pm$ 0.102 & Spectroscopically confirmed \\
% &  & 49 & 21.1 & 0.256 & 0.164 & Not consistent \\
% &  & 80 & 20.5 & 0.471 & 0.048 & Not consistent \\
 
% \hline
%\end{tabular}
%   \label{tab:append_2}
%\end{table*}

%\begin{table*}
%\ContinuedFloat
%\caption{Continued.}
%\begin{tabular}{cccccccc}
%\hline
15 & J2127+0827 &0.4392 &J212727.09+082724.21  &  8 & 20.5 & 0.306$\pm$0.145 & Spectroscopically confirmed \\
% &  & 23 & 19.7 &  &  & Star \\
% &  & 33 & 19.4 &  &  & Star \\
% &  & 43 & 22.4 & 0.713 & 0.08 & Not consistent \\
% &  & 63 & 21.7 & 0.727 & 0.053 & Not consistent \\
% &  & 70 & 23.3 & 0.656 & 0.173 & Not consistent \\
% &  & 75 & 20 &  &  & Star \\
% &  & 88 & 23.5 & 0.869 & 0.145 & Not consistent \\
% &  & 89 & 20.8 & 0.908 & 0.233 & Not consistent \\
16 & J2203-0022&0.4381 & J220329.98-002209.56 &  11 & 22.2 & 0.455$\pm$0.147 & Consistent photo-z but no nebular emission detected \\
& &  & J220329.76-002215.03  &  31 & 20 & 0.358$\pm$0.081 & Spectroscopically confirmed \\
 &  & & J220329.70-002217.06 &  42 & 23.5 & 0.965$\pm$0.611 & Consistent photo-z \\
% &  & 93 & 23 & 1.05 & 0.155 & Not consistent \\
17 & J2207-0901 & 0.5623 & J220702.92-090128.50 &  37 & 22 & 0.371$\pm$0.071 & Spectroscopically confirmed \\
% &  & 45 & 23.4 & 0.887 & 0.238 & Not consistent \\
 & &  & J220702.86-090132.85 &  45 & 22.9 & 0.568$\pm$0.187 & Consistent photo-z (Spectroscopically confirmed) \\
& &  & J220703.09-090125.14 &  56 & 20.9 & 0.551$\pm$0.029 & Consistent photo-z \\
& &  & J220702.02-090132.81 &  58 & 20.5 & 0.505$\pm$0.094 & Consistent photo-z \\
% &  & 60 & 22.2 & 0.421 & 0.08 & Not consistent \\
 & &  &J220703.22-090136.62  &  88 & 23.3 & 0.582$\pm$0.235 & Consistent photo-z \\
18 & J2301-0212 &0.5367 & J230100.83-021202.97 & 47 & 22.2 & 0.627$\pm$0.238 & Consistent photo-z \\
 %&  & 58 & 22.9 & 0.946 & 0.132 & Not Consistent \\
19 & J2326+0021 &0.5624 & J232653.20+002148.10 &  33 & 19.6 & 0.485$\pm$0.037 & Spectroscopically confirmed \\
 %&  & 45 & 20.1 & 0.261 & 0.064 & Not Consistent \\
 & & & J232652.87+002135.83 &  53 & 22.3 & 0.479$\pm$0.101 & Consistent photo-z \\
% &  & 65 & 20.4 & 0.633 & 0.042 & Not Consistent \\
% &  & 69 & 22.6 & 0.764 & 0.126 & Not Consistent \\
& &  & J232653.10+002156.14 &  85 & 22.5 & 0.591$\pm$ 0.083 & Consistent photo-z \\
% &  & 88 & 21 & 0.377 & 0.083 & Not Consistent \\
% &  & 92 & 18.2 & 0.189 & 0.124 & Not Consistent \\
20 & J2335-0237&0.5081 & J233548.79-023734.79  &  16 & 20.6 & 0.54$\pm$0.075 & Spectrosccopically confirmed \\
 & & & J233548.45-023738.64 &  30 & 22.9 & 0.693$\pm$0.339 & Consistent photo-z \\
 & & & J233549.09-023742.99 &  68 & 21.6 & 0.482$\pm$ 0.236 & Consistent photo-z \\
% &  & 73 & 23 & 1.098 & 0.184 & Not Consistent \\
% &  & 76 & 23.2 & 0.764 & 0.196 & Not Consistent \\
% &  & 78 & 22.6 & 0.8 & 0.139 & Not Consistent \\
% &  & 79 & 23.2 & 1.115 & 0.437 & Not Consistent \\
% &  & 80 & 21.5 & 0.563 & 0.056 & Not Consistent \\
 & & & J233549.48-023730.00  &  84 & 23.5 & 0.969$\pm$0.592 & Consistent photo-z \\
21 & J2338-0056 &0.4801&J233818.44-005610.48  &  16 & 22.2 & 0.516$\pm$0.077 & No nebular emission\\
 &&  & J233817.985-005617.62&  49 & 22.9 & 0.62$\pm$0.19 & Consistent photo-z \\
% &  & 56 & 23.4 & 0.793 & 0.138 & Not Consistent \\
 &&  & J233817.58-005609.86 &  60 & 22 & 0.497$\pm$0.037 & No nebular emission\\
 &&  & J233819.09-005605.87 &  79 & 22.3 & 0.449$\pm$0.082 & Spectroscopically confirmed \\
% &  & 86 & 21.8 & 0.718 & 0.177 & Not Consistent \\
 & & & J233817.38-005604.47 &  86 & 21.6 & 0.523$\pm$0.103 & Consistent photo-z \\
% &  & 88 & 23.4 & 0.872 & 0.1 &  Not Consistent \\
 \hline

    \end{tabular}
%    \caption{Caption}
\label{tab:decals}
\end{table*}

\section{$\rm{W^{Ca\textsc{ii}}_{3935}}$ vs $\rm{E(B-V)}$}

\begin{figure}
    \centering
    \includegraphics[width=0.495\textwidth]{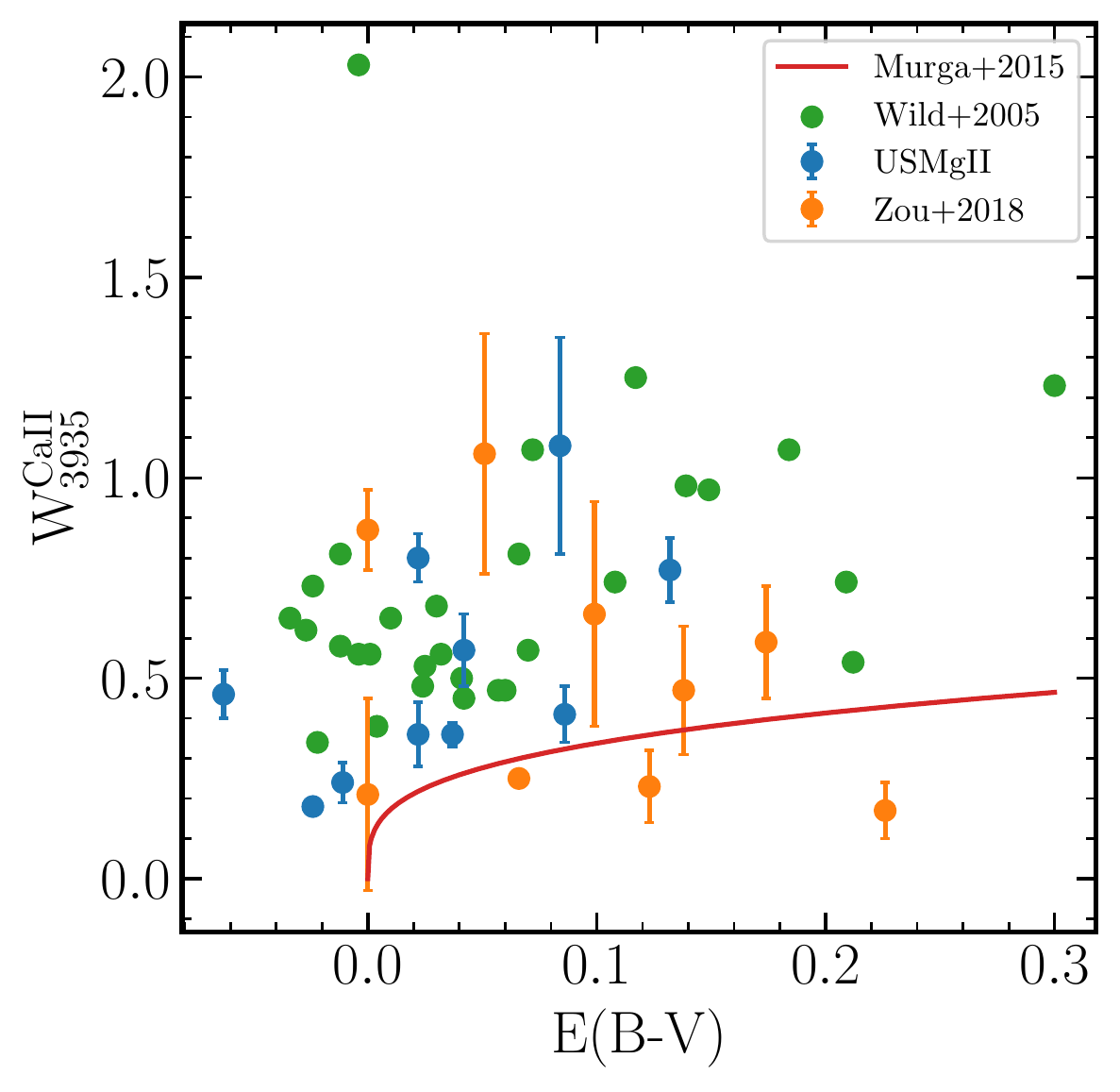}
    \caption{\CaII\ $\lambda$ 3935 rest equivalent width plotted against the color excess of the background quasars for our \usmg sample (blue), high redshift \CI\ absorbers (orange) \citep{Zou2018}, and the \CaII\ absorbers from SDSS DR3 in the redshift range $0.84\leqslant z_{abs}\leqslant 1.3$ \citep{Wild2005}. The red solid line corresponds to the relation obtained for our Milky way galaxy. }
    \label{fig:wca2_vs_dust}
\end{figure}

In Figure~\ref{fig:wca2_vs_dust} we plot rest equivalent width of Ca~{\sc ii}$\lambda$3936 against E(B-V) for our sample, C~{\sc i} absorbers from ~\citet{Zou2018} and DLAs from \citet{Wild2005}. The curve shown in this figure is for our Galactic sightlines found by \citet{Murga2015}. \citet{Wild2005} reported a possible correlation between the two quantities. Note that the sample spans a large E(B-V) for a given $\mathrm{W^{CaII}_{3935}}$. However, such a correlation was not clearly evident in the case of \citet{Zou2018}. Our data points roughly follow the trend shown by the sample of \citet{Wild2005}, however the error in our equivalent width measurements are large. Also our sample lacks systems with large E(B-V) and Ca~{\sc ii} absorption detections.

\section{$L_{\OIII}$ / $L_{\OII}$ vs $W_{2796}$}
\label{sec:o3bo2vswmg2}
\begin{figure}
    \centering
    \includegraphics[width=0.495\textwidth]{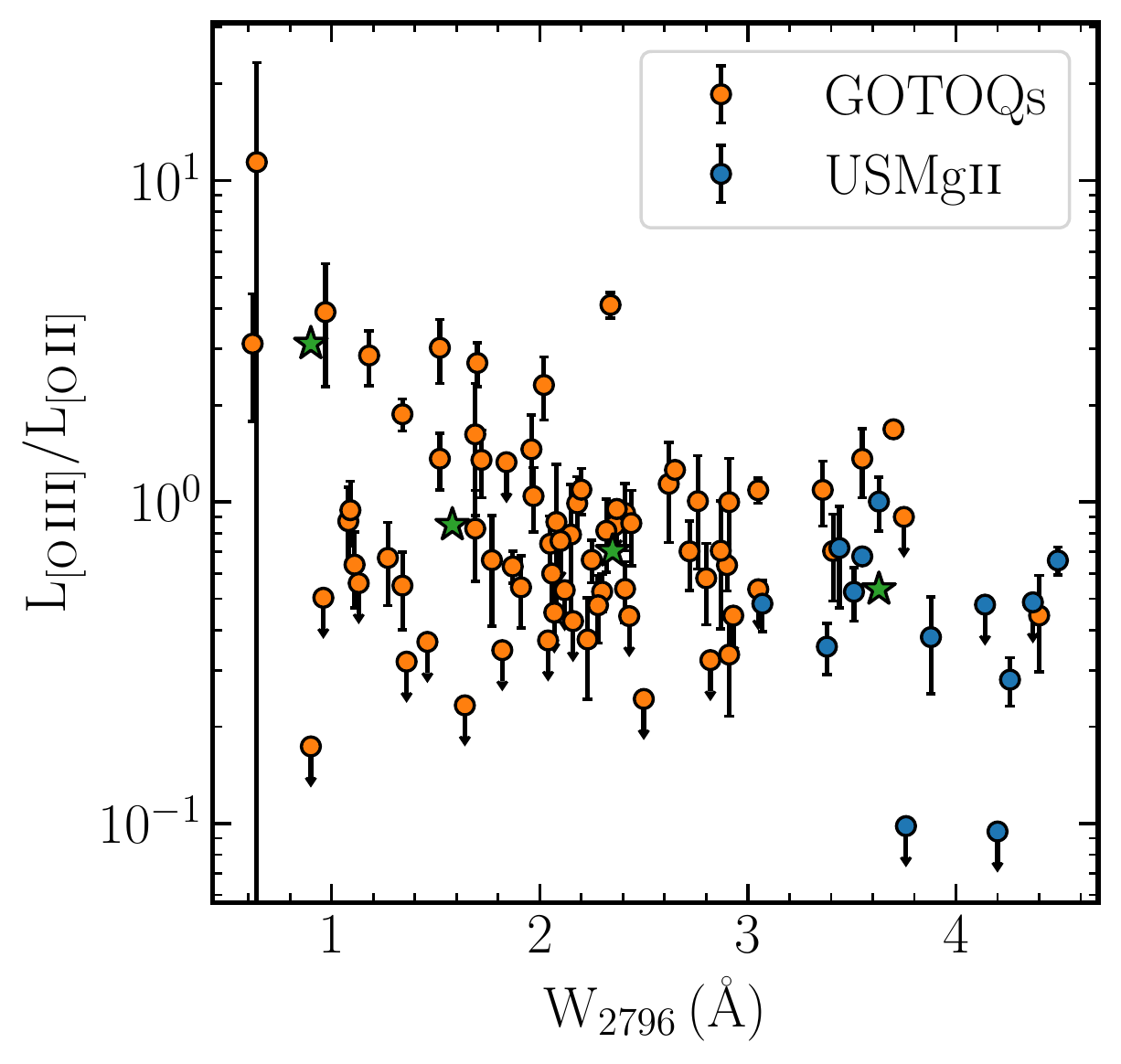}
    \caption{$L_{\OIII}$ / $L_{\OII}$ ratio vs the rest equivalent width of \MgII\ for both GOTOQs (orange) and our \usmg (blue) sample.Green stars give median values of the ratio for different $W_{2796}$ bins.}
    \label{fig:o3bo2vswmg2}
\end{figure}

In Figure~\ref{fig:o3bo2vswmg2} we plot the ratio of \OIII/\OII\ as a function of $\mathrm{W_{2796}}$ for GOTOQs and our \usmg\ absorbers.  Green stars show the average value of this ratio for 4 different $\mathrm{W_{2796}}$ bins. We do see a tendency for the ratio to be smaller for systems with larger $\mathrm{W_{2796}}$. Only for the GOTOQs, the Spearman rank correlation coefficient is -0.094 (p-value = 0.436) while for the combined sample (GOTOQS + Our \usmg sample) the Spearman rank correlation coefficient is -0.251 (p-value = 0.021). 

\begin{comment}

\section{Parameter estimations for the nearby potential \usmg host galaxy candidates}
As mentioned in the main text, we performed the SED fitting exercise for the galaxies identified in SDSS within a maximum radius of $\rm{40^{\prime\prime}}$ to infer the photometric redshift, using the SDSS $\rm{ugriz}$ photomertic measurements. If the photometric redshift is consistent within $1\sigma$ with that of the \usmg absorption, we consider it as a candidate responsible for the \usmg absorption. Taking the redshift same as that of the \usmg absorption, we perform the SED fitting once again to obtain their properties like the stellar masses and rest frame B-band luminosities to identify if they differ by any means from the \usmg host galaxies already identified by SALT spectroscopic observations.
\end{comment}

\clearpage
\twocolumn
\newpage

\section{USMg\textsc{ii} systems from the Literature}
\label{append :usmg_litterature}

%In this section, we list all the \usmg systems, that have already been studied in the literature.
In this section, we concentrate on various \MgII\ absorption surveys that have been conducted over the years
and in particular if they contain any \usmg absorption and the physical properties of the associated host galaxy. For this exercise we consider the following surveys : (i) MAGG \citep{Dutta2020}, (ii) MEGAFLOW \citep{Schroetter2019}, (iii) \Magiicat \citep{Nielsen_2013}, (iv) SIMPLE \citep{Bouche2007, Schroetter2015}, (v) \citet{Nestor_2011}, (vi) \citet{Gauthier2013} and, (vii) \citet{Huang2021}. 

As mentioned before in the text, MAGG \citep{Dutta2020} is a IFU based galaxy-centric \MgII\ absorption systems survey containing a total of 228 galaxies around 28 quasar fields in the redshift range $0.8 \leqslant z \leqslant 1.5$ using MUSE. This survey is about 90\% complete down to the SFR  $\sim 0.1\, M_\odot yr^{-1}$ and a projected distance of about 250-350 kpc from the background quasars. 

The MEGAFLOW survey \citep{Zabl_2019}, on the other hand, aimed to study the gas flows and the galactic winds in the galaxies, is a IFU based absorber-centric study using the same instrument, MUSE. This survey primarily contains 22 quasar fields with at least three strong ($W_{2796} \geqslant 0.5${\AA}) \MgII\ absorption systems along individual line of sight and a few more relatively weak ($0.3\text{\AA} \leqslant W_{2796} \leqslant 0.5\text{\AA}$) \MgII\ absorption systems. As a result, the MEGALFOW survey contains 79 \MgII\ absorption systems ($W_{2796} \geqslant 0.3\text{\AA}$) towards 22 different quasar sightlines in the redshift range $0.51 \leqslant z \leqslant 1.45$. 

The \Magiicat\ survey is a compilation  of the \MgII\ absorption systems in the redshift range $0.07 \leqslant z \leqslant 1.12$ based on the studies previously done and it is comprised of both absorber-centric and galaxy-centric inspections.

The SIMPLE survey is again an absorber-centric survey with $W_{2796} \geqslant 2\text{\AA}$ around the redshift $\sim 1$. The sample consists of 21 such \MgII\ absorption systems, out of which, for 14 cases the host galaxies are identified within an impact parameter of 54 kpc. The galaxies are detected using the  H$\alpha$ nebular  emission line with the 3$\sigma$ detection threshold is given by $f_{H\alpha} \geqslant 1.2\times 10^{-17}\, ergs\, s^{-1}\, cm^{-2}\,(\sim 0.5 M_\odot\, yr^{-1})$.

\citet{Nestor_2011} studied a pair of \usmg systems around the redshift $z \sim 0.7$ and identified a pair of \usmg host galaxies within the impact parameters of 61 kpc for both the cases.

\citet{Gauthier2013} also studied a particular \usmg absorption which is basically a part of our \usmg sample and identified a galaxy group containing 4 galaxies being responsible for the \usmg absorption observed against the background quasar.

\citet{Huang2021} studied the gas around a total of 380 galaxies using 156 background quasars in the redshift range $0.10 \leqslant z \leqslant 0.48$ upto an impact parameter of 500 kpc.

\begin{table*}
\caption{Details of \usmg\ from the literature}
\begin{tabular}{llcccccccc}
\hline
\multicolumn{1}{c}{Sample}   & \multicolumn{1}{c}{Quasar}    & \zabs    & $\rm{W_{2796}}$ & \zem  &  D    &  $\rm{L_{[O\textsc{ii}]}}$        & SFR                          & Environment\\
\multicolumn{1}{c}{Reference}              &           &          & ($\rm{\AA}$)    &       & (kpc) &  ($\rm{10^{40}\, ergs\, s^{-1}}$) & ($\rm{M_\odot yr^{-1}}$)     &            \\
\hline
\hline
 \citet{Gauthier2013} & J220702.64-090117.8 & 0.5624   & 4.18$\pm$0.30 & 0.5623 & 55   & 3.6 $\pm$ 0.4       & 0.5 $\pm$ 0.2                    & Group \\
    &                      &                              &             & 0.5621 & 38   & 18 $\pm$ 1 	     & 2.5 $\pm$ 0.8                    &       \\
                          &                     &          &             & 0.5623 & 209  & 18 $\pm$ 2          & 2.5 $\pm$ 1.1                    &       \\
                          &                     &          &             & 0.5604 & 246  & $\le$ 9             & $\le$ 1.3                        &       \\

 \citet{Nestor_2011}  & J074707.62+305414.9 & 0.7646   & 3.60$\pm$0.11 & 0.7660 & 36   & 83.12$\pm$0.81      & 11.64 $\pm$ 3.32                 & Pair\\
                        &                     &          &             & 0.7643 & 61   & 35.82$\pm$0.82      & 5.01$\pm$ 1.44 &  \\

 \citet{Nestor_2011}  & J141751.84+011556.1 & 0.6681   & 5.47$\pm$0.18 & 0.6671 & 29   & 112.3$\pm$0.58      & 15.73$\pm$4.49                 & Pair\\
    &                                           &          &             & 0.6678 & 58   & 36.00$\pm$0.39      & 5.04$\pm$1.44                  &                      \\
 
 \citet{Bouche2007} \&   &                      &          &             &         &      &                  &                              &                      \\   
 \citet{Schroetter2015} & J014717.76+125808.8  & 1.0391 & 4.025 & 1.0389 & 17.9 & -- & 10                             & Isolated \\

                    & 2QZJ022620.4-285751    & 1.0208 & 4.515       & 1.0223 &$\le$ 2 & --                & 8                             & Isolated \\

          & J044821.8+095051.7     & 0.8392 & 3.169       & 0.8391 & 13.7 & --                & 16                              & Isolated \\

                  & J094309.66+103400.6    & 0.9967 & 3.525       & 0.9956 & 24.3 & --                & 17                              & Isolated \\

                     & J142253.31-000149.0    & 0.9097 & 3.185       & 0.9096 & 12.7 & --                & 5                               & Isolated \\

                    & J233551.10+151453.2    & 0.8557 & 3.308       & 0.8557 & 17   & --                & 2.3                             & Isolated \\

 \citet{Dutta2020}   & J162116.92-004250.8    & 1.13351 & 3.229$\pm$ 0.002 & 1.13364 & 18 & 27$\pm$0.25     & 3.78$\pm$0.01                & Isolated \\

 \citet{Schroetter2019} & J010332.30+133233.5  & 1.0481 & 3.09$\pm$ 0.17   & 1.0483  & 9.1 & --    & 2.9$^{+2.1}_{-1.4}$                   & Isolated \\

                       & J103936.67+071427.3   & 0.8193 & 3.03$\pm$0.27    & 0.8192  & 24.5 & --    & 3.2$^{+2.4}_{-1.5}$                   & Isolated \\

                       & J110735.25+175731.4   & 1.0630 & 4.05$\pm$0.35    & 1.0637  & 22.1 & --    & 2.5$^{+1.8}_{-1.2}$                  & Isolated \\

 \citet{Nielsen_2013}  & J000448.11-415728.8   & 0.8366 & 4.422$\pm$0.002  & 0.8400  & 53.8 & --    &                                      & Isolated \\
\hline
\end{tabular}
\end{table*}

\section{$\mathrm{W_{2796}}$ vs. D correlations}
\label{wvsrhofit}
\begin{figure}
\includegraphics[width=0.5\textwidth]{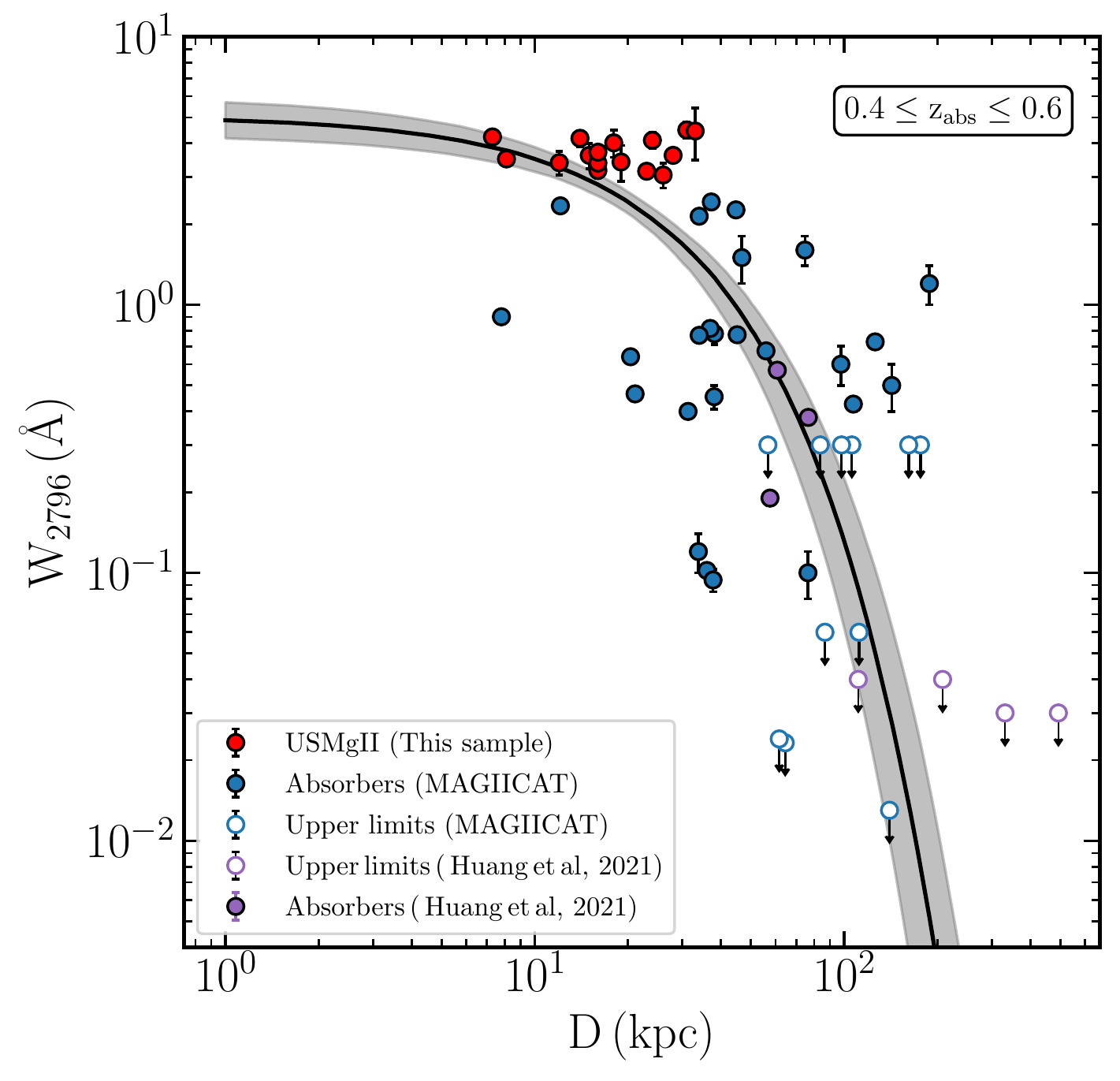}
    \caption{
    The figure shows the plot of $W_{2796}$ against the impact parameter $D$ for absorbers at $0.4\le z \le 0.6$ in different samples. The red, blue, and, purple corresponds to the \usmg, \Magiicat and, \citet{Huang2021} samples  respectively.}
    \label{fig:w_vs_rho_appendix}
\end{figure}

\begin{figure}
    \centering
    \includegraphics[width=0.495\textwidth]{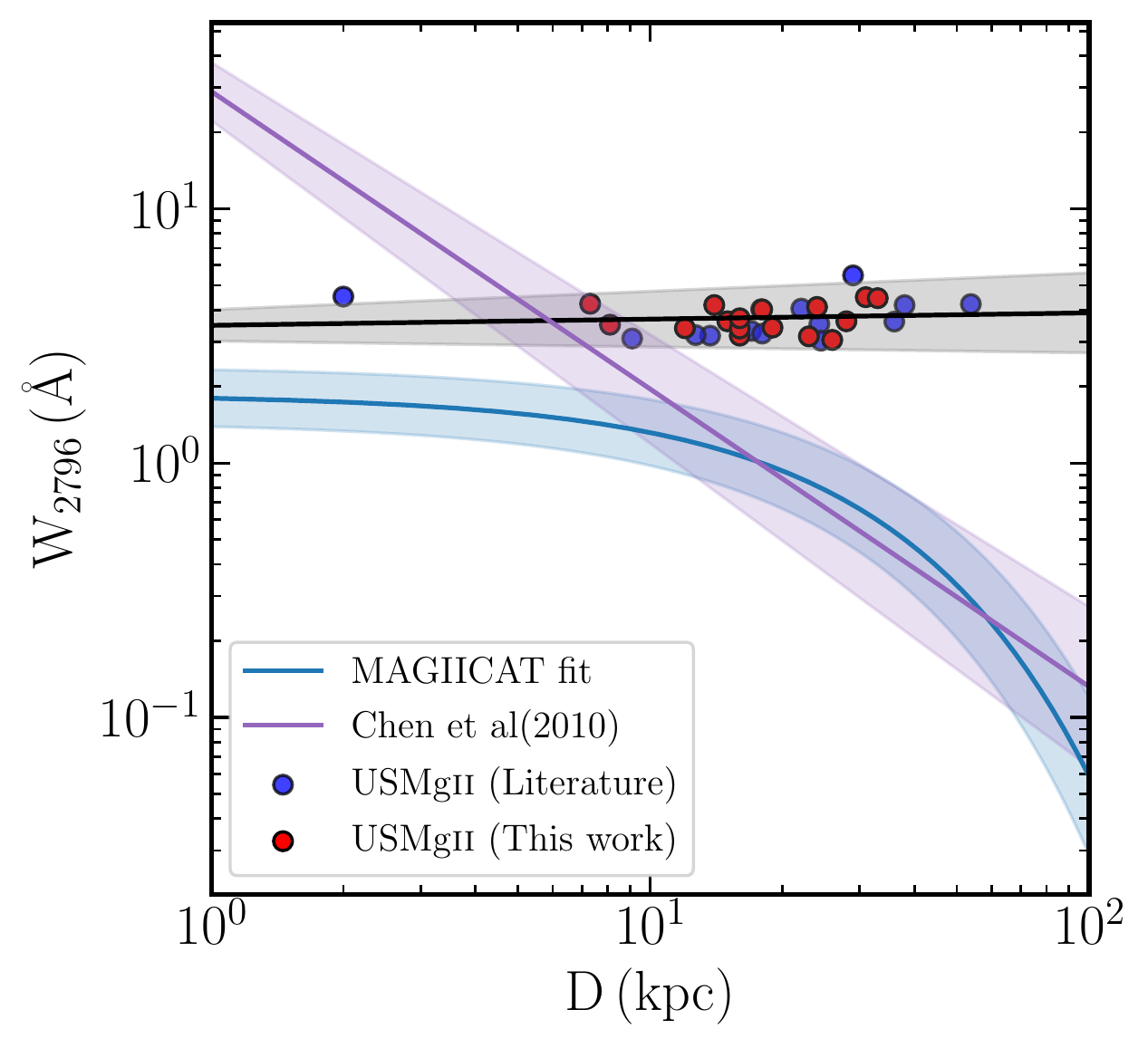}
    \caption{$\rm{W_{2796}}$ versus the impact parameter for the \usmg host galaxies from our sample (red) as well as from the literature (blue). The black line corresponds to a power law fit with the gray shaded region showing the $\rm{1\sigma}$ error to the fit.}
    \label{fig:usmg_w_vs_rho}
\end{figure}

To improve the characterisation of the impact parameter versus the $\rm{W_{2796}}$ relationship for strong \MgII\ absorbers at smaller impact parameters with our observations, we fit a log-linear model to our data along with other data available in the literature assuming a log-linear dependence between $\rm{W_{2796}}$ and $\rm{D}$ of the from
\begin{align}
\label{eqn:fitting_model}
    \rm{\log W_{2796}\,(\ang)\, =\, \alpha\, \times\, D\,(kpc) +\,  \beta}
\end{align}

Following \citet{Chen_2010}, \citet{Rubin2018} and \citet{Dutta2020}, we define the likelihood function as follows
%\begin{align}
\begin{eqnarray}
    \mathcal{L}(W) &= & \Bigg({ \displaystyle \prod_{i=1}^{n} \frac{1}{\sqrt{2\pi\sigma_i^2}} \exp{\Bigg\{ -\frac{1}{2} \Bigg[\frac{W_i - W(D_i)}{\sigma_i} \Bigg]^2 \Bigg\}}}\Bigg) \nonumber \\
    & & \times ~\Bigg{(}{\int ^{W_i}_{-\infty} \displaystyle\prod_{i=1}^{m} \frac{dW^{\prime}}{\sqrt{2\pi\sigma_i^2}} \exp{\Bigg\{ -\frac{1}{2} \Bigg[\frac{W^\prime - W(D_i)}{\sigma_i} \Bigg]^2 \Bigg\}}}\Bigg{)}
%\end{align}
\end{eqnarray}
where the total error for each measurement, $\sigma_i$ is given by,
\begin{align*}
    \sigma_i^2 = \sqrt{\sigma_c^2 + \sigma_{mi}^2}
\end{align*}
where $\sigma_c$ is the intrinsic scatter and $\sigma_{mi}$ is the measurement error in $W_{2796}$ for each measurement. To obtain the fitting parameters, $\alpha$, $\beta$, and the $\sigma_c$, we derive posterior probability distributions using the
UltraNest \footnote{\url{https://johannesbuchner.github.io/UltraNest/}} package \citep{Buchner2021}. This gives the best fitting parameters for all the \MgII\ absorbers from the literature including our \usmg absorbers as $\alpha = -0.019\pm0.002$, $\beta = 0.464\pm 0.039$ and $\sigma_c = 0.910\pm0.044$. In the left panel of Figure \ref{fig:w_vs_rho}, the best fit is shown in solid black line whereas the $1\sigma$ error on that curve is shown in gray. In the right panel of Figure \ref{fig:w_vs_rho_appendix}, we do the similar analysis, but this time restricting to only the \MgII\ absorbers in the redshift range $0.4\leqslant z_{abs} \leqslant 0.6$. This time the fitting parameters are $\alpha = -0.016\pm0.003$, $\beta = 0.702\pm 0.067$ and , $\sigma_c = 1.18\pm0.13$. Note that for both the cases, the \usmg host galaxies populate a region that is significantly different from what is expected from the fits. To understand how the $W_{2796}$ varies with impact parameter, we compiled a sample of \usmg absorbers from the literature and using our sample, and fitted the distribution with a simple model of the form Eqn. \ref{eqn:fitting_model}. The resultant fitting parameters are given by, $\alpha = 0.023\pm 0.047$ and $\beta = 0.543\pm 0.061$. The best fit and the associated $1\sigma$ error is shown in Figure \ref{fig:usmg_w_vs_rho}.
 
\section{Absolute B-band magnitude of \usmg\ host galaxies} 
\label{sec:MB_dist}
\begin{figure}
        \centering
        \includegraphics[viewport=8 8 380 390, width=0.45\textwidth,clip=true]{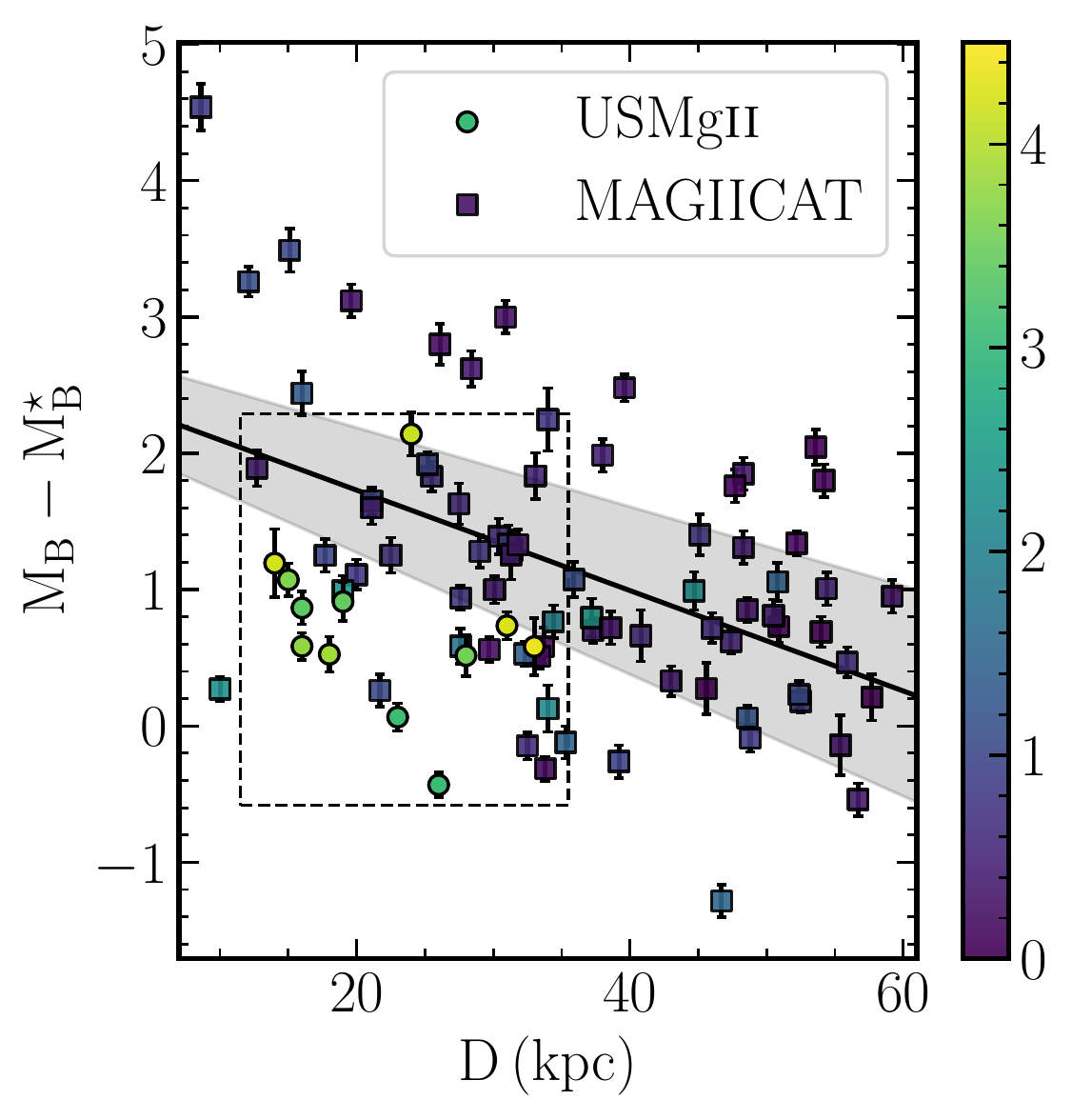}
    \caption{
     Absolute B-band magnitude of the \MgII\ host galaxies plotted against the impact parameter. The solid lines and shaded regions in both the plots are the best fit for the \Magiicat\ sample and the 1$\sigma$ error range in the fit respectively. The \usmg host galaxies span the region in the plot marked with a dashed rectangle. Clearly, for a given impact parameter, the \usmg galaxies are brighter and more massive on average compared to the \Magiicat\ galaxies.}
    \label{fig:MB_dist}
\end{figure} 

In Figure \ref{fig:MB_dist}, we have plotted the absolute B-band magnitude $\mathrm{M_B}$ of both the \usmg host galaxies and the \Magiicat host galaxies scaled by characteristic B-band magnitude ($M_B^\star$) at the galaxy redshift \citep{Faber2007}. 
Once again, we find a correlation between the impact parameters of \Magiicat\ host galaxies and the relative B band magnitude, that can be characterized by a linear fit of the form, $M_B - M_B^\star = (-0.037\pm0.008)D + (2.445\pm0.302)$. The solid black line and the grey shaded region around it in the figure corresponds to this fit and the associated $\rm{1\sigma}$ uncertainty to it.  Except for one case, the \usmg\ host galaxies tend to be brighter at a given impact parameter (and beyond the $1\sigma$ region) compared to the \Magiicat\ host galaxies.

% Don't change these lines
\bsp	% typesetting comment
\label{lastpage}
\end{document}